\documentclass[12pt, draftclsnofoot, onecolumn]{IEEEtran}
\usepackage{graphicx, wrapfig, comment}
\usepackage[noadjust]{cite}
\usepackage{amssymb}
\usepackage[noend]{algpseudocode}
\usepackage{amsmath,amssymb, mathabx, algorithm}
\usepackage{subcaption}
\usepackage{fancyvrb, dsfont}
\usepackage{hyperref}
\usepackage{dsfont, multicol}
\usepackage[all]{hypcap}
\usepackage{lineno}
\usepackage{setspace}
\usepackage{placeins}
\usepackage[font=small,labelfont=bf]{caption}
\usepackage[dvipsnames]{xcolor}

\newlength{\bulletwidth}

\newcommand{\qed}{\nobreak \ifvmode \relax \else
      \ifdim\lastskip<1.5em \hskip-\lastskip
      \hskip1.5em plus0em minus0.5em \fi \nobreak
      \vrule height0.75em width0.5em depth0.25em\fi}

\IEEEoverridecommandlockouts
\newtheorem{theorem}{Theorem}

\newcommand{\upperRomannumeral}[1]{\uppercase\expandafter{\romannumeral#1}}
\begin{document}

\newif\ifappendix
\appendixtrue

\title{Real-time Cooperative Communication for \\
\vspace{-10pt}Automation over Wireless}

\author{\normalsize{
    \IEEEauthorblockN{Vasuki~Narasimha~Swamy\IEEEauthorrefmark{1}, Sahaana~Suri\IEEEauthorrefmark{3}, Paul~Rigge\IEEEauthorrefmark{1}, Matthew Weiner\IEEEauthorrefmark{4}, \\Gireeja~Ranade\IEEEauthorrefmark{1}\IEEEauthorrefmark{2}, Anant~Sahai\IEEEauthorrefmark{1}, Borivoje~Nikoli\'{c}\IEEEauthorrefmark{1}}\\
    \IEEEauthorblockA{\IEEEauthorrefmark{1}University of California, Berkeley, CA, USA}
    \IEEEauthorblockA{\IEEEauthorrefmark{2}Microsoft Research, Redmond, WA, USA}\\
    \IEEEauthorblockA{\IEEEauthorrefmark{3}Stanford University, CA, USA}
    \IEEEauthorblockA{\IEEEauthorrefmark{4}Cohere Technologies, CA, USA}
    }
    \vspace{-12mm}
}

\maketitle

\begin{abstract}

High-performance industrial automation systems rely on tens of simultaneously active sensors and actuators and have stringent communication latency and reliability requirements. Current wireless technologies like WiFi, Bluetooth, and LTE are unable to meet these requirements, forcing the use of wired communication in industrial control systems.
This paper introduces a wireless communication protocol that capitalizes on multiuser diversity and cooperative communication to achieve the ultra-reliability with a low-latency constraint.


Our protocol is analyzed using the communication-theoretic delay-limited-capacity framework and compared to baseline schemes that primarily exploit frequency diversity. For a scenario inspired by an industrial printing application with thirty nodes in the control loop, $20$B messages transmitted between pairs of nodes and a cycle time of $2$ ms, an idealized protocol can achieve a cycle failure probability (probability that any packet in a cycle is not successfully delivered) lower than $10^{-9}$ with nominal SNR below $5$ dB in a $20$MHz wide channel.

\end{abstract}
\vspace{-3mm}

\bstctlcite{BIBcontrol}
\VerbatimFootnotes

\vspace{-3mm}
\section{Introduction}
\label{sec:intro}
High-speed ultra-reliable wireless communication networks are critical for developing near-real-time machine-to-machine networks and applications such as industrial automation, immersive virtual reality (VR) and the ``tactile internet." An interactive Internet-of-Things (IoT) will require require both ubiquitous sensing and simultaneous actuation of numerous connected devices. The latency requirements on the control loop 
for these applications will be in the low tens of milliseconds, with reliability requirements of one-in-a-million errors~\cite{FettweisTactile}. These specifications mirror those of industrial automation today~\cite{Weiner, SERCOSreliability}. However, these are unattainable using WiFi, Bluetooth and Long-Term Evolution (LTE),  and as a result industrial automation relies heavily on wired connections. 

This paper\footnote{This paper expands upon a conference version \cite{swamy2015cooperative} that contained early forms of these results.} introduces  a communication protocol framework for  industrial control and IoT applications that is designed to meet stringent Quality-of-Service (QoS) requirements. The protocol relies on multi-user diversity to achieve reliability 
without relying on time or frequency diversity created by natural motion, multipath or frequency selectivity. Futhermore, combining this with simultaneous relaying allows strict latency requirements to be met at reasonable signal-to-noise ratios (SNR).
In Section~\ref{sec:related} we first review 
communication for industrial control, then discuss cooperative communication and wireless diversity techniques, and finally place this work within the context of 5G research and contrast it with the complementary research on the co-design of control and communication systems. During this review, we also argue why time diversity cannot be reliably harnessed for low-latency communication, why frequency diversity cannot be counted on when ultra-reliability is desired, and why randomized carrier sense multiple access with collision avoidance (CSMA/CA) style approaches cannot be used reliably. Then, Section~\ref{sec:protocol} describes our multi-user-diversity-based protocol framework. Section~\ref{sec:results} compares the performance of our protocol to hypothetical frequency-diversity-based schemes as well as to schemes that do not leverage simultaneous transmissions. Finally, Section~\ref{sec:opt} examines the impact of fine-tuning the protocol parameters and explores duty-cycling to reduce power consumption and what this suggests for implementation.
 \ifappendix
All the formulas used to generate the plots are included in the Appendix.
 \else
All the formulas used to generate the plots are included in an online techreport \cite{fullversion}.
\fi

\subsection{Main results}
\begin{wrapfigure}{r}{0.4\textwidth}
\vspace{-30pt}
\begin{center}
\includegraphics[width = 0.38\textwidth]{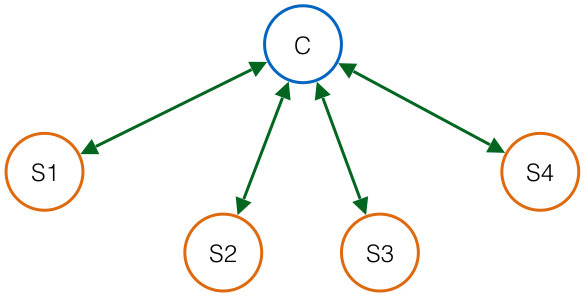}
\vspace{-10pt}
\caption{Star message flow topology}
\label{fig:star}
\end{center}
\vspace{-40pt}
\end{wrapfigure}
\begin{wrapfigure}{r}{0.4\textwidth}
\vspace{-30pt}
\begin{center}
\includegraphics[width = 0.38\textwidth]{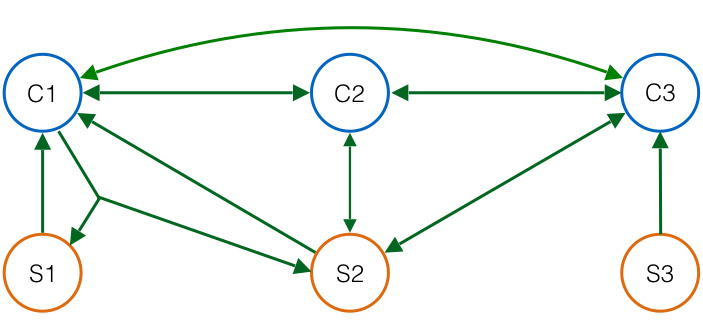}
\caption{An example of a generic non-star message flow topology. Notice that one of the message streams originating at C1 has two subscribers: S1 and S2. }
\label{fig:non-star}
\end{center}
\vspace{-30pt}
\end{wrapfigure}

The protocols in this paper (``Occupy CoW\footnote{OCCUPYCOW is an acronym for ``Optimizing Cooperative Communication for Ultra-reliable Protocols Yoking Control Onto Wireless.'' The name also evokes the similarity between our scheme and the ``human microphone'' implemented by demonstrators during the ``Occupy Wall Street'' movement.},'') target a local wireless domain where nominally all nodes are in range, but fading might cause a pair of nodes to be unable to hear each other. The traffic patterns (what we deem the ``information topology'') we are interested in consist of steady streams of messages, each originating at possibly different nodes within the network, and each stream subscribed to by some (possibly different) subset of nodes within the network. Within a short period of time, deemed a ``cycle time,'' every stream needs to deliver one packet reliably to its subscribers. The information topology can be arbitrary -- something naturally centralized like a star topology as shown in Fig.~\ref{fig:star} (e.g.~with a central controller talking to many sensor/actuators collecting streams of measurements and sending streams of commands) or something more generic as in Fig.~\ref{fig:non-star}.

The potential for bad fading is what frustrates the simple strategy of just giving each message stream its own time-slot. For example in Fig.~\ref{fig:star}, if the communication link from `C' to `S3' were badly faded, we would have a failure. We combat this by employing cooperative communication -- if any of the other nodes can hear C, it can relay the relevant message streams. As reviewed in Section~\ref{subsec:multi-user}, cooperative communication has been well studied in the wireless literature. In this paper, we specifically adapt it to the ultra-reliability low-latency regime.

\begin{wrapfigure}{r}{0.5\textwidth}
  \vspace{-30pt}
  \begin{center}
    \includegraphics[width=0.48\textwidth]{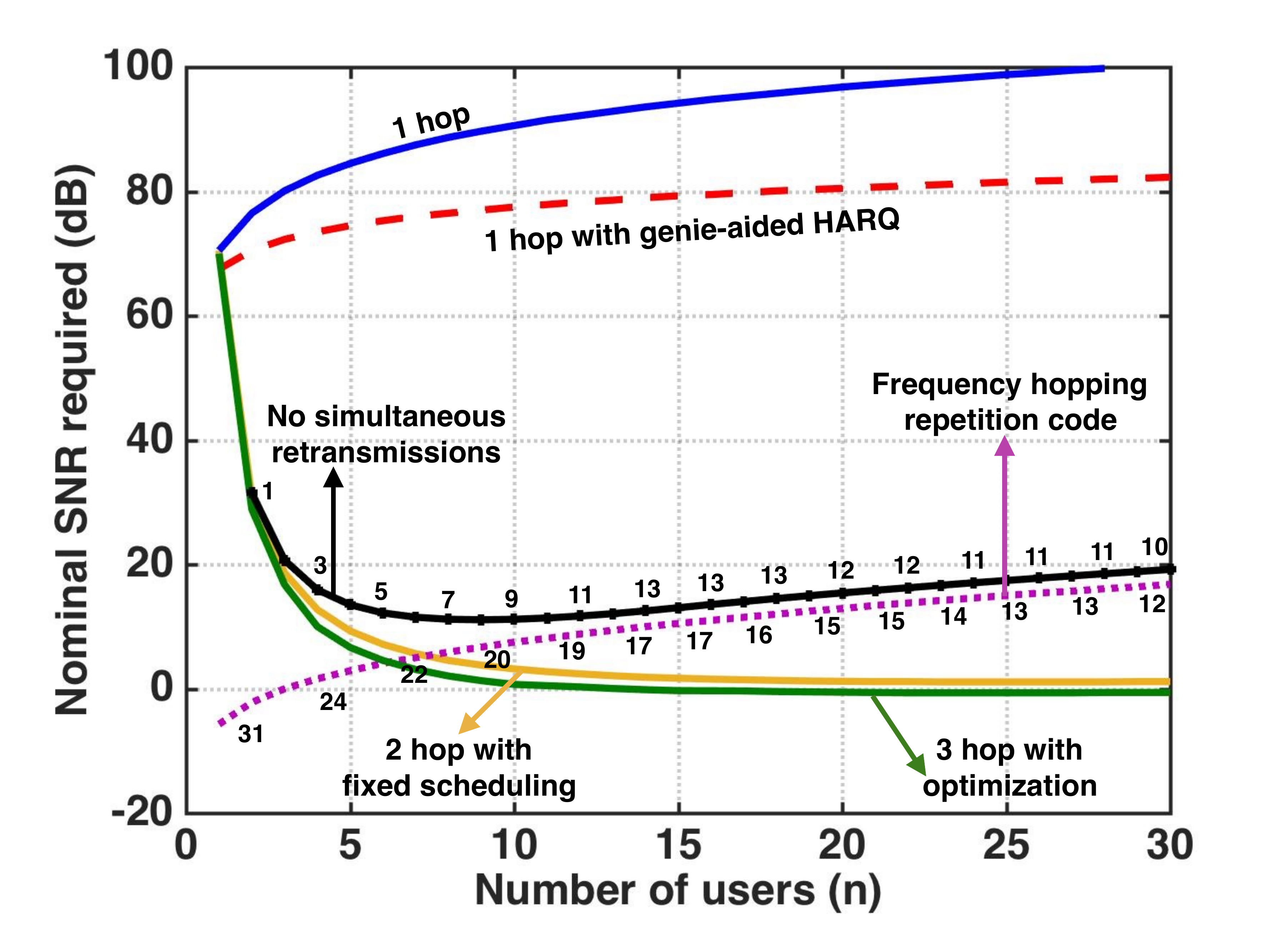}
  \end{center}
  \vspace{-20pt}
  \caption{{The performance of Occupy CoW for a star information topology compared with reference schemes for $m=160$ bit messages, $n=30$ active nodes, and a $2$ms cycle time, aiming at $10^{-9}$ probability of failure for a $20$MHz channel. The numbers next to the frequency-hopping scheme show the frequency diversity needed and those  next to the non-simultaneous retransmission scheme show the optimal number of relays per message stream.}}
  \label{fig:hockeystick}
  \vspace{-20pt}
\end{wrapfigure}


We define the cycle failure probability as the probability that any packet transmitted during the cycle was unsuccessful.
Our key findings are shown in Fig.~\ref{fig:hockeystick}, where the minimum SNR required to achieve a cycle failure probability of $10^{-9}$ is used to compare different protocols as the size of the network grows. We will revisit this in detail in Sec.~\ref{sec:res}. We see that a one-hop scheme, one that does not use cooperative communication (the top blue line), requires an unreasonably high nominal SNR. Even idealized hybrid automatic repeat request (HARQ --- the dashed red line) cannot eliminate the need to use high power to overcome a one-in-a-billion fading event. Harnessing diversity is required to do better. The purple dotted curve 
shows what idealized frequency hopping could achieve, assuming that nature has guaranteed sufficient frequency diversity (the number of independently faded subchannels required is labeled along the curve). For large enough networks, the black line in Fig.~\ref{fig:hockeystick} shows that only slightly worse performance is available by using cooperative communication where a subset of the nodes (the number of active relays is marked on the curve) take turns to relay messages that they have heard. This does not count on the multipath environment guaranteeing a lot of frequency diversity and instead harnesses the spatial diversity that independently located nodes bring. A further 20dB of gain is possible by moving to the OccupyCoW protocol described in this paper that combines relaying with simultaneous transmission of messages by relays, and this is what is shown by the yellow and green curves. The difference between the yellow and green curves is what can be gained by optimizing the protocol parameters, and is not as significant by comparison.

\vspace{-3mm}
\section{Related Work}
\label{sec:related}

\subsection{Industrial automation}
\label{subsec:indus_control}
Industrial control systems have traditionally been wired. Following trends in networking more broadly, proprietary point-to-point wired systems were replaced by \emph{fieldbus} systems such as SERCOS, PROFIBUS and WorldFIP that provide reliable real-time communication \cite{richardzurawski, SunitKumar, Willig05wirelesstechnology}. The desire to move to wireless communications to reduce bulk and installation costs \cite{ZandCDH12} led to an examination of wireless extensions of fieldbus systems~\cite{WilligPROFIBUS, MorelWirelessFIP}.
Unfortunately, these do not work in low-latency high-reliability settings since these wireless fieldbuses are largely derivative of wireless designs for non-critical consumer applications and incorporate features such as CSMA or ALOHA that can induce unbounded delays~\cite{CenaRealTime}. The IWLAN standard, which is based on a combination of PROFINET with 802.11n WLAN, attempts to resolve this issue by adding proprietary scheduled polling called iPCF (industrial Point Coordination Function) \cite{siemensIWLAN, trsek2016related, frotzscher2014requirements, lam2014novel}. To deal with a deep fade, IWLAN has to rely on handing over the faded node to a redundant access point, and this handover is sped up by using proprietary industrial-automation oriented enhancements to the 802.11n protocol. Even with these enhancements, handover can only occur at the time-scale of tens of milliseconds \cite{siemensIWLAN}.

Ideas from Wireless Sensor Networks (WSNs)~\cite{Akyildiz02, BoniventoIndustrialWireless, Yigitel} that provide high-reliability monitoring also cannot be easily adapted for very tight control loops because they inherently tolerate large latencies \cite{Willig07recentand}. The Wireless Interface for Sensors and Actuators (WISA)~\cite{ScheibleWISA} attempts to meet stringent real-time requirements, but the reliability of WISA ($\approx 10^{-4}$) does not  meet the desired specifications~\cite{GungorIndustrialWireless}. ZigBee PRO~\cite{ZigBeePRO} also fails to deliver high enough reliability~\cite{KimWirelessHART}.
Both ISA 100~\cite{ISA100} and WirelessHART~\cite{WHART} provide secure and reliable communication, but cannot meet the latency bounds as \emph{each} packet is $10$ms long. The median latency for a successful delivery is in the order of $100$ms and the protocols are heavily optimized for power consumption.
On the other hand, we are interested in packet sizes on the order of $10 \mu$s with maximum delivery latency of $2$ms~\cite{KimWirelessHART, akerberg2011future}.
There is a need for a fundamentally faster and more reliable protocol framework if we want to have a drop-in replacement for existing wired fieldbuses like SERCOS \upperRomannumeral{3}, which simultaneously provide a reliability of $10^{-8}$ {\em and} latency of $1$ ms when communicating among tens of concurrently active nodes.
\vspace{-10pt}
\subsection{Cooperative communication and multi-user diversity}
\label{subsec:multi-user}
The secret to getting reliability in wireless communication is to harness diversity~\cite{ViswanathTseBook}. Highly-reliable WSNs use techniques like channel hopping and  contention-based medium access control (MAC) to harvest time and frequency diversity, and multi-path routing as an indirect way to harvest spatial diversity~\cite{ZandCDH12}.
Unfortunately, low-latency applications like ours cannot use time diversity since the cycle times of single-digit milliseconds could very well be shorter than channel coherence times of tens of milliseconds. Techniques like Forward Error Correction (FEC) and Automatic Repeat Request (ARQ) also do not provide much advantage in the face of fading~\cite{Willig08}. Later in this paper, we demonstrate that frequency-diversity based techniques also fall short, especially when the required throughput pushes us to increase spectral efficiency. Even beyond the issue of poor performance, there is the issue of trust --- exploiting frequency diversity requires us to trust that nature will provide enough multipaths with a large enough delay spread to actually create frequency diversity~\cite{ViswanathTseBook}. Consequently, our protocol leverages spatial diversity instead.


The size of the networks targeted in this paper is moderate (10 - 100 nodes active at once). Therefore, there is an abundance of antennas in the system and we can harvest some resulting diversity.
Multi-antenna diversity is mainly of two types: a) sender diversity where multiple antennas transmit the same message through independent channels and b) receiver diversity where multiple receive antennas harvest  copies of the same signal received via independent channels.
Researchers have studied these techniques in great detail; so our treatment here is limited. Cooperation among distributed antennas can provide full sender-diversity without the need for physical arrays~\cite{Laneman}. Even with a noisy inter-user channel, multi-user cooperation increases capacity and leads to achievable rates that are robust to channel variations~\cite{UserDiversity}. The prior works in cooperative communication tend to focus on the asymptotic regimes of high SNR. By contrast, we are interested in low - moderate SNR regimes (around $10$ dB).

Multi-antenna techniques have been widely implemented in commercial wireless protocols like IEEE 802.11.
Sender-diversity harvesting techniques using relaying coupled with a time division multiple access (TDMA) based scheme have been explored for industrial control~\cite{Willig08, GirsRelayingLuby}.
Unfortunately, as we discuss later and can see in Fig.~\ref{fig:hockeystick}, strict TDMA for relays can scale badly with network size since relaying for one message consumes many slots to get enough diversity to attain high reliability. To scale better with network size, our protocol uses simultaneous transmission by many relays, using some distributed space-time codes (DSTCs) such as those in~\cite{Oggier, PVK, WuCDDHARQ}, so that each receiver can harvest a large diversity gain. 
While we do not discuss the specifics of space-time code implementation, recent work by Rahul et al.~\cite{Katabi} demonstrates that it is possible to implement schemes that harvest sender diversity using concurrent transmissions.

\vspace{-10pt}
\subsection{Recent developments in 5G protocols}
\label{subsec:5g_disc}
Latency and reliability have risen in importance as $5$G wireless is discussed, taking their place alongside a focus on increasing capacity and energy efficiency while also using mmWave frequencies~\cite{what_will_5g_be}. One important driver for very short round-trip time (RTT) latencies, of the order of $1$ms, is to support tactile feedback to wireless users, enabling immersive VR applications~\cite{2020_beyond_4G}.

Levanen et al.~\cite{levanen} concentrate on the proposed 5GETLA radio interface and show that latencies below $1$ms for payloads of size $50$kb are achievable provided a bandwidth of 100MHz is available. Though the targeted latency is of the same order as required by industrial control, they do not consider reliability guarantees or retransmissions.
A discussion of the feasibility, requirements, and design challenges of an OFDM based $5$G radio interface suitable for mission-critical machine type communication (MTC) 
concluded that multiple receive antennas are critical for interference mitigation~\cite{yilmaz2015analysis}. In similar spirit, coverage and capacity aspects concerning both noise-limited and interference-limited operations for MTC were considered in~\cite{brahmi2015deployment}. Various PHY and MAC layer solutions for mMTC (massive MTC) and uMTC (ultra-reliable MTC) are discussed in~\cite{bockelmann2016massive} where they conclude that higher-layer considerations ensure lean signaling by enabling longer sleep cycles and other techniques.
Efficient communication of short packets in the information-theoretic context was discussed in~\cite{DurisiKP15, Popovski14a,levanen} where the key insight is that when packets are short, the resources needed for metadata transmission should be considered. In this paper, we do not consider either metadata or interference and plan to address these in later work.

\vspace{-10pt}
\subsection{Control and communication co-design}
This paper's approach to enabling wireless industrial automation is to maximize the reliability of communication while simultaneously reducing latency. For completeness, we mention that a second approach towards achieving successful wireless industrial automation would be to adapt control algorithms to (the reliability and latency guarantees provided by) wireless communication and to co-design the two modules.
Fundamental limits for control and estimation of systems over both noiseless rate-limited channels~\cite{wong1997systems,tatikonda} and noisy channels~\cite{anytime} have been established.
A series of works~\cite{sinopoli2004kalman, schenato2007foundations, parkIntermittent} established the limits of control and estimation over packet dropping networks and it was recently shown that control and communication co-design could provide unbounded performance gains in such settings~\cite{ramnarayan2014side}.

The literature on adapting control to wireless communication has generally focussed on leveraging the optimization paradigms of control-theoretic synthesis.
Works like \cite{xiao2003joint,xiao2005joint,gupta2009data} combine data rate and quantization with performance optimization and dealing with packet drops.
A holistic view of network parameters including the placement of controller functionality has been studied in~\cite{park2011wireless,robinson2008optimizing}.
Finally, there are even more completely integrated approaches like the wireless control network idea proposed in Pajic et al.~\cite{pajic2011wireless} wherein the wireless network itself is modeled as the controller with the network topology providing an implementation constraint and the unreliability of the links viewed through the lens of robust control techniques~\cite{elia2005remote}. 

Our paper focuses exclusively on improving communication. This has two motivations. First, it follows the principle of layering as it allows unmodified control laws (which might not have been developed using any particular synthesis methodology or even stated performance criteria) to operate with a new communication layer~\cite{kawadia2005cautionary}. Second, it establishes a baseline upon which we can study the gains achievable through co-design, which warrants further investigation.

\vspace{-3mm}
\section{Protocol Design}
\label{sec:protocol}
The Occupy CoW protocol exploits multi-user diversity by using simultaneous relaying (i.e.~using diversity-oriented distributed space-time codes (DSTC)) to enable low-latency ultra-reliable communication between a set of nodes (say $n$ of them) within a ``cycle'' of length $T$. As described in the introduction, we assume that all nodes are in-range of each other and have a given nominal SNR. However, a bad fading event can cause transmissions to fail.
One could in principle wait for the bad fade to turn into a good fade. However, due to the coherence times being longer than the cycle time, channels do not change quickly enough. Therefore to reliably (meaning with low probability of failure) deliver packets, multiple paths to the destination need to be found.

The protocol is information-flow-centric rather than node-centric. There is an information topology (i.e.~a list of streams having sources and subscribers; where each stream generates one fresh fixed-size data packet at the start of each cycle that must reliably reach all its subscribers during that cycle) that is known to everyone in advance. 
Each packet gets dedicated time slots for transmission as well as relay retransmissions. We have two main versions of the protocol, as summarized in Algorithm~\ref{protocol-description}:
\begin{itemize}
\item Fixed schedule: Once an initial schedule (or order) of packets has been determined, all packets are transmitted once. Then, in the same order, all nodes that have the corresponding packet simultaneously retransmit it. This is a two hop version of the protocol. For three hops, all nodes that now have the corresponding packet simultaneously retransmit it again.
\end{itemize}
The inefficiency in the fixed schedule is that it dedicates slots to retransmit packets that were already successful. This forces all the retransmission slots to be shorter than they could have been. To avoid this, it seems like a good idea to adapt the retransmission schedule to concentrate only on the messages that need relaying. However, to achieve this, all the potential relays need to agree on which messages need to be retransmitted. This requires the reliable dissemination of scheduling information throughout the network, and acknowledgments (ACKs) from all of the network nodes are required before the retransmission of data packets.
\begin{itemize}
\item Adaptive schedule: Once an initial schedule (or order) of messages has been determined, all messages are transmitted once. All nodes then take turns broadcasting their own ACK packets where they indicate the messages that they have heard. These ACK packets are rebroadcast using the Fixed Schedule scheme above so that all nodes' ACK information gets reliably disseminated to everyone. Once all the ACKs are known, the data retransmission schedule is recomputed to include only those messages that have not yet reached all their subscribers and each data packet is in turn rebroadcast simultaneously by the nodes that have it. The data rates for retransmissions adapt so that the full cycle time is used.
\end{itemize}

{\setstretch{1.0}
\begin{algorithm}
\caption{Occupy CoW protocol}
\label{protocol-description}
\begin{algorithmic}[1]
\Procedure{Determine Schedule}{}
\State $\mathcal{S} \gets$ set of all nodes
\State $\mathcal{G} \gets$ ordered key-value pair table. Messages are the keys and lists of their subscribers are the values. Messages are transmitted as per their order in the table.
\State $scheme \gets$ fixed or adaptive
\State $hops \gets$ 2 or 3
\If {$scheme$ = fixed}
\Procedure{Fixed Schedule}{}
\State Phase 1:
\For{packet $g \in \mathcal{G}$}
{$g$ is broadcast in $g$'s pre-assigned slot. All other nodes listen.}
\EndFor

\State Phase 2:
\For{packet $g \in \mathcal{G}$}
{All nodes with $g$ simultaneously broadcast during $g$'s pre-assigned slot using a diversity-oriented DSTC. All others listen.}
\EndFor

\If {$hops$ = 3}
\State Phase 3:
\For{packet $g \in \mathcal{G}$}
{All nodes with $g$ simultaneously broadcast during $g$'s pre-assigned slot using a diversity-oriented DSTC. Interested subscribers listen.}
\EndFor
\EndIf
\EndProcedure
\Else
\Procedure{Adaptive Schedule}{}
\State Message Phase 1:
\For{packet $g \in \mathcal{G}$}
{$g$ is broadcast in $g$'s allocated slot. All other nodes listen.}
\EndFor

\For{node $s \in \mathcal{S}$}
{$a_s \gets$ ACK packet indicating the $g \in \mathcal{G}$ that $s$ has received.}
\EndFor

\State Scheduling Phase 1:
\For{node $s \in \mathcal{S}$}
{$s$ broadcasts $a_s$ in its pre-assigned slot. All others listen.}
\EndFor

\State Scheduling Phase 2:
\For{$s \in \mathcal{S}$}
{All nodes with ACK packet $a_s$ simultaneously retransmit it in its pre-assigned slot using a diversity-oriented DSTC. All others listen.}
\EndFor

\If {$hops$ = 3}
\State Scheduling Phase 3:
\For{$s \in \mathcal{S}$}
{All nodes with ACK packet $a_s$ simultaneously retransmit it in its pre-assigned slot using a diversity-oriented DSTC. All others listen.}
\EndFor
\EndIf

\State $\mathcal{G}_{s} \gets \emptyset$. This is the new adaptive schedule to be populated and is a subset of $\mathcal{G}$.

\For{packet $g \in \mathcal{G}$}
{If $g$ has not reached all its subscribers (as indicated by various $a_s$) then $\mathcal{G}_{s} \gets \mathcal{G}_{s} \bigcup g$.}
\EndFor

\State Message Phase 2:
\For{packet $g \in \mathcal{G}_s$}
{All nodes with $g$ simultaneously broadcast using a diversity-oriented DSTC during $g$'s slot according to the new schedule. All others listen.}
\EndFor

\If {$hops$ = 3}
\State Message Phase 3:
\For{packet $g \in \mathcal{G}_s$}
{All nodes with $g$ simultaneously broadcast using a diversity-oriented DSTC during $g$'s slot according to the new schedule. All interested subscribers listen.}
\EndFor
\EndIf
\EndProcedure
\EndIf
\EndProcedure
\end{algorithmic}
\end{algorithm}}

The protocol itself is information-topology independent, but star-topology examples will be used to explain fixed schedules in Sec.~\ref{sec:fixed-sched} and adaptive schedules in Sec.~\ref{sec:adaptive-sched}.

\subsection{Resource assumptions}
\label{sec:assumptions}
We make a few assumptions regarding the hardware and environment to focus on the conceptual framework of the protocol.
\begin{itemize}
\item All the nodes share a universal addressing scheme and order. Each node knows the initial order of messages being transmitted so there is no confusion about what is transmitted next. 
\item All nodes are half-duplex but can switch instantly from transmit mode to receive mode.
\item Clocks on each of the nodes are perfectly synchronized in both time and frequency. This could be achieved by adapting techniques from~\cite{HuangTiming}. Thus we can schedule time slots for specific packets and nodes can simultaneously transmit if so desired.
\item 
We assume that if $k$ relays simultaneously (with consciously introduced jitter\footnote{This jitter or explicit delay transforms spatial diversity into frequency-diversity as shown in \cite{WuCDDHARQ}. This scheme does not require a relay to know who else is relaying alongside it, and having a long enough OFDM symbol suffices.} or some other DSTC) transmit, then all receivers can extract signal diversity $k$ without having to know in advance who is relaying or how many simultaneous transmissions they are receiving.
\item (Only for adaptive-schedules) Nodes are capable of decoding variable-rate transmissions~\cite{VerduVariableLength}.
\end{itemize}

\subsection{Fixed Schedule Example}
\label{sec:fixed-sched}
\begin{figure}
\vspace{-10pt}
\centering
\includegraphics[width = \textwidth]{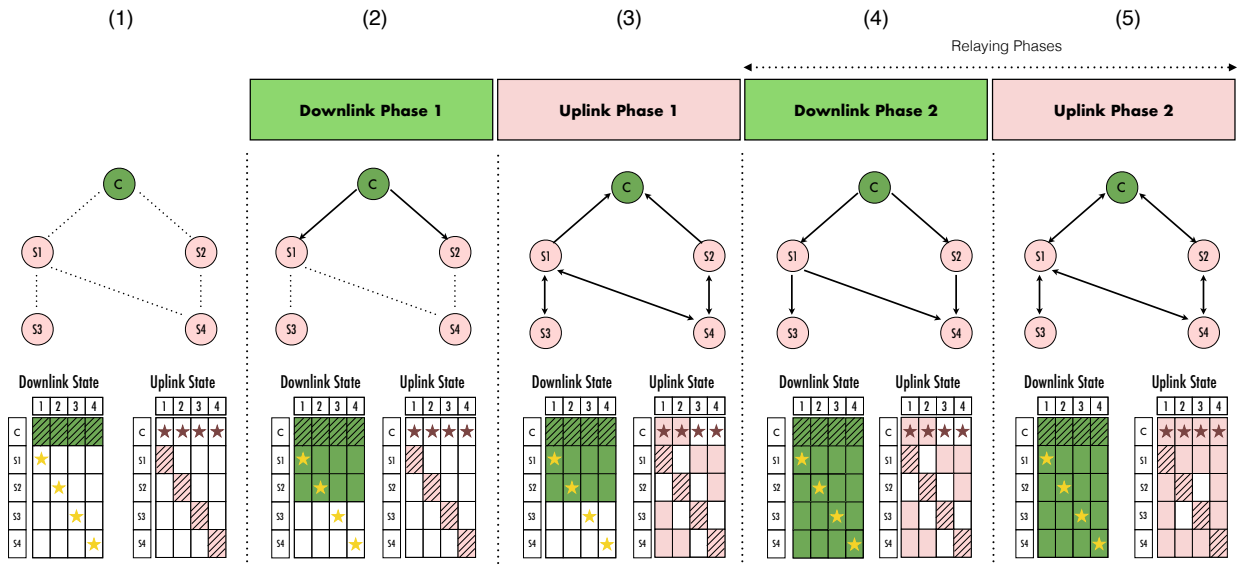}
\caption{Toy example with one controller and $4$ nodes. The graph illustrates which links are active during that phase. The downlink and uplink tables at each stage represent the information each node has at the end of that phase. Explained in detail in Sec.~\ref{sec:fixed-sched}.}
\label{fig:toy-example}
\vspace{-20pt}
\end{figure}
For simplicity we focus on a simple star information topology as in Fig.~\ref{fig:star}. A central controller (C) that must transmit $m$ distinct bits (downlink messages) to each of the four nodes. Each of nodes (S1-S4) must transmit $m$ distinct bits (uplink messages) to the controller. We define a cycle failure to be the event that at least one node fails to receive its downlink message, the controller fails to receive an uplink message from any of the nodes, or both. While there is no qualitative or quantitative difference between downlink and uplink packets, we use this terminology for ease of exposition. 

We will now run through a fixed schedule two-hop version of Occupy CoW on this network using Fig.~\ref{fig:toy-example}. Column (1) in the figure has two components in it --- the top figure shows the available communication links depicted by the dotted lines (the rest are faded out) and the bottom comprises two tables for the downlink and uplink information of each node. The table on the left is the downlink information state of each node (including the controller) and the table on the right is the uplink information state. Striped cells indicate message origins and starred cells indicate message destinations. For instance, since S1 is interested in downlink message 1 from the controller, the corresponding box in the downlink table is starred, and similarly for S2-S4. On the uplink, the controller is interested in the uplink packets from nodes S1 to S4, but these nodes do not care about each others packets, leading to stars only in the top row.  

Columns (2)-(4) indicate phases of the protocol. The graph shows directional links on which information is actively transmitted during the phase. As the nodes successfully hear different packets, the cells in the table are colored in. Initially, the cells corresponding to the controller's downlink state and S1 to S4's own uplink states are filled.

\subsubsection*{Phase \upperRomannumeral{1}}
In Phase \upperRomannumeral{1} each node transmits its messages in a predetermined order. In the schedule shown here, the controller first transmits the downlink packets for S1 through S4 in that order, and then S1 to S4 take turns transmitting their uplink packets. For illustration, we divide this Phase \upperRomannumeral{1} into two parts: Downlink Phase \upperRomannumeral{1} (Column (2)) and Uplink Phase \upperRomannumeral{1} (Column (3)). Since the controller can only reach S1 and S2 the links from C $\to$ S1 and C $\to$ S2 are active (bold directed lines) and the rest remain inactive. These two nodes hear downlink messages for all four nodes as shown in the Downlink table. S1 and S2 are thus possible relays for S3 and S4's downlink messages. Then, in  Uplink Phase \upperRomannumeral{1}, S1 transmits its message and C, S3 and S4 hear the message. When S2 transmits, C and S4 are able to hear the message. When S3 transmits only S1 is able to hear the message. When S4 transmits S1 and S2 are able to hear the message. The graph illustrates these links and cells corresponding to these received messages are filled. 

\subsubsection*{Phase \upperRomannumeral{2}}
In this phase (also divided into Downlink and Uplink), nodes \emph{simultaneously} transmit packets to help other nodes\footnote{Section~\ref{sec:round-robin} discusses a version of the protocol where relays take turns instead of simultaneously relaying. 
}. 

In Downlink Phase \upperRomannumeral{2} (column (4)), the first message to be relayed is the downlink packet of S1. Since C, S1 and S2 have this packet, they broadcast it using a DSTC. S3 and S4 can now decode S1's downlink packet. Similarly, S2's downlink packet is decoded by S3 and S4. 

The key point here is that S3's downlink packet is retransmitted by both C and S1 using a DSTC. Since S1 has a good channel to S3 (and S4) they both can decode this. The same ensues for S4's downlink packet. At this stage, all nodes (S1 to S4) have received their downlink packets.

The final phase is Uplink Phase \upperRomannumeral{2} (column (5)). 
S1's uplink packet is retransmitted by C, S1, S3 and S4 simultaneously using a DSTC and S2 is able to decode it. 
A similar procedure happens for S2's uplink packet and S1 and C are able to decode it. 

Again, S1 helps to transmit S3's uplink packet by simultaneously  broadcasting it along with S3. C and S4 are able to decode the message. A similar procedure happens for S4's packet. Once S4's uplink packet has been transmitted, the round is complete. In this instance, all messages have reached their subscribers since all the starred cells are filled. Notice however that S2 and S3 never hear each others' uplink messages.


\subsection{Adaptive Schedule Example}
\label{sec:adaptive-sched}
\begin{wrapfigure}{r}{0.37\textwidth}
  \vspace{-30pt}
  \begin{center}
    \includegraphics[width=0.35\textwidth]{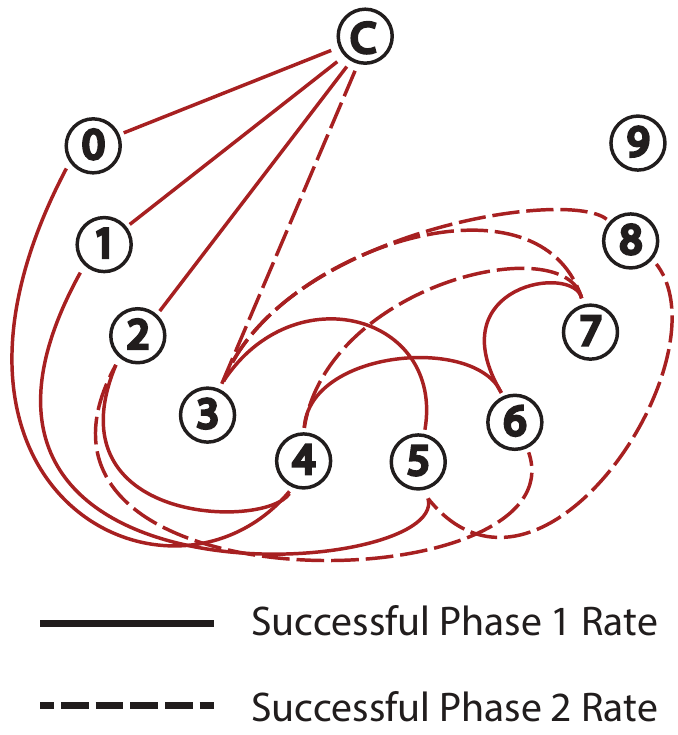}
  \end{center}
  \vspace{-20pt}
  \caption{{Network realization of the adaptive schedule example. The links that are present under different rates are depicted.}}
  \label{fig:layout}
  \vspace{-20pt}
\end{wrapfigure}
We again consider a star information topology for this example. There is one controller and $10$ nodes (S0 - S9). The controller has $m$ bits of information for each node and each node in turn has $m$ bits of information for the controller. In this example, we will consider an adaptive schedule three-hop protocol. 
The link realization of the network is shown in Fig.~\ref{fig:layout}. The controller (C) has direct links to nodes S0 - S2 at the rate of Phase \upperRomannumeral{1}. The rates of other phases depend on the number of nodes that succeeded in Phase \upperRomannumeral{1} -- thus links that were bad under the initial rate could be good under the new rate (for example the link between C and S4). Fig.~\ref{fig:protocol} walks through this example step-by-step and shows the information state at each node. For compactness, we have merged the two uplink and downlink tables for each node into a single table. During the downlink phases, the downlink part of the table is shown and during the uplink phases, the uplink part of the table is shown. For this example, we allocate time equally for all message phases (Downlink Phases \upperRomannumeral{1}, \upperRomannumeral{2} and \upperRomannumeral{3} and Uplink Phases \upperRomannumeral{1}, \upperRomannumeral{2} and \upperRomannumeral{3}) and by reciprocity assume that links present in Downlink Phase \upperRomannumeral{1} are present in Uplink Phase \upperRomannumeral{1} and so on.

\subsubsection*{Phase \upperRomannumeral{1}}
\begin{figure}[ht]
\centering
  \includegraphics[width = \textwidth]{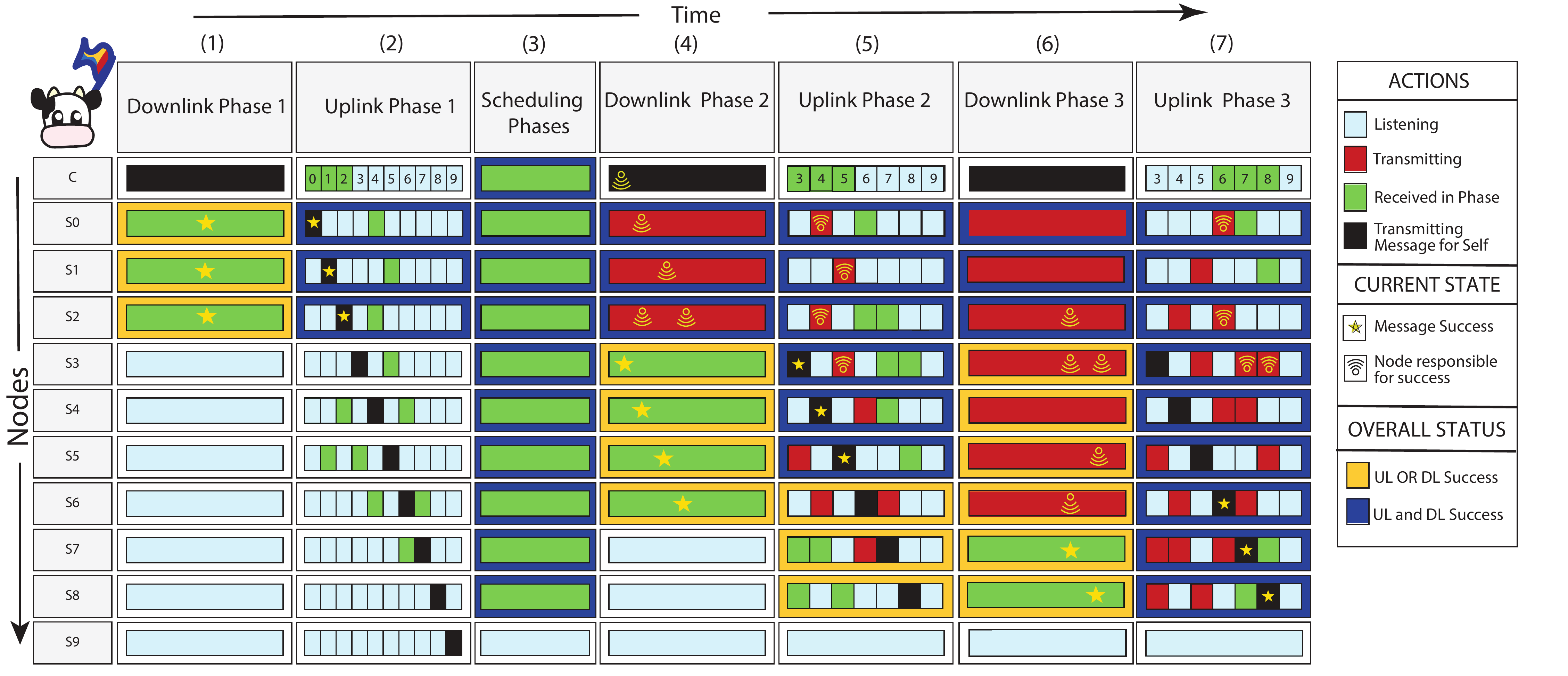}
  \caption{{The seven phases of the Occupy CoW protocol illustrated by a representative example. The table shows a variety of successful downlink and uplink transmissions using 0, 1 or 2 relays. S9 is unsuccessful for both downlink and uplink.}}
  \label{fig:protocol}
  \vspace{-25pt}
\end{figure}

This phase is just like its counterpart in the fixed-schedule case --- all messages get transmitted for the first time in their allotted slots. Phase \upperRomannumeral{1} is divided into two phases -- Downlink Phase \upperRomannumeral{1} (length of $T_{D_1})$) and Uplink Phase \upperRomannumeral{1} (length of $T_{U_1})$).
In Downlink Phase \upperRomannumeral{1}, the controller transmits the downlink packets of each of the nodes.
One can further optimize this by combining multiple packets from a single node into one larger packet for practical purposes (as shown in Fig.~\ref{fig:protocol}). The controller combines the individual messages into a single packet and broadcasts it at the rate $R_{D_1} = \frac{m \cdot n}{T_{D_1}}$. 
In the instance depicted in Fig.~\ref{fig:protocol}, Column 1, only S0, S1, and S2 successfully receive and decode the controller's packet. Note that these three ``direct links" to the controller are also depicted in Fig.~\ref{fig:layout}. At the end of Downlink Phase \upperRomannumeral{1}, S0, S1, and S2 have decoded both their individual messages as well as the messages intended for all of the other nodes. This is followed by Uplink Phase \upperRomannumeral{1}. In this phase the individual nodes transmit their uplink messages in separate packets in their assigned slots. 

In Fig.~\ref{fig:protocol}, Column 2, again only S0, S1, and S2 successfully transmit their messages to the controller. When a node is not transmitting, it is trying to listen for other messages -- thus S4 and S0 are able to hear each other, and so on. In Fig.~\ref{fig:layout}, we can also see the nodes which can hear each other even though they do not have anything to say to each other.
All successes thus far have been due to \emph{direct} connections between nodes and the controller. Due to this, we refer to these types of successes as ``one-hop" successes.

\subsubsection*{Scheduling Phases}
\label{subsec:schedule}
The scheduling phases are the key component in the adaptive scheduling scheme, since it is essential that all the nodes are aware of the packets that require retransmission. This allows them to compute the schedule according to which relays can help using the DSTC. 

During three scheduling phases (total length $T_{S}$ and each sub-phase of length $T_{S}/3$) the controller and the other nodes transmit short acknowledgments of 
$2n$ bits corresponding to $n$ downlink packets and $n$ uplink packets. Each phase is divided equally among the $n + 1$ nodes, resulting in a scheduling rate of $R_S = \frac{2n \cdot (n+1)}{T_S/3} = \frac{6n \cdot (n+1)}{T_S}$. 
In Scheduling Phase \upperRomannumeral{1}, all nodes take turns transmitting their acknowledgment (ACK) packets. For example, the controller could go first, then S0, then S1 and so on. In this example, the controller's ACK packet would be $1 1 1 1 1 1 1 1 1 1 1 1 1 0 0 0 0 0 0 0$ with the first $10$ ones corresponding to the downlink packets (known by assumption), the next $3$ ones indicate that the controller has the uplink packets of S0 - S2 and the rest of them are zero to indicate that the controller doesn't have those packets. 
Similarly S0's ACK packet is $1 1 1 1 1 1 1 1 1 1 1 0 0 0 1 0 0 0 0 0$ with the first $10$ ones corresponding to the downlink packets (as S0 has decoded all downlink packets), the next one is for its own uplink packet, the next three zeros for the uplink packets of S1 - S3 are followed by a one for S4's uplink packet and the rest are zeros corresponding to S5's - S9's uplink packets. 
After all ACK packets have been transmitted once, Scheduling Phase \upperRomannumeral{1} ends.

In Scheduling Phases \upperRomannumeral{2} and \upperRomannumeral{3}, these short ACK packets are retransmitted in a round-robin fashion by the nodes which have heard them using a DSTC in a fashion identical to fixed-schedule Occupy CoW Phase \upperRomannumeral{2}. For example, C's ACK packet is simultaneously transmitted by C, S0, S1, S2 and S3, S0's ACK packet is transmitted by C, S0 and S4 and so on. These ACK packets are relayed once again in Scheduling Phase \upperRomannumeral{3} so that all packets reach all nodes. At the end of Scheduling Phase \upperRomannumeral{3}, all nodes have `global ACK information' with high probability and are ready to adapt the retransmission schedule so that slots are not wasted on already successful data packets in Phases \upperRomannumeral{2} and \upperRomannumeral{3}. Fig.~\ref{fig:protocol}, Column 3 shows the information state of the nodes after the end of the scheduling phases. All nodes except S9 have received every ACK packet and know the schedule for the rest of the cycle. However, S9 has not received the scheduling information and therefore does not transmit anything for the rest of the cycle in order to avoid any interference to other packets.

\subsubsection*{Phase \upperRomannumeral{2}}
After the Scheduling Phases, we have Phase \upperRomannumeral{2} of data transmission. The messages that have already succeeded are the downlink and uplink packets of S0 - S2. Thus, the retransmission schedule only allocates time for the downlink packets of S3 - S9 and the uplink packets of S3 - S9. For illustrative purposes, we divide this phase further into two sub-phases -- Downlink Phase \upperRomannumeral{2} (length $T_{D_2}$) and Uplink Phase \upperRomannumeral{2} (length $T_{U_2}$). In general, if $a_D$ packets have succeeded in Downlink Phase \upperRomannumeral{1} and $a_U$ have succeeded in Uplink Phase \upperRomannumeral{1}, then the rates of transmission in these phases are: $R_{D_2} = \frac{m \cdot (n - a_D)}{T_{D_2}}$ and $R_{U_2} = \frac{m \cdot (n - a_U)}{T_{U_2}}$.
The relaying in these phases is the same as the relaying in Phase \upperRomannumeral{2} of the fixed schedule protocol -- except with a modified schedule. Because the schedule has adapted, it is possible that nodes that were initially unable to directly connect to the controller may now be able to, \emph{if} the rate during any of these phases is lower than that of the first. This may occur if enough nodes are successful in the first phase since fewer messages must now be sent or if the time allocated for the phases $T_{D_2}$ or $T_{U_2}$ is greater than $T_{D_1}$ or $T_{U_1}$ respectively resulting in a lower rate.

Downlink Phase \upperRomannumeral{2} is depicted in Fig.~\ref{fig:protocol}, Column 4. We see that node S3 gets its downlink message directly through the controller (due to reduced rate), and this is reflected in the dashed representation of the connection between node S3 and the controller in Fig.~\ref{fig:layout}.
As S0 and S2 are able to reach S4, it successfully receives the controller's message in two hops via S0 and S2 and so on.
At the end of Downlink Phase \upperRomannumeral{2}, nodes S0, S1, S2, S3, S4, S5, and S6 have successfully received their downlink messages. Uplink Phase \upperRomannumeral{2} is similar and is depicted in Fig.~\ref{fig:protocol}, Column 5 and the same set of nodes' uplink packets have successfully been delivered to the controller.

\subsubsection*{Phase \upperRomannumeral{3}}
Again, this phase is divided into Downlink Phase \upperRomannumeral{3} (length $T_{D_3}$) and Uplink Phase \upperRomannumeral{3} (length $T_{U_3}$) with rates $R_{D_3} = \frac{m \cdot (n - a_D)}{T_{D_3}}$ and $R_{U_3} = \frac{m \cdot (n - a_U)}{T_{U_3}}$ respectively. In these phases, `three-hop' successes occur.
For example, in Fig.~\ref{fig:protocol}, Column 6, S8 successfully receives its downlink packet through S5 (the full path is C $\to$ S1 $\to$ S5 $\to$ S8). The uplink counterpart is similar to downlink and at the end of Phase \upperRomannumeral{3}, all nodes except S9 have received their downlink packet and have successfully relayed their uplink message to the controller. The example depicted in Fig.~\ref{fig:layout} and~\ref{fig:protocol} is a failed instance of the protocol since the node S9 has not received its downlink message and the controller has not received S9's uplink packet.

\subsection{Information topology-dependent optimization} \label{sec:piggyback}
The adaptive schedule scheme can be  optimized for reduced implementation complexity when the information topology is a star. In particular, the scheduling phase can be shortened. 

For example, each node can piggyback a one bit ACK for their downlink packet onto their uplink message. Then the extra scheduling phases can be simplified to a single phase where the controller processes all the ACKs (received as well as not received) into a single packet that just lists which messages require retransmission. Then, all the nodes that can hear the controller get to know the schedule. These nodes can then modify the downlink packet (culling already successful messages and appending the global schedule to it) and simultaneously broadcast it. The nodes that can hear this first set of relays can then not only decode the downlink messages (despite not knowing the schedule) but also figure out the schedule itself so that they can help in the next phase. At this stage, nodes only reachable via three hops do not have the schedule and to propagate the information to them, we switch the order of Uplink Phase \upperRomannumeral{2} and Downlink Phase \upperRomannumeral{3} (Downlink Phase \upperRomannumeral{3} directly follows Downlink Phase \upperRomannumeral{2}).
The nodes reachable by two hops broadcast the downlink messages (with embedded schedule) again so that it can be heard by the nodes only reachable by three hops. Thus, even though all the nodes did not know the schedule at the beginning of Downlink Phase \upperRomannumeral{2}, they do get to know it by the end of Downlink Phase \upperRomannumeral{3} and that is sufficient for enabling Uplink Phases \upperRomannumeral{2} and \upperRomannumeral{3}. As you can see, this optimization exploits the star nature of the information topology to shorten the scheduling phase and furthermore, the total traffic dedicated to scheduling is substantially reduced. This optimized variation of the protocol is explored in detail in \cite{swamy2015cooperative}.

\vspace{-3mm}
\section{Analysis of Occupy CoW}
\label{sec:results}
We explore the Occupy CoW protocol with parameters in the neighborhood of a practical application, the industrial printer case described in Weiner et al.~\cite{Weiner}. In this particular scenario, the SERCOS \upperRomannumeral{3} protocol~\cite{SERCOS} supports the printer's required cycle time of $2$ ms with reliability of $10^{-8}$. Consequently, we target a $10^{-9}$ probability of failure for Occupy CoW. The printer has $30$ moving printing heads that move at speeds up to $3$ m/s over distances of up to $10$ m. Every $2$ ms cycle, each head's actuator receives $20$ bytes from the controller and each head's sensor transmits $20$ bytes to the controller.
The amount of information transmitted by the controller in a single cycle is $20 \times 30 = 600$ bytes. The total amount of sensor information transmitted by the heads to the controller in a single cycle is also $20 \times 30 = 600$ bytes. Therefore a total of $1200$ bytes or $9600$ bits of information is sent during a cycle of $2$ ms which corresponds to a desired goodput of $\frac{9600}{(2 \times 10^{-3})}$ bit/sec $= 4.8$ Mbit/sec.
If we assume access to a single dedicated $20$ MHz wireless channel, this $4.8$ Mbit/sec corresponds to an overall net spectral efficiency of $\frac{(4.8 \times 10^{6})}{(20 \times 10^{6})} = 0.24$ bits/sec/Hz.

\subsection{Behavioral assumptions for analysis}
\label{behav_assump}
The following behavioral assumptions are added to the resource assumptions in Sec.~\ref{sec:assumptions}.
\begin{itemize}
\item We assume a fixed nominal SNR and independent Rayleigh fading on each link.
\item We assume a single tap channel --- performance would improve if we reliably had more taps/diversity. Because the cycle-time is so short, the channels' coefficients do not change in a cycle and hence we use the delay-limited-capacity framework~\cite{HanlyTseDelayLimited,OzarowShamaiCellular}.
\item A link with complex fade $h$ and bandwidth $W$ is deemed good (no errors or erasures) if the rate of transmission $R$ is less than or equal to the link's capacity $C = W \log(1 + |h|^2\textit{SNR})$. Consequently, the probability of link failure is defined as
\begin{equation}
p_{\scriptscriptstyle link} = P(R > C) = 1 - \exp{\left(-\frac{2^{R/W} -1}{\textit{SNR}}\right).}
\label{eq:pfail_singlelink}
\end{equation}
From the above equation we see that if $R$ decreases, then the probability that the capacity $C$ is less than $R$ also decreases (the capacity $C$ did not change, only $R$ did). In other words, a channel which was unable to support a given rate might be able to support a lower rate.
\item We also assume channel reciprocity -- if a channel has fade $h$ between node A to B, then it is also $h$ from B to A as well.
\item If there are $k$ simultaneous transmissions\footnote{The cyclic-delay-diversity space-time-coding schemes we envision make the effective channel response longer. This can push the PHY into the ``wideband regime'', and a full analysis must account for the required increase in channel sounding by pilots to learn this channel~\cite{LozanoPorrat}. We defer this issue to future work but preliminary results suggest that it will only add $2-3$dB to the SNRs required at reasonable network sizes.}, then each receiving node harvests perfect sender diversity of $k$. For analysis purposes this is treated as $k$ independent tries for communicating the message that only fails if all the tries fail.
\item We do not consider any dispersion-style finite-block-length effects on decoding (justified in spirit by Yang et al.~\cite{YangDKP14}). A related assumption is that no transmission or decoding errors are undetected~\cite{forney} --- a corrupted packet can be identified (say using a $40$ bit hash) and is then completely discarded.
\end{itemize}
We derive the probability of failure for a two-hop downlink scheme in Sec.~\ref{sec:2-hop}, a union bound of the failure probability for a generic information topology in Sec.~\ref{sec:union-bound}, a similar bound on the probability of failure for relaying with non-simultaneous transmissions in Sec.~\ref{sec:round-robin} and the probability of failure for frequency hopping repetition coding in Sec.~\ref{sec:freq-hop}. These equations are used to derive results in Sec.~\ref{sec:res}.
\ifappendix
Additional derivations are found in the Appendix.
\else
Additional derivations are found in \cite{fullversion}.
\fi

\subsection{Two-hop downlink (star information topology)}
\label{sec:2-hop}
In a two-hop scheme, there are two shots at getting a message across. We derive the probability of protocol failure for both the fixed and adaptive schedule scheme. In both schemes, failure is the event that at least one of the $n$ nodes in the set $\mathcal{S}$ has not received its message by the end.
\subsubsection{Fixed schedule scheme}
In the fixed schedule two-hop scheme each message gets sent twice whether or not it was successful in the first try. The first phase's rate $R_{D_1} = \frac{n \cdot m}{T_{D_1}}$ and the corresponding probability of link failure is $p_1$. The second phase rate $R_{D_2} = \frac{n \cdot m}{T_{D_2}}$ (as all messages get sent two times) and the corresponding probability of link failure is $p_2$. Let the nodes successful in Phase \upperRomannumeral{1} be in a set $\mathcal{A}$ (with cardinality represented by the random variable $A$ and $a$ representing a specific size). The nodes in the set $\mathcal{S} \setminus \mathcal{A}$ succeed in Phase \upperRomannumeral{2} only if they connect to either the controller or at least one of the nodes in $\mathcal{A}$.
The Rayleigh fading assumption tells us that the probability that a link fails in Phase \upperRomannumeral{2} given it failed in Phase \upperRomannumeral{1} is given by $p_c = \min\left({\frac{p_{\scriptscriptstyle 2}}{p_{\scriptscriptstyle 1}}, 1}\right)$.
Then, the probability of not connecting to $\lbrace \text{controller} \bigcup \mathcal{A} \rbrace$ in Phase \upperRomannumeral{2} is $p_2^{a} \cdot p_c$.
The probability of 2-phase downlink system failure is thus:
\begin{align}
\text{P(fail)} = \sum\limits_{a = 0}^{n-1} \left(\binom{n}{a}(1 - p_{\scriptscriptstyle 1})^{a}(p_{\scriptscriptstyle 1})^{n - a}\right)  \left(1 - \left(1 - p_2^{a} \cdot p_c \right)^{n - a}\right).
\label{eq:2D_fixed}
\end{align}

\subsubsection{Adaptive schedule scheme}
\label{sec:adaptive-sched-analysis}
In the adaptive schedule two-hop scheme only the messages that were unsuccessful in Phase \upperRomannumeral{1} get sent again in Phase \upperRomannumeral{2}. The first phase is exactly like the fixed-schedule scheme.
The time allocated for Phase \upperRomannumeral{2} and the number of first phase successes $a$ dictate the Phase \upperRomannumeral{2} rate
$R_{D_2} = \frac{(n - a) \cdot m}{T_{D_2}}$. The corresponding probability of link failure is denoted $p_2(a)$ (the $(a)$ is used to indicate that it is a function of $a$).
As in the fixed-schedule case, the probability that the controller to node link fails in Phase \upperRomannumeral{2} given it failed in Phase \upperRomannumeral{1} is given by $p_c(a) = \min\left({\frac{p_{\scriptscriptstyle 2}(a)}{p_{\scriptscriptstyle 1}}, 1}\right)$. Then, the probability of not connecting to $\lbrace \text{controller} \bigcup \mathcal{A} \rbrace$ in Phase \upperRomannumeral{2} is $\left(p_2(a)\right)^a \cdot p_c(a)$.
The probability of 2-hop downlink system failure is thus:
\begin{align}
\text{P(fail)} = \sum\limits_{a = 0}^{n-1} \left(\binom{n}{a}(1 - p_{\scriptscriptstyle 1})^{a}(p_{\scriptscriptstyle 1})^{n - a}\right)  \left(1 - \left(1 - \left(p_2(a)\right)^a \cdot p_c(a) \right)^{n - a}\right).
\label{eq:2D_adaptive}
\end{align}
Notice that in the above derivation, we omitted the role of scheduling information even though it is actually crucial for adapting the rate of transmission in Phase \upperRomannumeral{2}. This is because we assume that the scheduling phases are allocated sufficient time such that the scheduling phase rate is lower than the rates of transmission in any of the other phases. Therefore any scheduling error in the protocol is also going to manifest as a delivery failure for a message packet. 
A property of ACK packets is that they want to reach all the nodes in the network. Therefore, a more detailed analysis for scheduling failure is as derived in the next subsection discussing the union bound, which is how we can upper bound the probability of failure for a generic topology (Sec.~\ref{sec:union-bound}).


\subsection{The union bound and generic information topologies}
\label{sec:union-bound}
Consider a generic network with $n$ nodes and $s$ message streams. 
Let's say that each stream has one origin and on average $d$ subscribers. For simplicity, the rates for all transmissions are kept constant at some rate $R$ with a corresponding probability $p$ of link failure. Consider a single message-destination pair. Let each message get two shots at reaching its subscribers -- directly from the source or through relays (say $j$ of them). Then the probability of the message reaching any specific destination is
\begin{align}
q_s 
&= \text{P(direct link)}\times \text{P}(\text{success} | \text{direct link}) + \text{P(no direct link)}\times \text{P}(\text{success} | \text{no direct link}) \nonumber\\
&= ((1 - p)\times 1) + \left(p \times \left( \sum_{j = 1}^{n-2} \binom{n-2}{j} (1 - p)^{j} p^{n - 2 - j} \left(1 - p^j\right) \right) \right). \label{eq:pre-union-bound}
\end{align}
Then the union bound on the probability of failure that even one of the $s$ messages did not reach one of its subscribers is:
\begin{equation}
\text{P(failure)} = s\times d \times (1 - q_s).
\label{eq:union-bound}
\end{equation}
As mentioned in Sec.~\ref{sec:adaptive-sched-analysis}, ACK information from each node has to disseminate throughout the network -- therefore if there are $n$ nodes in the network, there are $n$ ACK messages, and the number of subscribers for each is $n-1$. Consequently, the probability of ACK dissemination failure can be bounded by the union bound derived in this section.
\ifappendix
Equations for other error probabilities are derived similarly and can be found in the Appendix.
\else
Equations for other error probabilities are derived similarly and can be found in the extended version of this paper~\cite{fullversion}.
\fi

\subsection{Non-simultaneous relaying}
\label{sec:round-robin}
To tease apart the impacts of relaying and simultaneous transmission, it is useful to analyze relaying without simultaneous transmissions.
To have relays taking turns within a fixed-schedule scheme, the basic requirement is making the time-slots shorter. Suppose that we have $r$ potential relays for every data packet, designated in advance for every message stream. 
This means that if there are $k$ (either 2 or 3) hops, then each data packet will have a footprint of $1 + (k-1)r$ time slots. This means that for $s$ message streams, if the total cycle time is $T$ and each data packet is $m$ bits long, then the link data rate is $R = \frac{s \cdot m \cdot (1+(k-1)r)}{T}$.

Consider a single message stream and let $q_s(p,r)$ denote the probability of success to a single destination where $p$ is the probability of link error given the rate and SNR. The analysis for the two-hop case follows the union-bound case in \eqref{eq:pre-union-bound} with the number of potential relays $r$ playing the role of $n-2$ above. Consequently:
\begin{align}
q_s(p,r) &= ((1 - p)\times 1) + \left(p \times \left( \sum_{j = 1}^{r} \binom{r}{j} (1 - p)^{j} p^{r - j} \left(1 - p^j\right) \right) \right). \label{eq:with-relay-limits}
\end{align}

The union bound argument applies and so \eqref{eq:union-bound} continues to bound the probability of error, just with the slightly revised expression in \eqref{eq:with-relay-limits} for $q_s$.


\subsection{Frequency-hopping schemes}
\label{sec:freq-hop}
In Occupy CoW, during each message's transmission, the entire available bandwidth $W$ is used for coding at a link rate of $R$. In a frequency-hopping scheme, the available bandwidth $W$ is broken into $k$ sub-channels ($k > 1$) and each sub-channel carries the entire packet at the higher rate of $R_{sc}(k) = k \times R$. We assume that each sub-channel fades independently.
The analysis of the frequency hopping repetition coding scheme is very similar to the non-simultaneous relaying scheme. Let the probability of failure of a single sub-channel link at rate $R_{sc}(k)$ be $p_{sc}(k)$. Then the probability that a message was not successful is the probability that each of the sub-channels failed to deliver the message i.e., $\left(p_{sc}(k)\right)^k$.
Therefore, the failure probability of a frequency hopping repetition based scheme with $s$ streams each with a single destination is given by
\begin{equation}
\text{P(fail, } k\text{)} = 1 - \left(1 - \left(p_{sc}(k)\right)^k\right)^s.
\label{eq:freq-hop}
\end{equation}


\subsection{Results and comparison}
\label{sec:res}
Following Weiner et al.~\cite{Weiner} and the communication-theoretic convention, we use the minimum SNR required to achieve $10^{-9}$ reliability as our metric to compare fixed-schedule 2-hop and adaptive-schedule 3-hop Occupy CoW to four other baseline schemes. 
We calculate the minimum SNR required by various protocols to meet the specs in the following fashion. Assuming a fixed nominal SNR, we calculate the probability of failure for the protocol under consideration. Then, we search for the smallest value of nominal SNR that meets the reliability requirement of $10^{-9}$.
Fig.~\ref{fig:hockeystick} looks at performance (the minimum SNR required on the y-axis) for a star information topology with a central node sending $m=160$ bit messages to $n$ other nodes and receiving the same size messages from them. All this has to be completed within $2$ms and a bandwidth of $20$MHz.    Initially the minimum required SNR for Occupy CoW decreases with increasing $n$, even through the required throughput increases as $m \cdot n$, but the curves then flatten out. The gains of multi-user diversity eventually give way and the required SNR starts to increase for large $n$ as the required spectral efficiency increases.

The topmost blue solid curve in Fig.~\ref{fig:hockeystick} shows performance of the protocol restricted to just the first hop of Occupy CoW with one slot per message. The required SNR shoots off the figure for two reasons: (a) because the throughput increases linearly with the number of nodes and (b) to have the system probability of failure stay controlled with more messages to transmit, each individual message must be that much more reliable. The second scheme (red dashed curve) is purely hypothetical. It allows each message to use the entire $2$ms time slot for its own uplink and downlink message but without any relaying and thus also no diversity. This bounds what could possibly be achieved by using adaptive HARQ techniques and shows why harnessing diversity is essential. This is rising only because of effect (b) above.

The third reference scheme is the non-simultaneous relaying scheme described in Sec.~\ref{sec:round-robin} and plotted in Fig.~\ref{fig:hockeystick} by the black curve with markers. We see that this curve is always above the Occupy CoW lines --- showing the quantitative importance of simultaneous relaying. The curve is annotated with the best number of relays $r$ that minimizes the SNR required. As $r$ increases, the available spatial diversity increases, but the added message repetitions force the link data rates higher.

\begin{wrapfigure}{r}{0.45\textwidth}
  \vspace{-25pt}
  \begin{center}
    \includegraphics[width=0.42\textwidth]{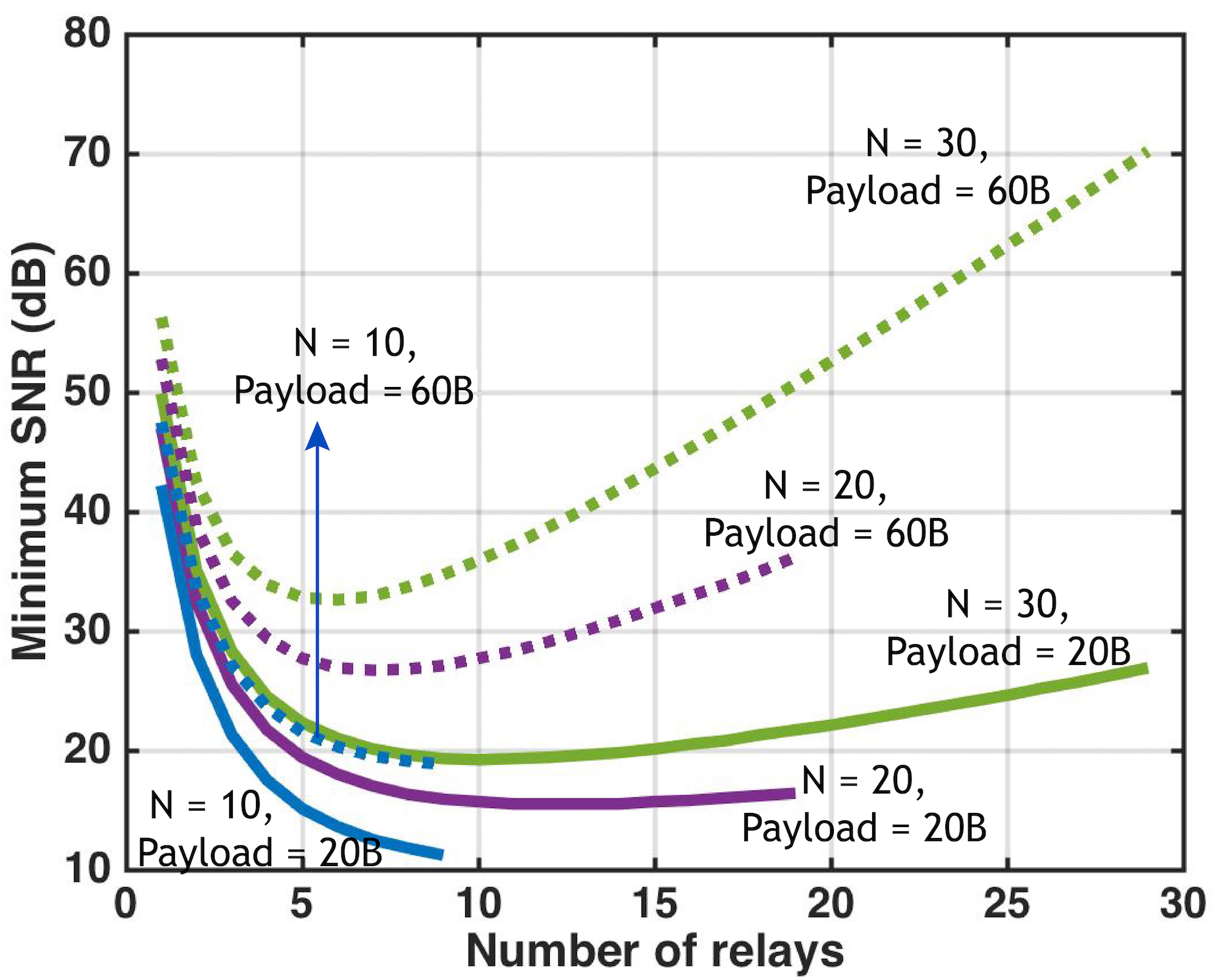}
  \end{center}
  \vspace{-20pt}
  \caption{{For non-simultaneous relaying, the minimum SNR required to achieve a $10^{-9}$ probability of system failure for different network and payload sizes as the number of nominated relays vary.}}
  \vspace{-15pt}
  \label{fig:relay_vs_snr}
  \vspace{-15pt}
\end{wrapfigure}

Fig.~\ref{fig:relay_vs_snr} explores the effect of the number of relays allocated on the required SNR for the scheme in Sec.~\ref{sec:round-robin}. 
For a network size of $n = 30$, a payload size of $60$B per message would select $r = 6$ as the optimal number of relays. Reducing the network size to $10$ makes $r = 9$ be the optimal number of relays. Compare this to a payload size of $20$B and $n=30$ --- not only is the optimal number of relays the same, the entire curve is very close to that for $n=10$ with payload $60B$. 
This is because for the same number of relays, the link data rates are the same and the factor of $3$ difference in the number of message streams demands a factor $3$ reduction in the probability of error per message --- which for nine relays is accomplished for less than $1$dB. 
Given a large enough network, the optimum number of relays seems to depend primarily on the aggregate rate. For a high-aggregate rate, we choose a smaller number of relays and for a lower one, we pick more relays.

The last reference scheme (the purple dotted line in Fig.~\ref{fig:hockeystick}) represents the hypothetical frequency-hopping described in Sec.~\ref{sec:freq-hop}. As the number $k$ of frequency hops increases, the available diversity increases, but the added message repetitions force the instantaneous link rates higher, just as additional relays do for non-simultaneous relaying. For low $n$ we prefer more frequency hops because of the diversity benefits. The SNR cost of doing this is not so high because the throughput is low enough (requiring a spectral efficiency less than $1.5$bits/s/Hz) that we are still on the cusp of the energy-limited regime of channel capacity.
For fewer than $7$ nodes, this says that using frequency-hopping is great --- as long as we can reliably count on $20$ or more guaranteed independently faded sub-channels to repeat across. After $7$ nodes, notice the frequency-hopping scheme is paralleling the non-simultaneous relaying scheme in Fig.~\ref{fig:hockeystick}. However, frequency hopping is optimized with more diversity and lower SNR because harnessing multiuser diversity requires the first-hop to actually reach enough relays to be able to use the reserved slots while frequency-diversity is just assumed to always be available.


Fig.~\ref{fig:hockeystick} also compares the fixed-schedule two-hop Occupy CoW protocol with equal phase lengths to an adaptive three-hop scheme optimized to minimize SNR. We see that these are very close to each other and the choice between these is not as as important as harnessing diversity and taking advantage of simultaneous transmissions. This is discussed in detail in the next Section~\ref{sec:opt}.

\begin{wrapfigure}{r}{0.5\textwidth}
  \vspace{-25pt}
  \begin{center}
    \includegraphics[width=0.48\textwidth]{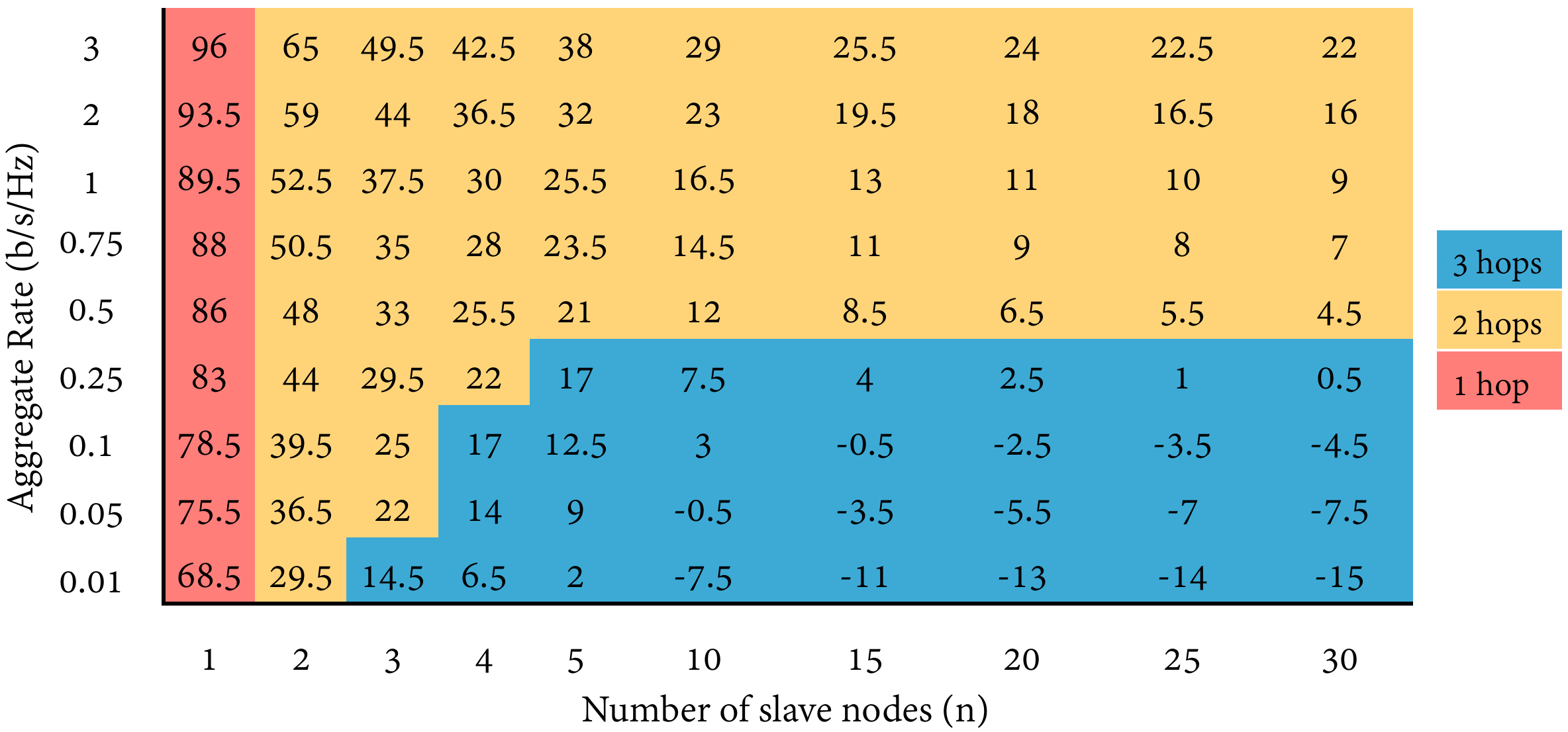}
  \end{center}
  \vspace{-20pt}
  \caption{{The number of hops and minimum SNR to be operating at to achieve a high-performance of $10^{-9}$ as aggregate rate and number of users are varied. Here, the time division within a cycle is unoptimized. Uplink and downlink have equal time, 2-hops has a 1:1 ratio across phases, and 3-hops has a 1:1:1 ratio for the 3 phases. The numbers here are for a star information topology but as the next figure shows, they would not be much different for generic topologies.}}
  \label{fig:dui}
  \vspace{-15pt}
\end{wrapfigure}
It turns out that the aggregate goodput required (overall spectral efficiency considering all users) is the most important parameter for choosing the number of relay hops in our scheme. This is illustrated clearly in Fig.~\ref{fig:dui}. This table shows the SNR required and the best number of hops to use for a given $n$.
With one node, clearly a $1$ phase scheme is all that is possible. As the number of nodes increases, we transition from $2$-phase to $3$-phase schemes being better. For $n \geq 5$, aggregate rate is what matters in choosing a scheme, since $3$-phase schemes have to deal with a $3\times$ increase in the instantaneous rate due to each phases' shorter time, and this dominates the choice. In principle, at high enough aggregate rates, even the one-hop scheme will be best with enough users. But when the target reliability is $10^{-9}$, this is at absurdly high aggregate rates\footnote{We estimate this is around aggregate rate $40$ --- that would correspond to $40$ users each of which wants to simultaneously achieve a spectral efficiency of $1$.}. In the practical regime, diversity wins.

\begin{wrapfigure}{r}{0.45\textwidth}[15]
  \vspace{-25pt}
  \begin{center}
    \includegraphics[width=0.42\textwidth]{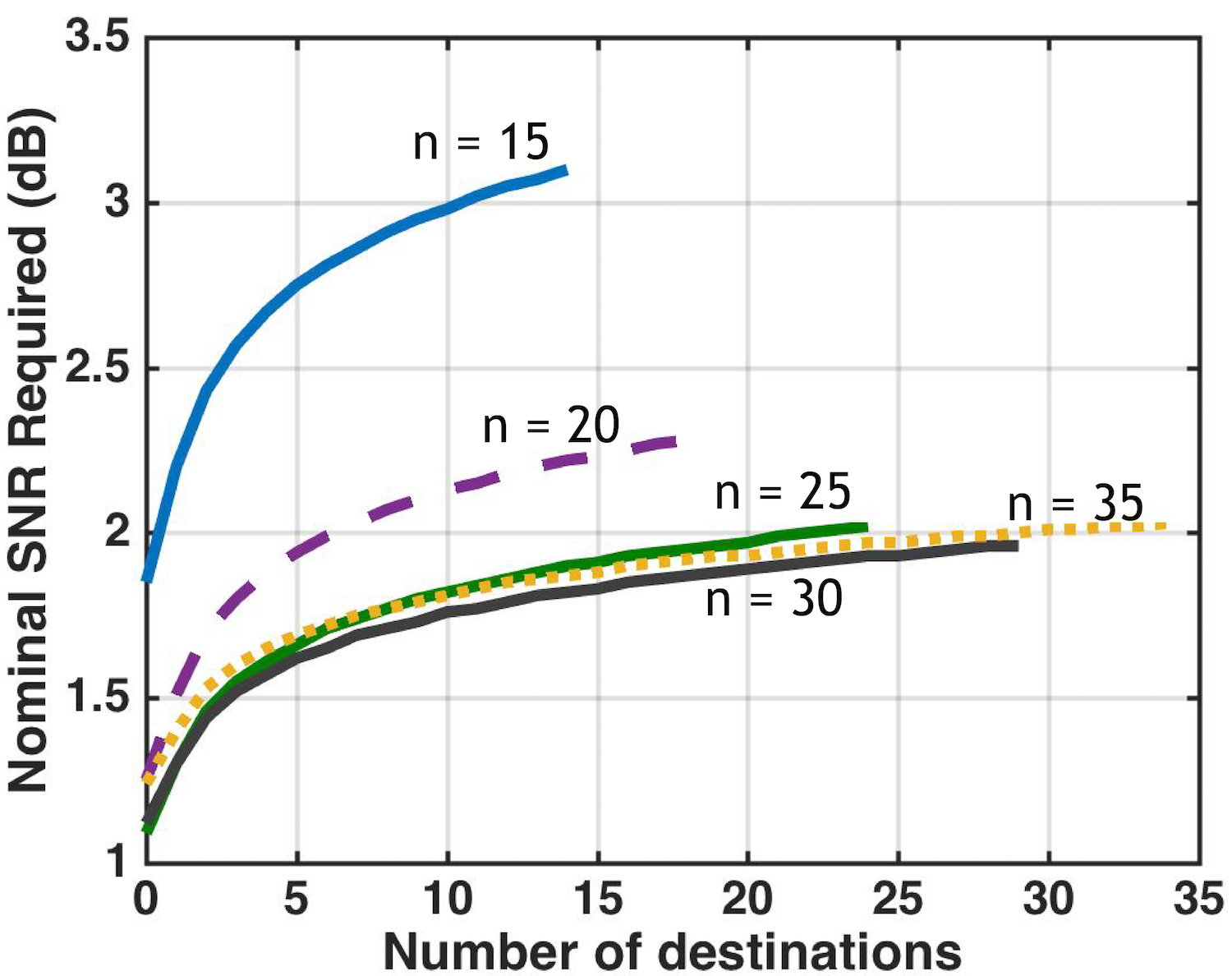}
  \end{center}
  \vspace{-20pt}
  \caption{{Number of destinations vs SNR required for different network sizes for $m=160$ bit messages and $n= \left(15, 20, 25, 30, 35\right)$ nodes with $20$MHz and a $2$ms cycle time, aiming at $10^{-9}$ probability of failure. The SNR values at ``$0$'' destinations represents the SNR required for a star information topology.}}
  \label{fig:dest_vs_snr}
  \vspace{-15pt}
\end{wrapfigure}
We now consider the case of a generic non-star topology using \eqref{eq:union-bound}. Figure~\ref{fig:dest_vs_snr} considers the SNR required for a varying number of destinations for different network sizes.
The number of destinations per message ranges from $1$ to $n - 1$. For comparison purposes, at ``$0$'' destinations, we have plotted the SNR required for the star information topology. There is an SNR `penalty' for each message having multiple destinations but even when everyone wants to hear everything, this penalty is quite modest. 
The case of $n-1$ destinations is similar to simply reducing the tolerable probability of failure by a factor of $1/n$. The extra SNR required is on the order of $1$dB for medium to large network sizes because of the ample diversity available.

\vspace{-3mm}
\section{Optimization of Occupy CoW}
\label{sec:opt}
\subsection{Phase-length Optimization}
The protocols we have described come with the choice of the number of phases (2 or 3) and a choice of fixed or adaptive schedule. Furthermore, there is the choice of the time allocated for different phases.
How does one pick the `right' parameters? Does it matter? To answer that, we compare the performance (minimum SNR required to achieve the specs) of a simple 2-hop fixed schedule scheme where the time available is equally divided among the phases and a 3-hop adaptive protocol with optimal phase lengths minimizing the SNR required.

We focus on a star information topology because it has both extremes of downlink (one source with separate messages for many destinations) and uplink (the vice-versa).
We consider downlink and uplink separately and look at the optimal allocation of time  for a three-hop protocol which minimizes the SNR required to meet the performance specifications. Here we used a simple brute force search over time allocations.
We find that the optimal phase-length allocations are far from even. We also find that the SNR savings that we achieve by having different lengths is minimal and believe that the implementation complexity of building a system which can code and decode at variable rates is a bigger deal and ultimately negates out the small SNR savings achieved by phase-length optimization and dealing with all the ACK information.

\begin{figure}[htb]
\centering
\begin{subfigure}[b]{0.49\textwidth}
\centering
\includegraphics[width=0.8\textwidth]{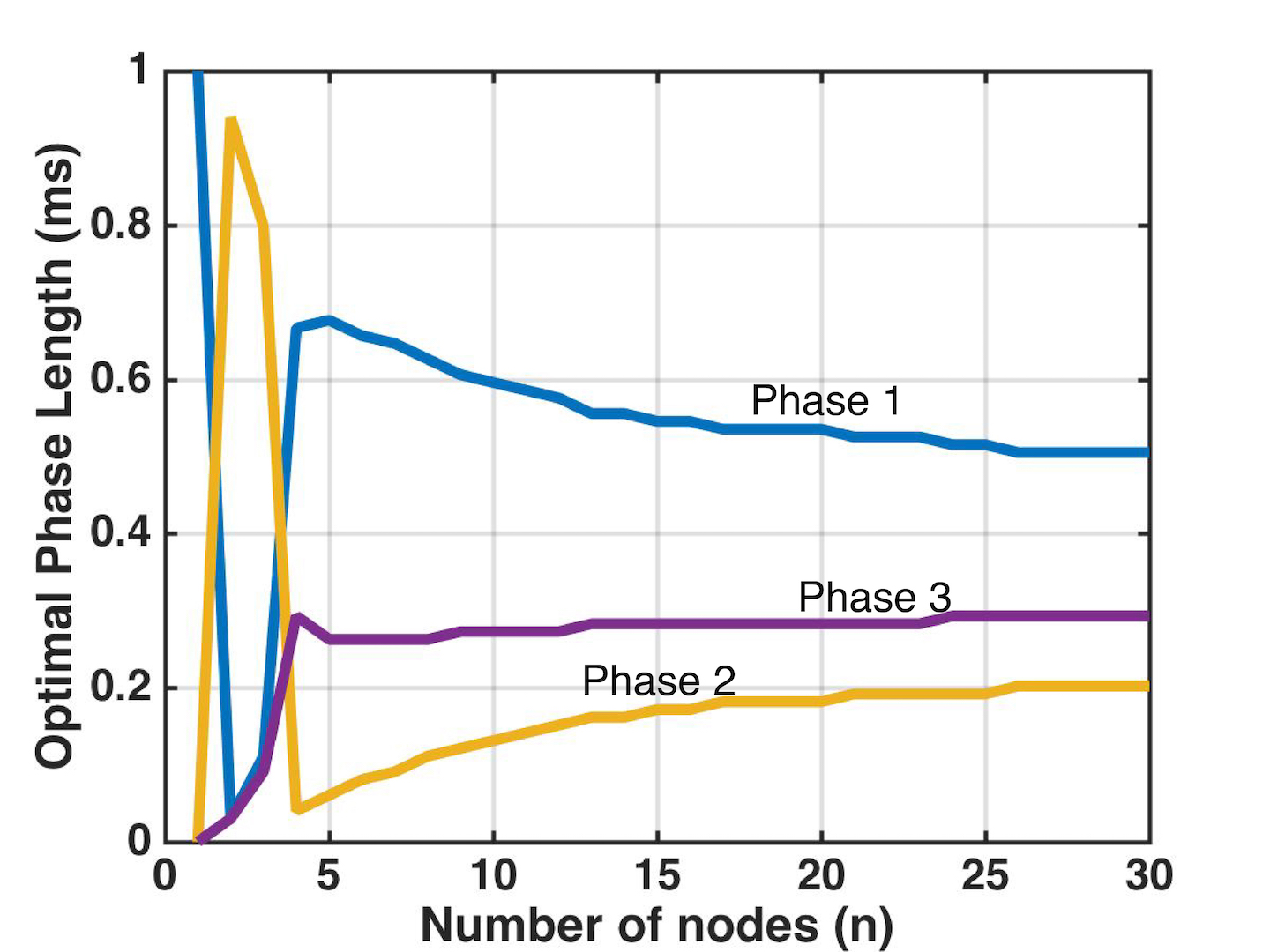}
\caption{Optimal fraction of time allocated for downlink phase \upperRomannumeral{1}, \upperRomannumeral{2} and \upperRomannumeral{3} in the three-hop protocol at the smallest SNR which meets the performance requirements.}
\label{fig:3-hop-minphase-downlink}
\end{subfigure}
~\begin{subfigure}[b]{0.49\textwidth}
\centering
\includegraphics[width=0.8\textwidth]{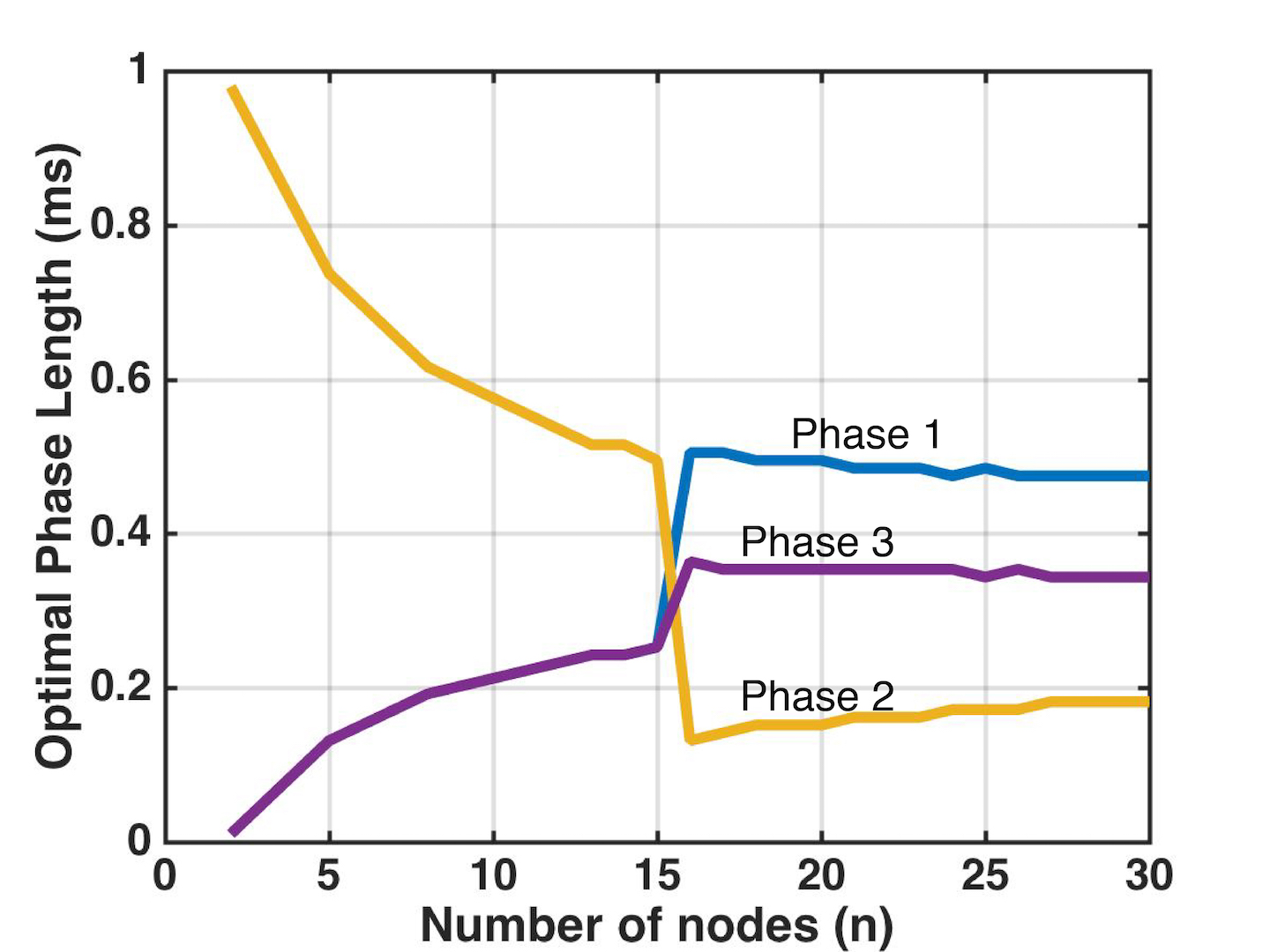}
\caption{Optimal fraction of time allocated for uplink phase \upperRomannumeral{1}, \upperRomannumeral{2} and \upperRomannumeral{3} in the three-hop protocol at the smallest SNR which meets the performance requirements.}
\label{fig:3-hop-minphase-uplink}
\end{subfigure}
\caption{Optimal phase allocation for three-hops with 160 bit messages, 30 users, $2\times 10^4$ total cycle length.}
\label{fig:3-hop-minphase}
\vspace{-20pt}
\end{figure}
Let us consider the adaptive scheduling protocol with a $2$ms cycle divided equally between Uplink and Downlink. How should we divide the times across phases for this? 
Assume that the ACK information is reliably delivered for free. 
Let the times allocated for Phase \upperRomannumeral{1}, \upperRomannumeral{2} and \upperRomannumeral{3} of downlink and uplink be $T_{D_1}$, $T_{D_2}$ and $T_{D_3}$ and $T_{U_1}$, $T_{U_2}$ and $T_{U_3}$respectively such that $T_{D_1} + T_{D_2} + T_{D_3} = 1$ms and $T_{U_1} + T_{U_2} + T_{U_3} = 1$ms.
Similarly, let the times allocated for phase \upperRomannumeral{1}, \upperRomannumeral{2} and \upperRomannumeral{3} of uplink be $T_{U_1}$, $T_{U_2}$ and $T_{U_3}$ respectively such that $T_{U_1} + T_{U_2} + T_{U_3} = 1$ms.

\noindent \textbf{Downlink:} Figure \ref{fig:3-hop-minphase-downlink} shows the optimal allocation of time for phase \upperRomannumeral{1}, \upperRomannumeral{2} and \upperRomannumeral{3} for downlink.
The optimization suggests that phase \upperRomannumeral{1} should be the longest, phase \upperRomannumeral{2} the shortest and phase \upperRomannumeral{3} in between (except for network sizes 1 and 2 where the optimal strategies are 1 hop and 2 hop respectively).
Phase \upperRomannumeral{3} is longer than Phase \upperRomannumeral{2} to make sure that the messages reach everyone possible as more links open up during phase \upperRomannumeral{3}.
Phase \upperRomannumeral{1} is longest to ensure that the messages are initially successfully decoded by enough nodes to ensure maximal spread.\\
\noindent \textbf{Uplink:} Figure \ref{fig:3-hop-minphase-uplink} shows the optimal allocation of time for phase \upperRomannumeral{1}, \upperRomannumeral{2} and \upperRomannumeral{3} for uplink. In uplink, the critical paths are the ones connecting to the controller rather than the inter-node links. Hence, phase \upperRomannumeral{3} is allocated more time than in the downlink Phase. 
\vspace{-10pt}
\subsection*{How much SNR does optimization save?}
\begin{wrapfigure}{r}{0.45\textwidth}
  \vspace{-25pt}
  \begin{center}
    \includegraphics[width=0.44\textwidth]{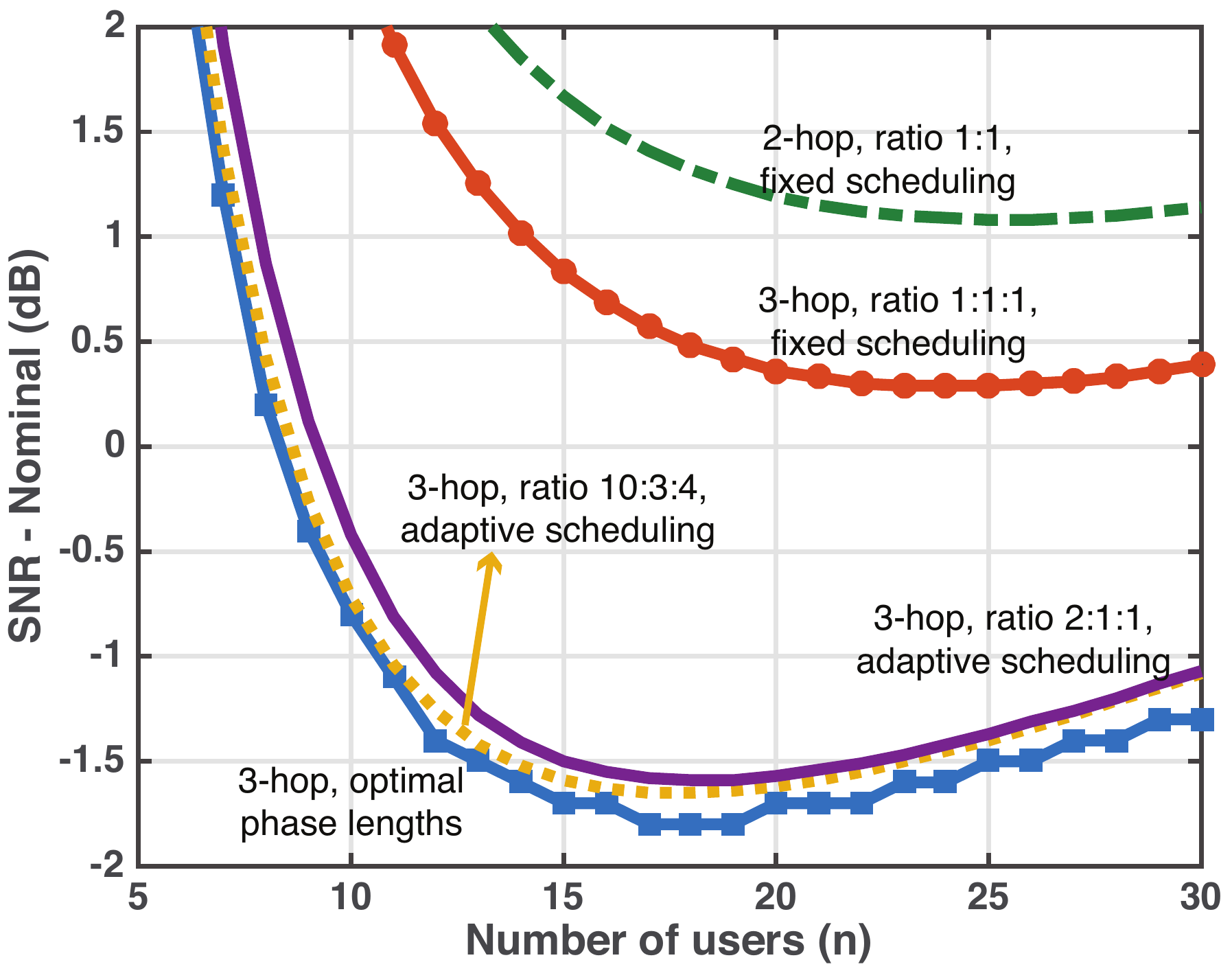}
  \end{center}
  \vspace{-10pt}
  \caption{Comparing the SNR required for optimum downlink phase length allocation and a few non-optimal allocations.}
 \label{fig:opt_downlink}
  \vspace{-20pt}
\end{wrapfigure}
For concreteness, let us consider the downlink side. (Uplink is similar.) Figure \ref{fig:opt_downlink} considers different phase length allocations including the optimal phase length allocation and several other suboptimal allocations. For a star network of $30$ nodes, we see that the difference between the various allocations and schemes is minimal.
Note that these results are for a star information topology which is the `best' case in terms of the SNR required. As the benefits of optimization are marginal in the best case, the benefits in a generic topology are even more negligible. Furthermore, adaptive scheduling is a harder problem in a non-star topology as one cannot mostly piggyback the relevant ACK or scheduling information onto packets that would be sent anyway, as discussed in Section~\ref{sec:piggyback}. Consequently, we conclude that though we have many knobs to turn which can optimize the performance of the protocol in terms of required SNR, the benefits for that metric are not going to be that substantial.
This is not to say that there might not be other reasons for wanting to use adaptive scheduling --- e.g.~to support additional best-effort traffic by harvesting time-slots that are not needed for relaying time-critical packets. However, that is beyond the scope of this paper.

\vspace{-10pt}
\subsection{Power consumption and the effect of duty cycling } \label{sec:power}
\begin{wrapfigure}{r}{0.45\textwidth}
  \vspace{-35pt}
  \begin{center}
    \includegraphics[width=0.44\textwidth]{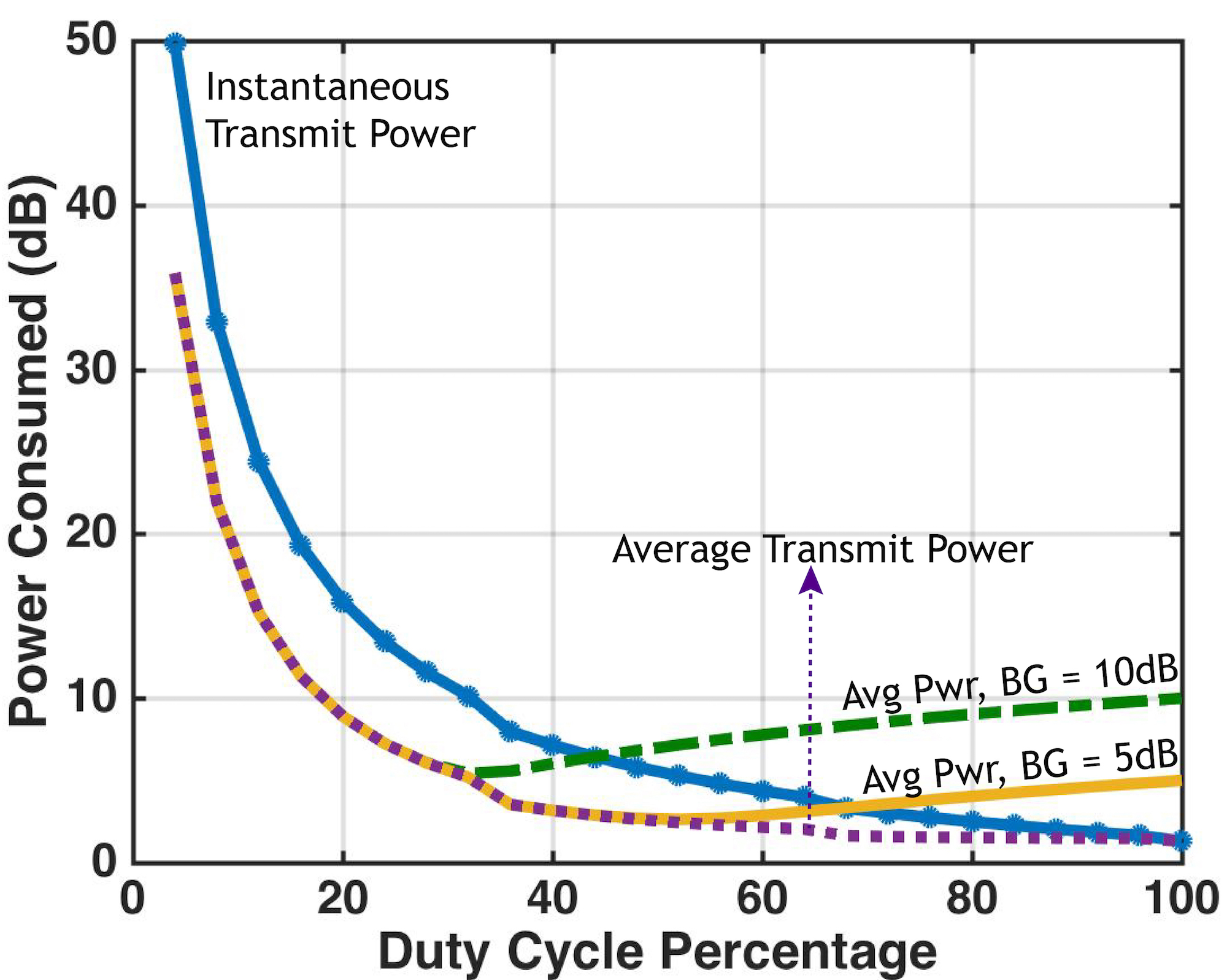}
  \end{center}
  \vspace{-10pt}
  \caption{Effect of duty cycling percentage (i.e. time awake) on the power required for different on-time percentages for $m=160$ bit messages and $n=30$ nodes with $20$MHz and a $2$ms cycle time, aiming at $10^{-9}$ probability of failure.}
 \label{fig:duty-cycle}
  \vspace{-30pt}
\end{wrapfigure}
The Occupy CoW protocol as described so far relies on all nodes being awake and listening at all times (when not transmitting). However, in most practical wireless systems, nodes are put to sleep often to conserve energy, even during active periods. If such duty-cycling is to be introduced, what percentage of time should the nodes be put to sleep? To answer this question, we first modify the protocol to handle duty-cycling by using the ideas used to understand non-simultaneous relaying.

We dedicate a percentage of nodes per message as pre-allocated potential relays (say $x \%$). 
These wake up during the message's transmission -- they either listen for the message or simultaneously re-transmit the message if they have it. The equation~\eqref{eq:pre-union-bound} can be modified so that the maximum number of relays is not $n-2$ but $r = \left\lceil \frac{x\times (n-2)}{100} \right\rceil$. Thus the probability of success of a single message-destination pair $q_{ds}(r)$ is:
\vspace{-5pt}
\begin{align}
q_{ds}(r) 
&= ((1 - p)\times 1) + \left(p \times \left( \sum_{j = 1}^{r} \binom{r}{j} (1 - p)^{j} p^{r - j} \left(1 - p^j\right) \right) \right)
\end{align}
Thus we have that the duty-cycled protocol's probability failure with $s$ message streams and $d$ average subscribers per stream is bounded by
\vspace{-5pt}
\begin{equation}
\text{P(failure)} \leq  s\times d \times (1 - q_{ds}(r)).
\label{eq:duty-cycle-union-bound}
\end{equation}

Fig.~\ref{fig:duty-cycle} shows the power consumed to reach the target reliability as a function of the time awake (duty cycle percentage). The blue curve plots the power consumed by a node when awake (in units of received SNR) in order to meet the required reliability. The purple dotted line takes into account the percentage of time the node is asleep, and plots the average transmit power used. Because this is minimized at 100\% duty cycle, if transmit power consumption were all that mattered, it would not be worth sleeping at all. To get a more refined answer, we recognize that there is some level of background power consumption in the wireless circuitry which accounts for listening and encoding/decoding processes whenever the node is awake~\cite{GoldsmithBahaiPower}. The green line is the average total power consumed assuming a background power consumption of $10$dB (i.e.~the background power is the same as what the transmit power would be to give a $10$dB SNR.) and the yellow line is for $5$dB background power consumption. These plots reveal an easy rule of thumb about the desired operating point -- operate with a duty-cycle percentage such that the transmit power required is equal to the background power.

\vspace{-3mm}
\section{Conclusions \& Future Work}
This paper proposed a wireless communication protocol framework for high-performance industrial-automation systems that demand ultra-high reliability and low-latency for many message streams within a network with many active nodes. The protocol framework is targeted to a single wireless local domain where all nodes are nominally in range of each other, but can handle any arbitrary information topology in terms of which node is subscribed to which message stream. Harnessing significant diversity is absolutely essential for ultra-reliability and cooperative communication using relaying can access multiuser diversity. To achieve low-latency, simultaneous transmission using a diversity-oriented distributed space-time code is important, especially when the payload sizes are such that spectral efficiency is a concern. This gives a significant SNR advantage over pure frequency-hopping approaches while also not demanding that nature guarantee a lot of frequency diversity. Time diversity is also not viable when the tolerable latency is shorter than the coherence time, leaving multiuser diversity as the only real choice. When the background power used for having the wireless subsystem turned on is significant, it is beneficial to have subsets of nodes go to sleep while relying on others to listen and relay messages. Although this increases the transmit power required, it reduces overall network power consumption. Simple phase length allocations and a fixed schedule are suffice to achieve our target reliability are reasonable SNR; optimized scheduling and phase lengths only provide marginal savings.

\ifappendix
\section*{Appendix}
In this appendix, we extend the union-bound for an arbitrary topology to include 3-hop success and then provide more nuanced calculations for the specialized star topology where we consider the downlink and uplink stages separately.
A downlink failure occurs when at least one node fails to receive its message from the controller in the downlink stage and an uplink failure is vice-a-versa.
The method of calculating the probability of error for uplink and downlink depends on the number of protocol hops. Finally, a union bound over the uplink and downlink phases is used to determine the overall probability of cycle failure.
This is a slightly conservative estimate, since in reality, each phase reuses channels from previous phases and iterations of the protocol.
For the generic topology, we calculate the bound for a fixed schedule (and fixed transmission rate) while we consider the adaptive schedule protocol for the star topology.

\subsection*{Notation:}
In order to effectively present the derived expressions, we provide a guide to the notation that will be used in the following sections.
Let a transmission over a single link be an ``experiment." A binomial distribution with $n$ independent experiments, probability of success $1-p$, and number of success $m$ will be referred to as
\begin{equation}
B(n,m,p) = \binom{n}{m} (1-p)^{m}p^{n-m}.
\label{eq:B}
\end{equation}
The probability of at least one out of $n$ independent experiments failing will be denoted as
\begin{equation}
F(n,p) = 1 - (1 - p)^{n}.
\label{eq:F}
\end{equation} 
The probability of a good link has already been described in \eqref{eq:pfail_singlelink}.
Following general convention, for each depicted set, the set itself will be represented in script font. The random variable representing the number of nodes in that set will be presented in uppercase letters. Finally, the instantiation of that random variable (the cardinality of the set), will be in lowercase letters.
We assume that if $R_i$ exceeds capacity, the transmission will surely fail (with probability 1). If $R_i$ is less than capacity, the transmission will surely succeed and decode to the right codeword.


\subsection{Union bound for 3-hop protocol:}
The union bound analysis for a 2-hop protocol for a generic topology was provided in Sec.~\ref{sec:union-bound}. In this section, we extend this to consider 3-hop successes. Consider a generic network with $n$ nodes and $s$ message streams.
Let's say that each stream has one origin and on average $d$ subscribers. For simplicity, the rates for all transmissions are kept constant at some rate $R$ with a corresponding probability $p$ of link failure as given by Eq.~\eqref{eq:pfail_singlelink}. Consider a single message-destination pair. Let each message get \emph{three} shots at reaching its subscribers -- directly from the source or through two or three hop relays. Then the probability of the message reaching any specific destination is
\begin{align}
q_s &= \text{P(success to a single destination)} \nonumber\\
&= \text{P(direct link)}\times \text{P}(\text{success} | \text{direct link}) + \text{P(no direct link)}\times \text{P}(\text{success} | \text{no direct link}) \nonumber
\end{align}

The probability of a direct link $\text{P(direct link)} = 1 - p$ (and $\text{P(no direct link)} = p$) and the probability of success given there is a direct link $\text{P}(\text{success} | \text{direct link}) = 1$.
There are two ways for a node to succeed indirectly either by connecting to the set of nodes that heard the message directly from the source (let that set be $\mathcal{I}$) or if it did not connect to $\mathcal{I}$ then, by connecting to the set of nodes that heard the message from $\mathcal{I}$ (let this set be $\mathcal{J}$).
If the node connects to $\mathcal{I}$, then the node succeeds in two hops (source $\to \mathcal{I} \to$ destination).
If the node did not connect $\mathcal{I}$ but connects to $\mathcal{J}$, then the node succeeds in three hops (source $\to \mathcal{I} \to \mathcal{J} \to$ destination). The probability of success when there is no direct link is then given by\\
$\text{P}(\text{success} | \text{no direct link})$
$$= \sum_{i = 1}^{n - 2} \left(B(n - 2, i, p) \left\lbrace \left( 1 -p^i \right) + p^i \sum_{j = 1}^{n - 2 - i}\left( B( n - 2 - i, j, p^i) \cdot \left( 1 - p^j\right) \right) \right\rbrace\right) $$

Then the union bound on the probability of failure that even one of the $s$ messages did not reach one of its subscribers is:
\begin{equation}
\text{P(failure)} = s\times d \times (1 - q_s).
\label{eq:union-bound}
\end{equation}

\subsection{Star Topology Analysis}
The crux of analysis for star-topology relies on partitioning each stage of the protocol into a number of distinct states. As we saw when stepping through Fig.~\ref{fig:protocol}, our protocol facilitates successful transmission via various different pathways. Successes and failures occur in many different ways. We account for all means of success by first enumerating all possible paths of success in each phase.
We then partition the set of all nodes, $\mathcal{S}$, into sets corresponding to those paths of success (if they succeed), and the set of nodes that fail, $\mathcal{E}$. We refer to any given instantiation of these sets as a state, and the probability of error is calculated by analyzing all possible instantiations of these sets. There are two main methods of analysis used to calculate the probability of error: by counting the number of failure states, or by calculating the probability of failing given a particular state.
We divide the analysis into three sections, corresponding to the one-hop, two-hop, and three-hop protocols. We derive the probabilities of error for the downlink and uplink stages for each.

Recall that when calculating the probability of cycle error, we partition the set of all nodes into various other sets corresponding to their method of success. Through the course of the analysis, we will be using the sets denoted in Fig.~\ref{fig:bubs} for both uplink and downlink.
\begin{figure}[htb]
\centering
\includegraphics[scale = 0.6]{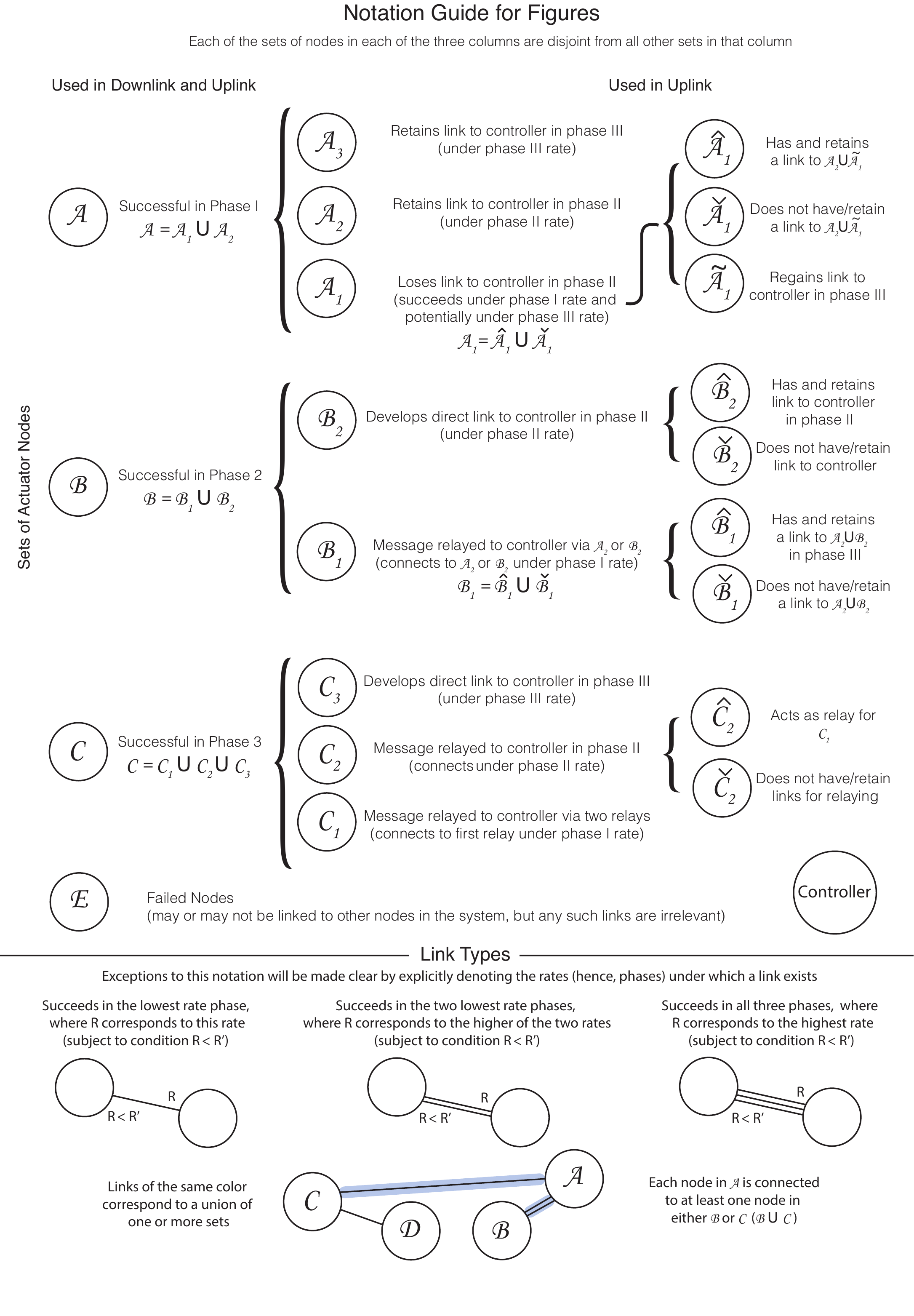}
\caption{This figure enumerates the various sets that we will be using throughout the analysis. In addition, how we represent various links in each of the protocol figures is also found here.}
\label{fig:bubs}
\end{figure}
\FloatBarrier

\subsection{One-Hop Protocol:}
Recall that in this framework the entire protocol consists of stages 1 and 2 of Fig.~\ref{fig:protocol}. The controller broadcasts messages, each of length $m$ bits for each node, to the $n$ nodes, and the nodes respond by transmitting their information as in Fig.~\ref{fig:protocol}. In this case, no relaying occurs at all. Downlink receives time $T_D$ and uplink receives time $T_U$, where $T_U + T_D = T$, the total cycle time.
\begin{figure}[htb]
  \begin{center}
    \includegraphics[width=0.47\textwidth]{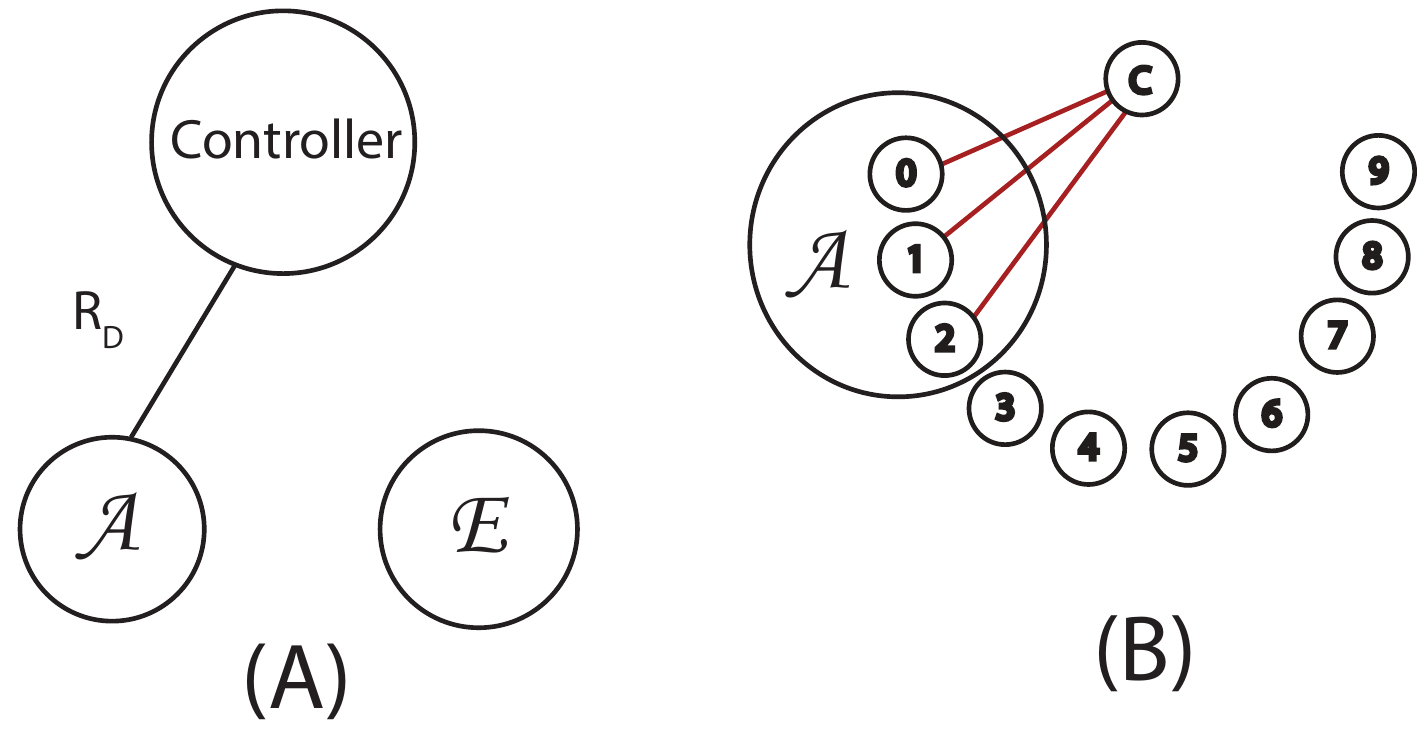}
  \end{center}
  \caption{We denote the set of nodes that have a direct link to the controller by $\mathcal{A}$. A node fails in one hop if it is not in Set $\mathcal{A}$. This case is the same for both downlink and uplink, but the rates of trasmission are $R_{D}$ and $R_U$, respectively. Just the downlink is depicted in this figure. Referring back to the original example used in the protocol section, nodes $S0$, $S1$, and $S2$ belong in Set $\mathcal{A}$, while the rest would fall under Set $\mathcal{E}$.}
  \label{fig:one_hop}
\end{figure}
\FloatBarrier
\subsubsection{One-Hop Downlink}
\begin{theorem}
Let the downlink time be $T_D$, the number of non-controller nodes be $n$, and the message size be $m$. The transmission rate is given by $R_D = \frac{m \cdot n}{T_D}$, and the corresponding probability of failure of a single link, denoted by $p_D$, is given by Eq.~\eqref{eq:pfail_singlelink}. The probability of cycle failure is then
\begin{equation}
P(\text{fail, 1D}) = F(n,p_D)
\label{eq:1D}
\end{equation}
\end{theorem}
\begin{IEEEproof}
The rate of transmission is $R_{D}=\frac{m \cdot n}{T_D}$. Hence, following Eq.~\eqref{eq:pfail_singlelink}, we can define probability $p_{D}$ of failure of a single link. The protocol succeeds only if all nodes receive their messages from the controller in a single transmission. Therefore their point-to-point links to the controller must all succeed (see Fig.~\ref{fig:one_hop}). Thus we get that the probability of failure for a one-hop downlink protocol is $P(\text{fail, 1D}) = F(n,p_{D})$.
\end{IEEEproof}

\subsubsection{One-Hop Uplink}
\begin{theorem}
Let the uplink time be $T_U$, the number of non-controller nodes be $n$, and the message size be $m$. The transmission rate is given by $R_U = \frac{m\cdot n}{T_U}$ and the corresponding probability of failure of a single link, denoted by $p_U$, is given by Eq.~\eqref{eq:pfail_singlelink}. The probability of cycle failure is then
\begin{equation}
P(\text{fail, 1U}) = F(n,p_U).
\label{eq:1U}
\end{equation}
\end{theorem}
\begin{IEEEproof}
For the uplink transmission rate of $R_{U}={\frac{m \cdot n}{T_U}}$, the probability of failure of a single link is denoted as $p_U$. Analogous to downlink, a one-hop uplink protocol succeeds if and only if all nodes get their information to the controller in a single transmission (see Fig.~\ref{fig:one_hop}).
Thus we get $P(\text{fail, 1U}) = F(n,p_{U})$.
\end{IEEEproof}

\subsection{Two-Hop Protocol}
In a two-hop protocol, both the controller and the nodes get two chances to get their messages across. 
Phases 5 and 7 in Fig.~\ref{fig:protocol} would not occur. Again we use the union bound to upper bound the total probability of cycle error by adding the probability of downlink failure and the probability uplink failure. If downlink wasn't successful, the nodes would not have the scheduling information thus leading to uplink failure as well.
Thus, we see that the union bound is a conservative estimate of the total probability of cycle failure.

\subsubsection{Two-Hop Downlink}
\begin{theorem}
Let the Phase \upperRomannumeral{1} downlink time be $T_{D_1}$, the Phase \upperRomannumeral{2} downlink time be $T_{D_2}$, the number of non-controller nodes be $n$, and the message size be $m$. The Phase \upperRomannumeral{1} transmission rate is given by $R_{D_1} = \frac{m\cdot n}{T_{D_1}}$ and the corresponding probability of a single link failure, $p_{D_1}$, is given by Eq.~\eqref{eq:pfail_singlelink}. The Phase \upperRomannumeral{2} transmission rate is given by $R_{D_2}^{(a)} = \frac{m\cdot (n-a)}{T_{D_2}} + \frac{2n}{T_{D_2}}$, where $a$ is the number of ``successful nodes'' in Phase \upperRomannumeral{1} and the corresponding probability of a single failure, $p_{D_2}^{(a)}$, is given by Eq.~\eqref{eq:pfail_singlelink} (the superscript $(a)$ is to indicate the dependence on $a$). The probability of downlink failure is then
\begin{equation}
P(\text{fail, 2D}) = \sum\limits_{a = 0}^{n-1} F\left(n - a, \left(p_{D_2}^{(a)}\right)^{a} \cdot p_{con}^{(a)}\right) B(n,a, p_{D_1})
\label{eq:2D}
\end{equation}
where, $p_{con}^{(a)} = \min\left({\frac{p_{D_2}^{(a)}}{p_{D_1}}, 1}\right)$.
\end{theorem}

\begin{IEEEproof}
A node can succeed by having a direct link to the controller in the first hop ($\mathcal{A}$), or by having a direct link to either the controller or set $\mathcal{A}$ in the second hop ($\mathcal{B}$). Note that it is possible for a node to not have a direct link to the controller under the initial rate, but have a direct link under the Phase \upperRomannumeral{2} rate. 
In Fig.~\ref{fig:2d}, we see that this list is exhaustive. We will now derive the probability that there exists at least one node that does not fall in Set $\mathcal{A}$ or $\mathcal{B}$.
The rate of transmission in Phase \upperRomannumeral{1}, $R_{D_1}$, is dictated by the time allocated for this phase, $T_{D_1}$, given by $\frac{m\cdot n}{T_{D_1}}$. Let $\mathcal{A}$ (cardinality $a$), be the set of successful nodes in Phase \upperRomannumeral{1}. The rate in Phase \upperRomannumeral{2}, $R_{D_2}^{(a)}$, depends on the realized $a$ and the time allocated for this phase, $T_{D_2}$. The result is $R_{D_2}^{(a)} = \frac{m\cdot (n-a)}{T_{D_2}} + \frac{2n}{T_{D_2}}$, where $\frac{2n}{T_{D_2}}$ is the rate of the scheduling message sent (1 bit for downlink acknowledgement and 1 bit for uplink acknowledgement).
\begin{figure}[htb]
  \begin{center}
    \includegraphics[width=0.4\textwidth]{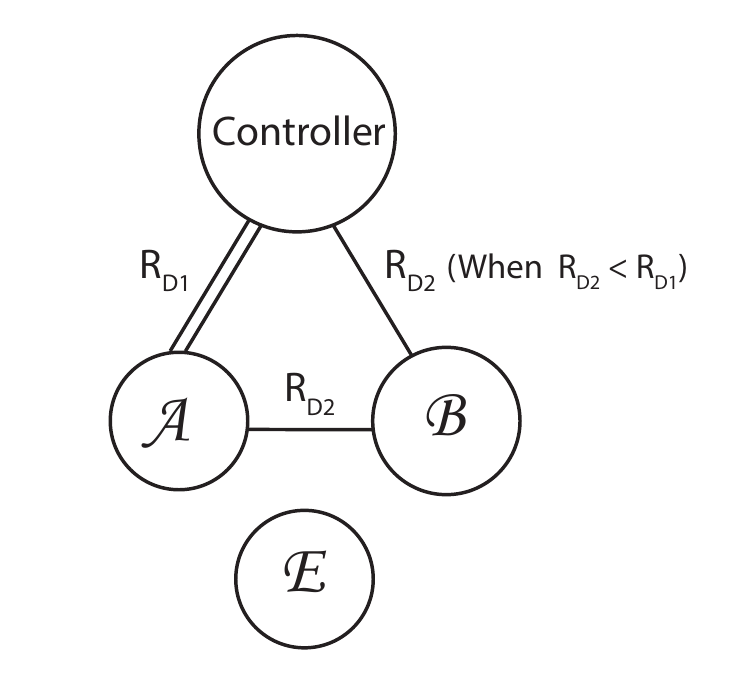}
  \end{center}
  \caption{The only ways to succeed in a two-hop protocol is by having a direct link to the controller to begin with (double line), or having a direct link under the new rate (single line) to either the controller or one of the nodes who heard the controller to begin with. }
 \label{fig:2d}
\end{figure}
\FloatBarrier

For ease of analysis, we make use of the fact that the scheduling phase effectively behaves as an extension of the downlink portion of the protocol.
Let the probability of link failure corresponding to $R_{D_1}$ and $R_{D_2}^{(a)}$ be defined as $p_{D_1}$ and $p_{D_2}^{(a)}$, respectively, by following Eq.~\eqref{eq:pfail_singlelink}).
As mentioned before, a link to the controller may improve in Phase \upperRomannumeral{2}. The probability that a controller-to-node link fails in phase \upperRomannumeral{2}, given it failed in phase \upperRomannumeral{1}, is given by\footnote{Recall that the fading distributions are assumed to be Rayleigh. Hence $p_{con}^{(a)} = P(R_{D_2}^{(a)} > C | R_{D_1} > C) = \frac{P(R_{D_2}^{(a)} > C \& R_{D_1} > C)}{P(R_{D_1} > C)} = \frac{P(C < \min{\lbrace R_{D_1},R_{D_2}^{(a)}}\rbrace)}{P(C < R_{D_1})}$. Then we use Eq.~\eqref{eq:pfail_singlelink} to get the final expression.} $p_{con}^{(a)} = P\left(R_{D_2}^{(a)} > C | R_{D_1} > C\right) = \min\left({\frac{p_{D_2}^{(a)}}{p_{D_1}}, 1}\right)$.

We decouple the two phases of the protocol. An error event can only occur if fewer than $n$ nodes succeed in Phase \upperRomannumeral{1} --- $\mathcal{A} < n$. The probability of a certain number of nodes succeeding in the first round, $P(A = a)$ can be modeled as a binomial distribution with probability of failure $p_{D_1}$, as a node must rely on just its link to the controller.
Thus, $P(A = a) = B(n,a, p_{D_1})$.

Conditioned on the number of nodes that succeeded in Phase \upperRomannumeral{1}, the probability of a node in $\mathcal{S}\backslash \mathcal{A}$ failing in Phase \upperRomannumeral{2} reduces to the probability of the node failing to reach any of the nodes in $\mathcal{A}$ and the controller under the new rate, $R_{D_2}^{(a)}$.
Each node in $\mathcal{S}\backslash \mathcal{A}$ has a probability ${\left(p_{D_2}^{(a)}\right)^{a}} \cdot p_{con}^{(a)}$ of failing in this way, where $p_{con}^{(a)}$ is the probability of failing to the controller under the new rate and $\left(p_{D_2}^{(a)}\right)^{a}$ is the probability of failing to reach any of the previously successful nodes.
Hence the probability that at least one of the remaining $n-a$ is unable to connect to the controller can be expressed with Eq.~\eqref{eq:F} as, $P(\text{fail}|A=a) = F\left(n - a, {\left(p_{D_2}^{(a)}\right)^{a}} \cdot p_{con}^{(a)}\right)$.

We then sum over all possible values of $a$ less than or equal to $n-1$, as a cycle failure only occurs when at least one node fails. The probability of failure of the 2-hop downlink protocol is then given by:
\begin{equation}
P(\text{fail, 2D}) = \sum\limits_{a = 0}^{n-1} P(\text{fail}|A=a) \cdot P(A = a) = \sum\limits_{a = 0}^{n-1} F\left(n - a, {\left(p_{D_2}^{(a)}\right)^{a}} \cdot p_{con}^{(a)}\right) B(n,a, p_{D_1})
\label{eq:2D}
\end{equation}
\end{IEEEproof}

\subsubsection{Two-Hop Uplink}
\begin{theorem}
Let the Phase \upperRomannumeral{1} uplink time be $T_{U_1}$, the Phase \upperRomannumeral{2} uplink time be $T_{U_2}$, the number of non-controller nodes be $n$ and the message size be $m$.
The Phase \upperRomannumeral{1} transmission rate is given by $R_{U_1} = \frac{(m+1)\cdot n}{T_{U_1}}$, and the corresponding probability of a single link failure, $p_{U_1}$, is given by Eq.~\eqref{eq:pfail_singlelink}.
The Phase \upperRomannumeral{2} transmission rate is given by $R_{U_2}^{(a)} = \frac{m\cdot (n-a)}{T_{U_2}}$, where $a$ is the number of ``successful nodes'' in Phase \upperRomannumeral{1} and the corresponding probability of a single failure, $p_{U_2}^{(a)}$, is given by Eq.~\eqref{eq:pfail_singlelink}. The probability of cycle failure is then
\begin{eqnarray}
\begin{aligned}
P(\text{fail, 2U}) &= \sum\limits_{a = 0}^{a_{0} - 1}\sum\limits_{a_2 = 0}^{a} F\left(M_U, {p_{\scriptscriptstyle U_1}^{a_2}}\right)  B\left(a, a_2, q^{(a)}\right)\cdot B(n,a,p_{U_1})\\
&+ \sum\limits_{a = a_{0}}^{n - 1}\sum\limits_{b_2 = 0}^{M_U -1 } F\left(M_U - b_2, {p_{\scriptscriptstyle U_1}^{a + b_2 }}\right) B\left(M_U, b_2, 1 - \widetilde{q}^{(a)}\right) B(n,a,p_{U_1})
\label{eq:2U}
\end{aligned}
\end{eqnarray}
where,
\begin{multicols}{2}
\begin{itemize}
\item $a_0 = \min \left(n \cdot \frac{T_{U_1} - T_{U_2}}{T_{U_1}}, 0\right)$
\item ${q}^{(a)} = P\left( C < R_{U_2}^{(a)} | C > R_{U_1}\right) = \frac{p_{U_2}^{(a)} - p_{U_1}}{1 - p_{U_1}}$
\item ${\tilde{q}^{(a)}} = P\left(R_{U_2}^{(a)} < C | R_{U_1} > C\right) = 1- \frac{p_{U_2}^{(a)}}{p_{U_1}}$
\item $M_U = n - a$
\end{itemize}
\end{multicols}
\end{theorem}

\begin{IEEEproof}
The derivation of the two-hop uplink error is a little more involved. For the two-hop uplink, the rate of transmission in Phase \upperRomannumeral{1}, $R_{U_1}$, is dictated by the time allocated for this phase, $T_{U_1}$ and is equal to $\frac{(m+1)\cdot n}{T_{U_1}}$. Let the nodes that were successful in Phase \upperRomannumeral{1} be in Set $\mathcal{A}$ (cardinality $a$). The rate in Phase \upperRomannumeral{2}, $R_{U_2}^{(a)}$, depends on the realization of $a$, and the time allocated for this phase, $T_{U_2}$.
The result is $R_{U_2}^{(a)} = \frac{m\cdot (n-a)}{T_{U_2}}$. This means there are two distinct cases to consider, one where the new rate has increased, and one where it has decreased.

\noindent\textbf{Case 1: $R_{U_2}^{(a)} \geq R_{U_1}$}\\
If the second phase rate is higher, the means of success can be as depicted in Fig.~\ref{fig:2U_1}. We will now derive the probability of error for this case.
When $R_{U_2}^{(a)} \geq R_{U_1}$, some initially successful links will no longer exist as the link between nodes may not be capable of tolerating a higher rate (the rate of transmission may become larger than capacity). In order to enter this case, there exists a threshold, $a_{0}$, of how many users must fail in Phase \upperRomannumeral{1}. The threshold is derived from the condition for having $R_{U_2}^{(a)} \geq R_{U_1}$, as $a_0 = \min \left(n \cdot \frac{T_{U_1} - T_{U_2}}{T_{U_1}}, 0\right)$.

\begin{figure}
  \begin{center}
    \includegraphics[width=0.4\textwidth]{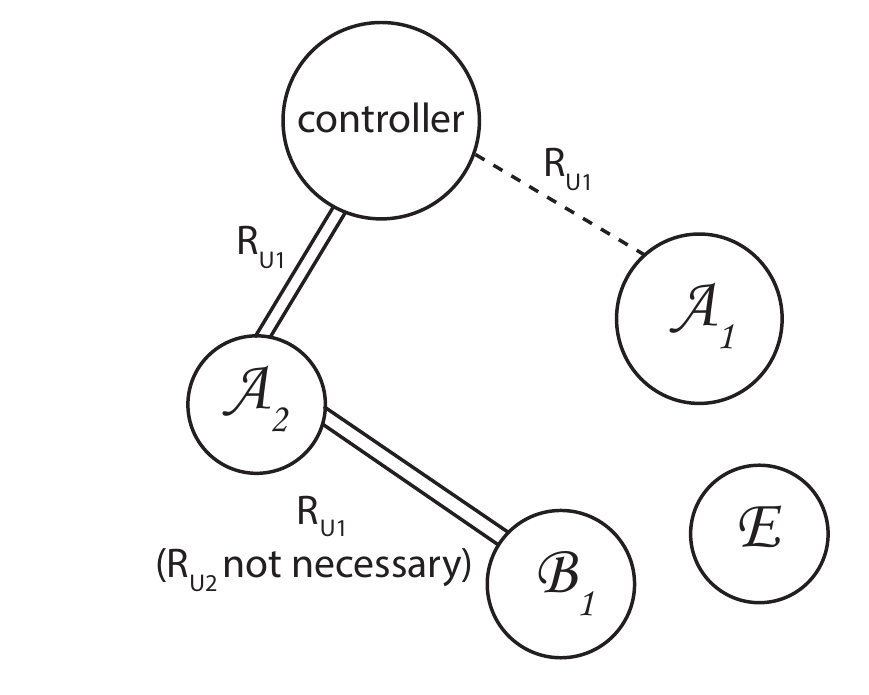}
  \end{center}
  \caption{This figure depicts the possible means of success in a two-hop uplink protocol when $R_{U_2}^{(a)} \ge R_{U_1}$. The paths are: only having a direct link to the controller under $R_{U_1}$ (dashed line), having a direct link under $R_{U_1}$ \& $R_{U_2}^{(a)}$ (double lines) to either the controller or one of the nodes who retained their link to the controller under $R_{U_2}^{(a)}$.}
 \label{fig:2U_1}
 \vspace{-20pt}
\end{figure}
There exist three methods of success in a two-hop uplink protocol with potentially increased rate.

\begin{itemize}
\item A node can have a direct link to the controller in the first phase, and in the second phase as well, under the higher rate. Let $\mathcal{A}_2$ (cardinality = $a_2$) be the nodes in $\mathcal{A}$ that retain their connection to the controller in both phases.

\item A node can simply have a link to the controller in the first phase, and lose its connection in the second phase. Let the probability of a successful link (in Phase \upperRomannumeral{1}) failing in Phase \upperRomannumeral{2} be denoted as\footnote{Recall that the fading distributions are assumed to be Rayleigh. Hence $q = P( C < R_{U_2}^{(a)} | C > R_{U_1}) = \frac{P(R_{U_1} < C < R_{U_2}^{(a)})}{P(C < R_{U_1})}$. Then we use Eq.~\eqref{eq:pfail_singlelink} to get the final expression.} ${q}^{(a)} = P( C < R_{U_2}^{(a)} | C > R_{U_1}) = \frac{p_{U_2}^{(a)} - p_{U_1}}{1 - p_{U_1}}$. The nodes that lose their links are in Set $\mathcal{A}\backslash \mathcal{A}_2 = \mathcal{A}_1$.

\item A node can succeed in two-hops if, in the \emph{first} phase, it connected to a node in $\mathcal{A}_2$, so its message can be relayed in the second phase. These nodes are denoted by $\mathcal{B}_1$ in Fig.~\ref{fig:2U_1}. This method is the only means of succeeding in the second phase, as we are in the case where the rate can only increase, so no new links will be formed.

\end{itemize}
We now derive the probability that a node is not in any of the above sets. We first expand the quantity we wish to compute into a form that is simpler to work with.
\begin{eqnarray*}
\begin{aligned}
P(\text{fail, 2U case 1}) &= P(\text{fail 2U \textbar case 1}) \cdot P(\text{case 1}) = \sum\limits_{a = 0}^{a_{0} - 1} P(\text{fail 2U}| A = a) \cdot P(A = a) \\
& = \sum\limits_{a = 0}^{a_{0} - 1}\sum\limits_{a_2 = 0}^{a} P(\text{fail to reach } \mathcal{A}_2 | A = a, A_2 = a_2) \cdot P(A_2 = a_2 | A = a) \cdot P(A = a )
\end{aligned}
\label{eq:2U_2}
\end{eqnarray*}

Conditioned on the events that occurred in Phase \upperRomannumeral{1}, i.e., given some realization of $A$ and $A_2$, a failure occurs when a node in $S \backslash A$ fails to reach any of the nodes in $A_2$ under $R_{U_1}$.
This can be expressed with Eq.~\eqref{eq:F}, as $P(\text{fail to reach } \mathcal{A}_2 | A = a, A_2 = a_2) = F(M_U, {p_{\scriptscriptstyle U_1}^{a_2}})$ where $M_U = n - a$.
Given that $A = a$ nodes succeeded in the first phase, we can calculate the probability of $A_2 = a_2$ by treating the probability of a given link failing as being distributed Bernoulli($1 -q$). Using Eq.~\eqref{eq:B}, we get $P(A_2 = a_2 | A = a) = B\left(a, a_2, q^{(a)}\right)$.
The probability that $A = a$ is then distributed as a binomial distribution, just as $A = a$ in the downlink case, meaning $P(A = a) = B(n,a,p_{U_1})$.
This gives us the first portion of Theorem 4, the probability of failure in a two-hop uplink scheme:
\begin{eqnarray*}
\begin{aligned}
P(\text{fail 2U, case 1})
& = \sum\limits_{a = 0}^{a_{0} - 1}\sum\limits_{a_2 = 0}^{a} \big\lbrace F\left(M_U, {p_{\scriptscriptstyle U_1}^{a_2}}\right)  B\left(a, a_2, q^{(a)}\right)\cdot B(n,a,p_{U_1})
 \big\rbrace
\end{aligned}
\label{eq:2U_2a}
\end{eqnarray*}
where $M_U = n - a$.

\noindent\textbf{Case 2: $R_{U_2}^{(a)} < R_{U_1}$}\\
We are interested in the event that $R_{U_2}^{(a)} < R_{U_1}$. This case arises when  $A = a > a_{0}$. Here, some new links may have been added to the system with probability\footnote{Recall that the fading distributions are assumed to be Rayleigh. Hence $\tilde{q}^{(a)} = P\left(R_{U_2}^{(a)} < C | R_{U_1} > C\right) = \frac{P\left(R_{U_2}^{(a)} < C < R_{U_1}\right)}{P(C < R_{U_1})}$. Then we use Eq.~\eqref{eq:pfail_singlelink} to get the final expression.} ${\tilde{q}^{(a)}} = P\left(R_{U_2}^{(a)} < C | R_{U_1} > C\right) = 1- \frac{p_{U_2}^{(a)}}{p_{U_1}}$.
Let $\mathcal{B}_2$ (cardinality $b_2$) be the nodes in ${S} \backslash \mathcal{A}$ that can directly reach the controller in Phase \upperRomannumeral{2}.
Fig.~\ref{fig:2U_2} portrays all possible paths of success. In order to succeed, a node must fall under one of three categories.
\begin{itemize}
\item A node may succeed directly in the first hop (is in $\mathcal{A}$). In this case, links cannot go bad, so no node in $\mathcal{A}$ loses connection to the controller.

\item A node may also succeed in the second phase by being able to connect to the controller under the new, lower rate (is in $\mathcal{B}_2$), even if it did not connect to the controller under the first rate.

\item A node can succeed in two-hops by reaching any other node in $\mathcal{A}_2$ or $\mathcal{B}_2$ in the first hop, and having its message relayed to the controller in the second hop (is in $\mathcal{B}_1$ in Fig.~\ref{fig:2U_2}).

\end{itemize}

We derive the probability that a node does not connect to the controller in any of the above ways. We first expand the quantity we wish to compute into a form that is simpler to work with.
\begin{eqnarray*}
\begin{aligned}
P(\text{fail 2U, case 2}) & = P(\text{fail 2U \textbar case 2}) \cdot P(\text{case 2}) = \sum\limits_{a = a_{0}}^{n - 1} P(\text{fail 2U}| A_2 = a) \cdot P(A_2 = a)\\
& = \sum\limits_{a = a_{0}}^{n - 1}\sum\limits_{b_2 = 0}^{M_U -1 } P(\text{fail to reach } \{\mathcal{A}_2, \mathcal{B}_2\} | A_2 = a, B_2 = b_2) \cdot P(B_2 = b_2 , A_2 = a)\\
\end{aligned}
\label{eq:2U}
\end{eqnarray*}
where $M_U = n - a$.

\begin{figure}
  \begin{center}
    \includegraphics[width=0.39\textwidth]{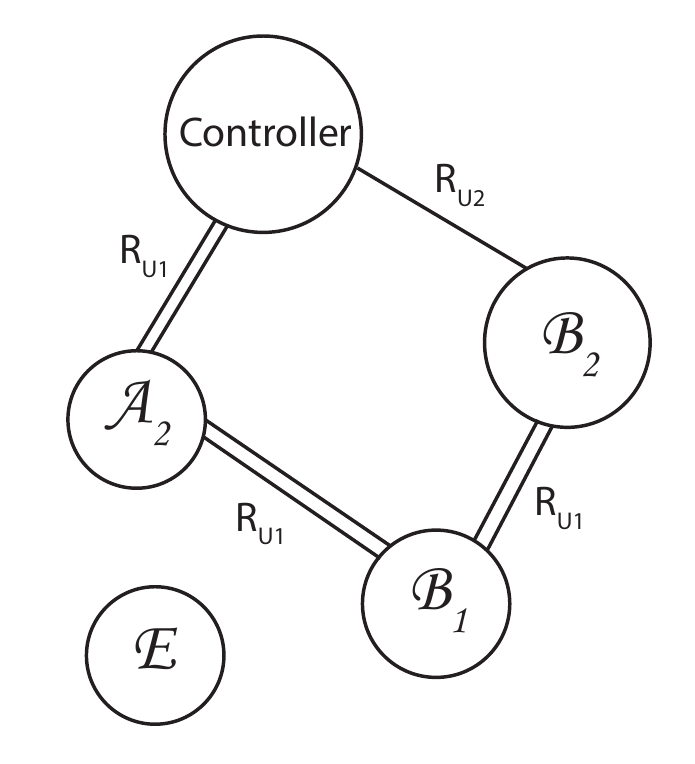}
  \end{center}
  \caption{This figure depicts the only ways to succeed in two-hop uplink, given that $R_{U_2} < R_{U_1}$. They are: to have a direct connection to the controller under any of the two rates, or to have connected, in phase \upperRomannumeral{1} (double lines), to a node that can succeed via a direct link to the controller.}
 \label{fig:2U_2}
 \vspace{-20pt}
\end{figure}
The first term in the final expression corresponds to failing to reach a previously successful node in Phase \upperRomannumeral{1}. Given some instantiation of $A_2$ and $B_2$, the probability that a node fails to reach the controller is the probability that it failed to reach any of the nodes in set $\mathcal{A}_2$ and $\mathcal{B}_2$ under the first rate. This is distributed Bernoulli with parameter $p_{U_1}^{a + b_2}$, so the probability that at least one node failed to reach the controller after two-hops can be expressed with Eq.~\eqref{eq:F} as $P(\text{fail to reach } \{\mathcal{A}_2, \mathcal{B}_2\} | A_2 = a, B_2 = b_2) = F(M_U - b_2, {p_{\scriptscriptstyle U_1}^{a + b_2 }}) $.

The probability of a node succeeding directly to the controller under $R_{U_2}^{(a)}$ given it was not in $\mathcal{A}_2$ is ${\tilde{q}}^{(a)}$, so the probability that $B_2 = b_2$ given $A_2 = a$ can be written with Eq.~\eqref{eq:B} as $P(B_2 = b_2 | A_2 = a)  =  B\left(M_U, b_2, 1 - \widetilde{q}^{(a)}\right)$.
The probability that $A_2=a$ is exactly as in the first case, as Set $\mathcal{A}_2$ is the set of nodes that were able to successfully transmit their message to the controller in Phase \upperRomannumeral{1}. This gives us $ B(n,a,p_{U_1})$, completing the second portion of Theorem 4 as follows.
\begin{eqnarray*}
\begin{aligned}
P(\text{fail 2U, case 2})
& = \sum\limits_{a = a_{0}}^{n - 1}\sum\limits_{b_2 = 0}^{M_U -1 }  F(M_U - b_2, {p_{\scriptscriptstyle U_1}^{a + b_2 }}) B\left(M_U, b_2, 1 - \widetilde{q}^{(a)} \right) B(n,a,p_{U_1})
\end{aligned}
\label{eq:2U}
\end{eqnarray*}
where $M_U = n - a$.
The probability of failure of the two-hop uplink protocol is then given by Eq.~\eqref{eq:2U}, where the first term comes from case 1, and the second is from case 2.
\begin{eqnarray}
\begin{aligned}
P(\text{fail 2U}) &= \sum\limits_{a = 0}^{a_{0} - 1}\sum\limits_{a_2 = 0}^{a} F(M_U, {p_{\scriptscriptstyle U_1}^{a_2}})  B\left(a, a_2, q^{(a)}\right)\cdot B(n,a,p_{U_1})\\
&+ \sum\limits_{a = a_{0}}^{n - 1}\sum\limits_{b_2 = 0}^{M_U -1 } F(M_U - b_2, {p_{\scriptscriptstyle U_1}^{a + b_2 }}) B\left(M_U, b_2, 1 - \widetilde{q}^{(a)}\right) B(n,a,p_{U_1})
\label{eq:2U}
\end{aligned}
\end{eqnarray}
where $M_U = n - a$.
\end{IEEEproof}

\subsection{Three-Hop Protocol}
The failed protocol depicted in Fig.~\ref{fig:protocol} is a three-hop protocol, where both the controller and nodes get three chances to get their message across.  The total time for downlink and uplink are optimally divided between the three phases to minimize the SNR required to attain a target probability of error. 

\subsubsection{Three-Hop Downlink}
\begin{theorem}
Let the Phase \upperRomannumeral{1}, Phase \upperRomannumeral{2} and Phase \upperRomannumeral{3} downlink time be $T_{D_1}$, $T_{D_2}$ and $T_{D_3}$ respectively, number of non-controller nodes be $n$, and message size be $m$.
The Phase \upperRomannumeral{1} transmission rate is given by $R_{D_1} = \frac{m \cdot n}{T_{D_1}}$, and the corresponding probability of a single link failure, $p_{D_1}$, is given by Eq.~\eqref{eq:pfail_singlelink}.
The Phase \upperRomannumeral{2} and Phase \upperRomannumeral{3} transmission rate is given by $R_{D_2}^{(a)}  = \frac{m\cdot (n-a)}{T_{D_2}} + \frac{2n}{T_{D_2}}$, and $R_{D_3}^{(a)} = \frac{m\cdot (n-a)}{T_{D_3}} + \frac{2n}{T_{D_3}}$ where $a$ is the number of ``successful nodes'' in Phase \upperRomannumeral{1}, and the corresponding probability of a single failure, $p_{D_2}$ and  $p_{D_3}$, is given by Eq.~\eqref{eq:pfail_singlelink}. The probability 3-hop downlink failure is then
\begin{equation}
P(\text{fail, 3D}) = \sum\limits_{a = 0}^{n-1} \sum\limits_{b = 0}^{M_D - 1} B(n, a, p_{D_1}) B\left(M_D, b, \left(p_{D_2}^{(a)}\right)^a q_{21}^{(a)}\right) F\left(M_D - b, \left(p_{D_3}^{(a)}\right)^b \left(q_{32}^{(a)}\right)^a q_{321}^{(a)}\right)
\label{eq:3D}
\end{equation}
where, $M_D = n - a$, $q_{21}^{(a)} = P\left(C < R_{D_2}^{(a)} | C < R_{D_1}\right) = \min\left(\frac{p_{D_2}^{(a)}}{p_{D_1}}, 1 \right)$, $q_{32}^{(a)} = P\left(C < R_{D_3}^{(a)} | C < R_{D_2}^{(a)}\right) = \min\left(\frac{p_{D_3}^{(a)}}{p_{D_2}^{(a)}}, 1 \right)$ and $q_{321}^{(a)} = P\left(C < R_{D_3}^{(a)} | C < \min(R_{D_1}, R_{D_2}^{(a)})\right) = \min\left( \max \left(\frac{p_{D_3}^{(a)}}{p_{D_1}}, \frac{p_{D_3}^{(a)}}{p_{D_2}^{(a)}}\right), 1 \right)$.
\end{theorem}
\begin{IEEEproof}
The rate of transmission in Phase \upperRomannumeral{1}, $R_{D_1}$, is determined by the time allocated for this phase, $T_{D_1}$. Let the nodes who were successful in Phase \upperRomannumeral{1} be in Set $\mathcal{A}$ (cardinality $a$). The rate in Phase \upperRomannumeral{2}, $R_{D_2}^{(a)}$ and Phase \upperRomannumeral{3}, $R_{D_3}^{(a)}$ depends on the realization of $a$, and the time allocated for the phase, $T_{D_2}$ and $T_{D_3}$. As before, $R_{D_2}^{(a)} = \frac{m\cdot (n-a)}{T_{D_2}} + \frac{2n}{T_{D_2}}$, $R_{D_3}^{(a)} = \frac{m\cdot (n-a)}{T_{D_3}} + \frac{2n}{T_{D_3}}$.
The probabilities of link error corresponding to each rate $R_{D_1}$, $R_{D_2}^{(a)}$ and $R_{D_3}^{(a)}$ are $p_{D_1}$, $p_{D_2}^{(a)}$ and $p_{D_3}^{(a)}$ respectively.
Fig.~\ref{fig:3D} displays an exhaustive list of ways to succeed in a three-hop downlink protocol.
\begin{itemize}
\item A node can succeed directly from the controller in the first hop under rate $R_{D_1}$ (Set $\mathcal{A}$).

\item A node can succeed in phase \upperRomannumeral{2} of the protocol by either directly connecting to the controller under the new rate, $R_{D_2}^{(a)}$, or by connecting to one of the nodes in Set $\mathcal{A}$ (is in Set $\mathcal{B}$).

\item A node can succeed in the third phase from any of the nodes in Set $\mathcal{B}$ or Set $\mathcal{A}$ (if $R_{D_3}^{(a)} < R_{D_2}^{(a)}$) or directly from the controller (if $R_{D_3}^{(a)} < \min(R_{D_2}^{(a)}, R_{D_1})$).

\end{itemize}

In order to calculate the probability of error of a three-hop downlink protocol, we will unroll the state space in a manner similar to the two-hop derivations. To calculate the overall probability of failure in 2-hop downlink, we sum over all possible instantiations of the sets of interest that result in failure. In this case, we are interested in the event that at least one node, which does not fall in Sets $\mathcal{A}$ and $\mathcal{B}$, is also not in $\mathcal{C}$ (fails given the instantiations of set $\mathcal{A}$ and $\mathcal{B}$).
$$P(\text{fail, 3D})  =  \sum\limits_{a = 0}^{n-1} \sum\limits_{b = 0}^{M_a-1} P(\text{fail}|A = a, B = b) P(B = b | A = a) P(A = a) \text{ where } M_D = n-a.$$
\begin{figure}[htb]
  \begin{center}
    \includegraphics[width=0.4\textwidth]{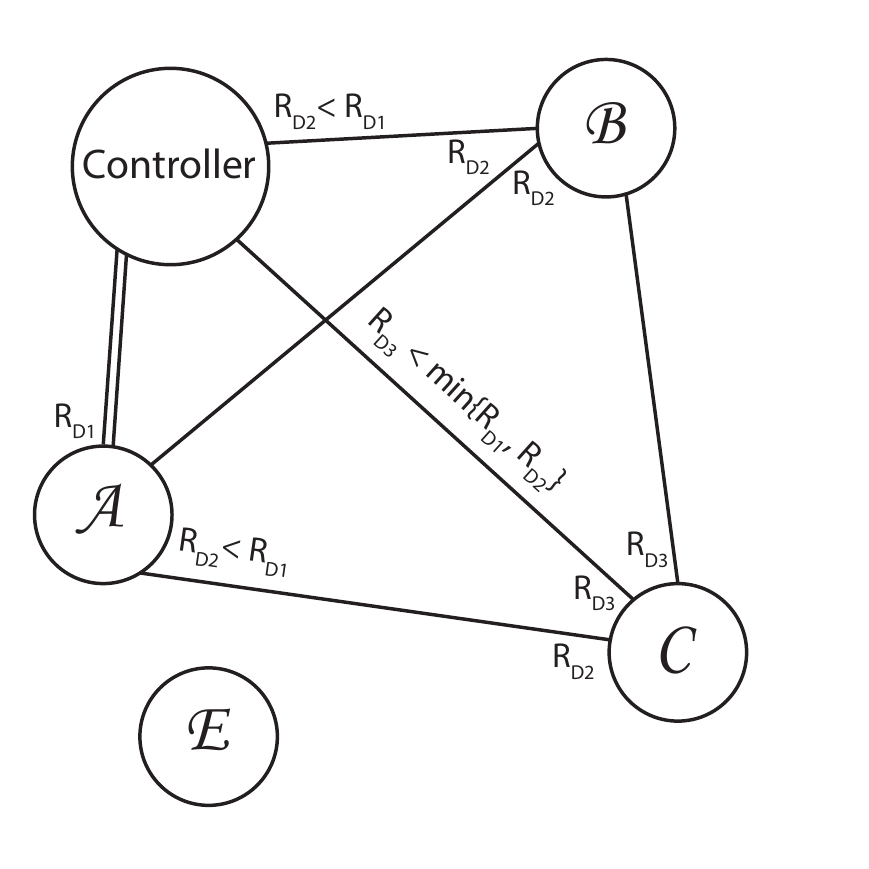}
  \end{center}
  \caption{The only ways to succeed in a three-hop downlink protocol are displayed. A node can succeed in the first phase directly from the controller, in Phase \upperRomannumeral{2} from either the controller or someone who succeeded in Phase \upperRomannumeral{1}, and in Phase \upperRomannumeral{3} from someone who succeeded in Phase \upperRomannumeral{2}. Please refer to Fig.~\ref{fig:bubs} to recall the exact meaning of each set name.}
 \label{fig:3D}
\end{figure}
\FloatBarrier
Given $B = b$ and $A = a$, the probability of a node (not in $\mathcal{A}$ or $\mathcal{B}$) failing after three-hops is the probability that it cannot receive its message from either a node in Set $\mathcal{B}$ or Set $\mathcal{A}$ (if $R_{D_3}^{(a)} < R_{D_2}^{(a)}$) or directly from the controller (if $R_{D_3}^{(a)} < \min(R_{D_2}^{(a)}, R_{D_1})$). This is distributed Bernoulli $\left(p_{D_3}^{(a)}\right)^b \cdot \left(q_{32}^{(a)}\right)^a \cdot q_{321}^{(a)}$, and can be written with Eq.~\eqref{eq:F} as
$F\left(n - (a + b), \left(p_{D_3}^{(a)}\right)^b \cdot \left(q_{32}^{(a)}\right)^a \cdot q_{321}^{(a)}\right)$\\
$ = F\left(M_D - b, \left(p_{D_3}^{(a)}\right)^b \cdot \left(q_{32}^{(a)}\right)^a \cdot q_{321}^{(a)}\right)$.
Given $A = a$, we can calculate the probability of a node not succeeding in Phase \upperRomannumeral{2} as $\left(p_{D_2}^{(a)}\right)^a q_{21}^{(a)}$, as it must fail to receive its message from all of the nodes in Set $\mathcal{A}$, and from the controller under the phase \upperRomannumeral{2} rate. Hence we calculate the probability that $B = b$ using a binomial distribution with parameter $\left(p_{D_2}^{(a)}\right)^a \cdot q_{21}^{(a)}$ as $B\left(M_D, b, \left(p_{D_2}^{(a)}\right)^a \cdot q_{21}^{(a)}\right)$.
The probability of $A = a$ is exactly the same as we have seen before, at it relies on just point to point links to the controller, each of which fails with probability $p_{D_1}$ (we use Eq.~\eqref{eq:pfail_singlelink}). This gives us $B(n, a, p_{D_1})$.
Therefore, the probability of failure of the 3-phase downlink protocol is given by
\begin{eqnarray*}
\begin{aligned}
P(\text{fail, 3D}) & =  \sum\limits_{a = 0}^{n-1} \sum\limits_{b = 0}^{M_D-1} P(A = a) P(B = b | A = a) P(\text{fail}|A = a, B = b)\\
& = \sum\limits_{a = 0}^{n-1} \sum\limits_{b = 0}^{M_D-1} B(n, a, p_{D_1}) B\left(M_D, b, \left(p_{D_2}^{(a)}\right)^a q_{21}^{(a)}\right) F\left(M_D - b, \left(p_{D_3}^{(a)}\right)^b \left(q_{32}^{(a)}\right)^a q_{321}^{(a)}\right)
\end{aligned}
\label{eq:D}
\end{eqnarray*}
where $M_D = n-a$.
\end{IEEEproof}

\subsubsection{Three-Hop Uplink}
\begin{theorem}
Let the Phase \upperRomannumeral{1}, Phase \upperRomannumeral{2} and Phase \upperRomannumeral{3} uplink time be $T_{U_1}$, $T_{U_2}$ and $T_{U_3}$ respectively, number of non-controller nodes be $n$, and message size be $m$.
The Phase \upperRomannumeral{1} transmission rate is given by $R_{U_1} = \frac{(m+1) \cdot n}{T_{U_1}}$. The Phase \upperRomannumeral{2} and Phase \upperRomannumeral{3} transmission rate is given by $R_{U_2}^{(a)}  = \frac{m\cdot (n-a)}{T_{U_2}}$, and $R_{U_3}^{(a)} = \frac{m\cdot (n-a)}{T_{U_3}}$ where $a$ is the number of ``successful nodes'' in Phase \upperRomannumeral{1}. The probability of cycle failure is then
\begin{eqnarray}
\begin{aligned}
P(\text{fail, 3U}) &= \sum_{a = 0}^{n - 1} \Bigg[ \left(  \sum_{b_2 = 0}^{n - a - 1} \sum_{b_1 = 0}^{n - a - b_2 - 1} \sum_{c_3 = 0}^{n - a - b - 1} \sum_{c_2 = 0}^{n - a - b - c_3 - 1}  P(\text{fail}_1) \right) \mathds{1}\left(R_{U_1} \geq R_{U_2} > R_{U_3}\right)  \\
&+  \left( \sum_{b_2 = 0}^{n - a - 1} \sum_{b_1 = 0}^{n - a - b_2 - 1} \sum_{\widehat{b}_2 = 0}^{b_2} \sum_{\widehat{b}_1 = 0}^{b_1} \sum_{c_2 = 0}^{n - a - b -1}  P(\text{fail}_2) \right) \mathds{1}\left(R_{U_1} > R_{U_3} \geq R_{U_2}\right)\\
&+  \left( \sum_{a_3 = 0}^{a} \sum_{b_2 = 0}^{n - a - 1} \sum_{b_1 = 0}^{n - a - b_2 - 1} \sum_{\widehat{b}_1 = 0}^{b_1} \sum_{c_2 = 0}^{n - a - b - 1} P(\text{fail}_3) \right) \mathds{1}\left(R_{U_3} \geq R_{U_1} > R_{U_2}\right) \\
&+  \left( \sum_{a_2 = 0}^{a} \sum_{a_3 = 0}^{a_2} \sum_{\widehat{a}_1 = 0}^{a - a_2} \sum_{b_1 = 0}^{n - a - 1} \sum_{\widehat{b}_1 = 0}^{b_1} P(\text{fail}_4) \right) \mathds{1}\left(R_{U_3} > R_{U_2} \geq R_{U_1}\right)\\
&+  \left( \sum_{a_2 = 0}^{a} \sum_{\widetilde{a} = 0}^{a - a_2} \sum_{\widehat{a}_1 = 0}^{a - a_2 - \widetilde{a}_1} \sum_{b_1 = 0}^{n - a - 1} \sum_{\widehat{b}_1 = 0}^{b_1} P(\text{fail}_5) \right) \mathds{1}\left(R_{U_2} \geq R_{U_3} > R_{U_1}\right)\\
&+  \left( \sum_{a_2 = 0}^{a} \sum_{b_1 = 0}^{n - a - 1} \sum_{\widehat{b}_1 = 0}^{b_1} \sum_{c_3 = 0}^{n - a - b_1 - 1} \sum_{c_2 = 0}^{n - a - b_1 - c_3 - 1} \sum_{\widehat{c}_2 = 0}^{c_2} P(\text{fail}_6) \right) \mathds{1}\left(R_{U_2} > R_{U_1} \geq R_{U_3}\right) \Bigg]\\
\label{eq:3U}
\end{aligned}
\end{eqnarray}
where
\begin{eqnarray*}
\begin{aligned}
P(\text{fail}_1) &= F\left(n - a - b - c_2 - c_3, p_1^{b_1 + c_2}\right) \times B\left(n - a - b - c_3, c_2, q_{21}^{a + b_2 + c_3}\right) \times \\
& \hspace{10pt} \times B\left(n - a - b, c_3, q_{32}\right) \times B\left(n - a - b_2, b_1, p_1^{a + b_2}\right) \times B\left(n - a, b_2, q_{21}\right) \times B(n, a, p_1)
\end{aligned}
\end{eqnarray*}
is the probability of failure of the 3-hop uplink protocol if the relationship between the rates is $R_{U_1} \geq R_{U_2} > R_{U_3}$,
\begin{eqnarray*}
\begin{aligned}
P(\text{fail}_2) &= F\left(n - a - b - c_2, p_1^{\widehat{b}_1 + c_2}\right) \times B\left(n - a - b, c_2, q_{21}^{a + \widehat{b}_2}\right) \times B(n - a - b_2, b_1, p_{1}^{a + b_2}) \times \\
& \hspace{10pt} \times B\left(b_1, \widehat{b}_1, s_{22}[a + \hat{b}_2, a + b_2]\right) \times B\left(b_2, \hat{b}_2, r_{32}\right) \times B\left(n - a, b_2, q_{21}\right) \times B(n, a, p_1)
\end{aligned}
\end{eqnarray*}
is the probability of failure of the 3-hop uplink protocol if the relationship between the rates is $R_{U_1} > R_{U_3} \geq R_{U_2}$,
\begin{eqnarray*}
\begin{aligned}
P(\text{fail}_3) &= F\left(n - a - b - c_2, p_1^{\widehat{b}_1 + c_2}\right) \times B\left(n - a - b, c_2, q_{21}^{a_3}\right) \times B\left(b_1, \widehat{b}_1, s_{22}[a_3, a + b_2] \right) \times \\
& \hspace{10pt} \times B\left(n - a - b_2, b_1, p_1^{a+b_2}\right) \times B\left(a, a_3, r_{31}\right) \times B\left(n-a, b_2, q_{21}\right) \times B(n, a, p_1)
\end{aligned}
\end{eqnarray*}
is the probability of failure of the 3-hop uplink protocol if the relationship between the rates is $R_{U_3} \geq R_{U_1} > R_{U_2}$,
\begin{eqnarray*}
\begin{aligned}
P(\text{fail}_4) &= F\left(n - a - b_1, p_1^{\widehat{a}_1 + \widehat{b}_1}\right) \times B(a_1, \widehat{a}_1, p_2^{a_3}) \times B\left(b_1, \widehat{b}_1, s_{21}[a_3, a_2]\right) \times B\left(n - a, b_1, p_1^{a_2}\right) \times  \\
& \hspace{10pt}  \times B\left(a_2, a_3, r_{32}\right) \times B\left(a, a_2, r_{21}\right) \times B(n, a, p_1)
\end{aligned}
\end{eqnarray*}
is the probability of failure of the 3-hop uplink protocol if the relationship between the rates is $R_{U_3} > R_{U_2} \geq R_{U_1}$,
\begin{eqnarray*}
\begin{aligned}
P(\text{fail}_5) &= F\left(n - a - b_1, p_1^{\widehat{a}_1 + \widehat{b}_1}\right) \times B\left(a - \widetilde{a}_1 - a_2, \widehat{a}_1, p_{2}^{\widetilde{a}_1 + a_2}\right) \times B\left(b_1, \widehat{b}_1, s_{21}[a_2, a_2]\right) \times\\
& \hspace{10pt} \times B\left(n - a, b_1, p_1^{a_2}\right) \times B\left(a - a_2, \widetilde{a}_1, m_{312}\right) \times B\left(a, a_2, r_{21}\right) \times B(n, a, p_1)
\end{aligned}
\end{eqnarray*}
is the probability of failure of the 3-hop uplink protocol if the relationship between the rates is $R_{U_2} \geq R_{U_3} > R_{U_1}$,
\begin{eqnarray*}
\begin{aligned}
P(\text{fail}_6) &= F\left(n - a - b - c_2 - c_3, p_1^{\widehat{b}_1 + \widehat{c}_2}\right) \times B\left(c_2, \widehat{c}_2, s_{21}[a + c_3, a + c_3]\right) \times B\left(b_1, \widehat{b}_1, s_{21}[a_2, a_2]\right) \times \\
& \hspace{10pt} \times B(n - a - b_1, c_3, q_{31}) \times B(n - a - b - c_3, c_2, p_{1}^{a_1 + c_3})  \times\\
& \hspace{10pt} \times B\left(n - a, b_1, p_1^{a_2}\right) \times B\left(a, a_2, r_{21}\right) \times B(n, a, p_1)
\end{aligned}
\end{eqnarray*}
is the probability of failure of the 3-hop uplink protocol if the relationship between the rates is $R_{U_2} > R_{U_1} \geq R_{U_3}$, where:
\begin{multicols}{2}
\begin{itemize}
\item $p_1 = p_{U_1} = P( C < R_{U_1} )$
\item $p_2 = p_{U_2}^{(a)} = P( C < R_{U_2}^{(a)} )$
\item $p_3 = p_{U_3}^{(a)} = P( C < R_{U_3}^{(a)} )$
\item $q_{21} = P( C < R_{U_2}^{(a)} | C < R_{U_1} )$
\item $q_{31} = P( C < R_{U_3}^{(a)} | C < R_{U_1} )$
\item $q_{32} = P( C < R_{U_3}^{(a)} | C < R_{U_2}^{(a)} )$
\item $r_{21} = P( C < R_{U_2}^{(a)} | C > R_{U_1} )$
\item $r_{31} = P( C < R_{U_3}^{(a)} | C > R_{U_1} )$
\item $r_{32} = P( C < R_{U_3}^{(a)} | C > R_{U_2}^{(a)} )$
\item $m_{312} = P(C < R_{U_3}^{(a)} | R_{U_1} < C < R_{U_2}^{(a)})$
\item $s_{ij}[f, g] = {(1 - p_{i}^f)}/{(1 - p_{j}^g)}$ where $f$ and $g$ are cardinalities of sets $F$ and $G$.
\item $b = b_1 + b_2$
\end{itemize}
\end{multicols}

\end{theorem}
\begin{IEEEproof}
We will now deal with each case one-by-one to understand all the subtle effects that occur in the uplink case.\\
\noindent\textbf{Case 1: $R_{U_1} \geq R_{U_2} > R_{U_3}$}\\
The rate of transmission in Phase \upperRomannumeral{1}, $R_{U_1}$, is determined by the time allocated for this phase, $T_{U_1}$. Let the nodes who were successful in Phase \upperRomannumeral{1} be in Set $\mathcal{A}$ (cardinality $a$). The rate in Phase \upperRomannumeral{2}, $R_{U_2}^{(a)}$ and Phase \upperRomannumeral{3}, $R_{U_3}^{(a)}$ depends on the realization of $a$, and the time allocated for the phase, $T_{U_2}$ and $T_{U_3}$. As before, $R_{U_2}^{(a)} = \frac{m\cdot (n-a)}{T_{U_2}}$, $R_{U_3}^{(a)} = \frac{m\cdot (n-a)}{T_{U_3}}$.
The probabilities of link error corresponding to each rate $R_{U_1}$, $R_{U_2}^{(a)}$ and $R_{U_3}^{(a)}$ are $p_{U_1}$, $p_{U_2}^{(a)}$ and $p_{U_3}^{(a)}$ (abbreviated to $p_1$, $p_2$ and $p_3$) respectively.
Fig.~\ref{fig:3U_1} displays an exhaustive list of ways to succeed in case 1 of the three-hop uplink protocol.

\begin{figure}[htb]
\centering
\includegraphics[width = 0.6\textwidth]{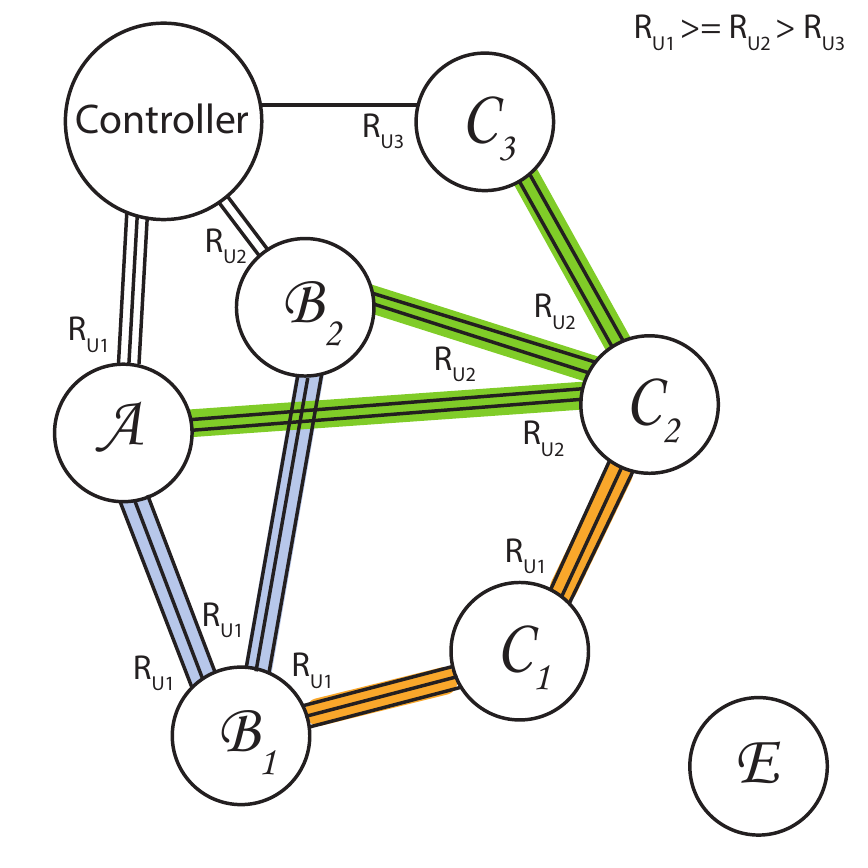}
\caption{{Case 1: $R_{U_1} \geq R_{U_2} > R_{U_3}$. The only ways to succeed in the 1st case of 3-hop uplink protocol are displayed. A node can succeed in Phase \upperRomannumeral{1} directly, in Phase \upperRomannumeral{2} by connecting to the controller or a node which can succeed in Phase \upperRomannumeral{2}, and in Phase \upperRomannumeral{3} by directly connecting to the controller or connecting to the nodes which have connections to the controller in Phase \upperRomannumeral{2} (thus succeeding in 2 hops) or connecting via 2 hops to the nodes which have connections to the controller (thus succeeding in 3 hops).}}
\label{fig:3U_1}
\end{figure}
\FloatBarrier
\begin{itemize}
\item A node can succeed directly to the controller in the first hop under rate $R_{U_1}$ (is in set $\mathcal{A}$).

\item A node can succeed in the second phase of the protocol by connecting directly to the controller under the new rate, $R_{U_2}^{(a)}$ (is in set $\mathcal{B}_2$).

\item A node can succeed in the second phase of the protocol by connecting in the first phase (is in set $\mathcal{B}_1$) to one of the nodes in the set $\mathcal{A} \bigcup \mathcal{B}_2$ (the set of nodes which can communicate to the controller in phase \upperRomannumeral{2}).
This ensures that the nodes which can connect to the controller in the second phase already have the message.

\item A node can succeed in the third phase of the protocol by connecting directly to the controller under the new rate, $R_{U_3}^{(a)}$ (is in set $\mathcal{C}_3$).

\item A node can succeed in the third phase in a two-hop fashion by connecting to the set $\mathcal{A} \bigcup {\mathcal{B}}_2 \bigcup \mathcal{C}_3$ under the lower phase two rate $R_{U_2}^{(a)}$ (is in set $\mathcal{C}_2$). The set $\mathcal{A} \bigcup {\mathcal{B}}_2 \bigcup \mathcal{C}_3$ is the set of nodes which can connect to the controller in the third phase. Connecting to $\mathcal{A} \bigcup {\mathcal{B}}_2 \bigcup \mathcal{C}_3$ in phase \upperRomannumeral{2} ensures that the message to be conveyed in phase \upperRomannumeral{3} has been conveyed to the relays by phase \upperRomannumeral{2}.

\item A node can succeed in the third phase in a three-hop fashion by connecting to the set $\mathcal{C}_2 \bigcup {\mathcal{B}}_1$ in the first phase under rate $R_{U_1}$ (is in set $\mathcal{C}_1$). The set $\mathcal{C}_2 \bigcup {\mathcal{B}}_1$ is the set of nodes which can connect to the set $\mathcal{A} \bigcup {\mathcal{B}}_2 \bigcup \mathcal{C}_3$ (they can connect to the controller in the third phase) in the second phase. Connecting to the set $\mathcal{C}_2 \bigcup {\mathcal{B}}_1$ in the first phase ensures that the message to be conveyed in the third phase has been conveyed to the right relays by the second phase.
\end{itemize}

To calculate the probability of error of the three-hop uplink protocol, we will unroll the state space in a manner similar to the three-hop downlink derivations and sum over all possible instantiations of the sets of interest that result in failure. In this case, we are interested in the event that at least one node which does not fall in sets $\mathcal{A}$, $\mathcal{B} = \mathcal{B}_1 \bigcup \mathcal{B}_2$ and $\mathcal{C}_2 \bigcup \mathcal{C}_3$, is also not in $\mathcal{C}_1$ (fails given the instantiations of set $\mathcal{A}$, $\mathcal{B}$, $\mathcal{C}$).

The probability of $A = a$ is exactly the same as we have seen before, as it relies on just point to point links to the controller, each of which fails independently with probability $p_1 = p_{U_1}$ (we use Eq.~\eqref{eq:pfail_singlelink}). This gives us $B(n, a, p_1)$.
Given $A = a$, we can calculate the probability of a node not being able to gain a connection to the controller in the second phase given there was no connection in the first phase as $q_{21} = P(C < R_{U_2}^{(a)} | C < R_{U_1}^{(a)}) = (p_2)/( p_1)$. $\mathcal{B}_2$ is the set which can connect to the controller in the second phase. Hence we calculate the probability that ${B}_2 = {b}_2$ using a binomial distribution with parameter $q_{21}$ as $B(n - a, b_2, q_{21})$.

Given $A = a$ and $B_2 = b_2$, we can calculate the probability of a node not succeeding in Phase \upperRomannumeral{2} in two hops as $p_{1}^{a + b_2}$, as it must fail to connect to $\mathcal{A} \bigcup \mathcal{B}_2$ in the first phase.
Hence we calculate the probability that $B_1 = b_1$ using a binomial distribution with parameter $p_{1}^{a + b_2}$ as $B(n - a - b_2, b_1, p_{1}^{a + b_2})$.

Given $A = a$, $B_2 = b_2$ and $B_1 = b_1$, we can calculate the probability of a node not being able to gain a connection to the controller in the third phase given there was no connection in the first two phases as $q_{32} = P(C < R_{U_3}^{(a)} | C < R_{U_2}^{(a)}) = (p_3)/( p_2)$. $\mathcal{C}_3$ is the set which can connect to the controller in the third phase. Hence we calculate the probability that ${C}_3 = {c}_3$ using a binomial distribution with parameter $q_{32}$ as $B(n - a - b, c_3, q_{32})$.

Given $A = a$, $B_1 = b_1$, $B_2 = b_2$ and $C_3 = c_3$, we can calculate the probability of a node not succeeding in Phase \upperRomannumeral{3} in two hops as $q_{21}^{a + {b}_2 + c_3}$, as it must fail to connect to $\mathcal{A} \bigcup {\mathcal{B}}_2 \bigcup \mathcal{C}_3$ in the second phase having failed to connect in the first phase already.
Hence we calculate the probability that $C_2 = c_2$ using a binomial distribution with parameter $q_{21}^{a + {b}_2 + c_3}$ as $B(n - a - b- c_3, c_2, q_{21}^{a + {b}_2 + c_3})$.
Given $A = a$, $B_1 = b_1$, $B_2 = b_2$, $C_3 = c_3$ and $C_2 = c_2$, the probability of a node (not in $\mathcal{A} \bigcup \mathcal{B}_1 \bigcup \mathcal{B}_2 \bigcup \mathcal{C}_2 \bigcup \mathcal{C}_3$) failing after three-hops is the probability that it cannot connect to $\mathcal{C}_2 \bigcup {\mathcal{B}}_1$ in the first phase. This is distributed Bernoulli $p_1^{{b}_1 + c_2}$, and can be written with Eq.~\eqref{eq:F} as $F(n - a - b - c_2 - c_3, p_1^{{b}_1 + c_2})$.

Thus we have that given the realization $A = a$, the probability that the protocol fails under case 1: $R_{U_1} \geq R_{U_2} > R_{U_3}$ is given by
$$ P(\text{fail} | \text{Case 1}, A = a) = \left(  \sum_{b_2 = 0}^{n - a - 1} \sum_{b_1 = 0}^{n - a - b_2 - 1} \sum_{c_3 = 0}^{n - a - b - 1} \sum_{c_2 = 0}^{n - a - b - c_3 - 1}  P(\text{fail}_1) \right)$$

where
\begin{eqnarray*}
\begin{aligned}
P(\text{fail}_1) &= F\left(n - a - b - c_2 - c_3, p_1^{b_1 + c_2}\right) \times B\left(n - a - b - c_3, c_2, q_{21}^{a + b_2 + c_3}\right) \times \\
& \hspace{10pt} \times B\left(n - a - b, c_3, q_{32}\right) \times B\left(n - a - b_2, b_1, p_1^{a + b_2}\right) \times B\left(n - a, b_2, q_{21}\right) \times B(n, a, p_1)
\end{aligned}
\end{eqnarray*}

\noindent\textbf{Case 2: $R_{U_1} > R_{U_3} \geq R_{U_2}$}\\
The rate of transmission in Phase \upperRomannumeral{1}, $R_{U_1}$, is determined by the time allocated for this phase, $T_{U_1}$. Let the nodes who were successful in Phase \upperRomannumeral{1} be in Set $\mathcal{A}$ (cardinality $a$). The rate in Phase \upperRomannumeral{2}, $R_{U_2}^{(a)}$ and Phase \upperRomannumeral{3}, $R_{U_3}^{(a)}$ depends on the realization of $a$, and the time allocated for the phase, $T_{U_2}$ and $T_{U_3}$. As before, $R_{U_2}^{(a)} = \frac{m\cdot (n-a)}{T_{U_2}}$, $R_{U_3}^{(a)} = \frac{m\cdot (n-a)}{T_{U_3}}$.
The probabilities of link error corresponding to each rate $R_{U_1}$, $R_{U_2}^{(a)}$ and $R_{U_3}^{(a)}$ are $p_{U_1}$, $p_{U_2}^{(a)}$ and $p_{U_3}^{(a)}$ (abbreviated to $p_1$, $p_2$ and $p_3$) respectively.
Fig.~\ref{fig:3U_2} displays an exhaustive list of ways to succeed in case 2 of the three-hop uplink protocol.

\begin{figure}[htb]
\centering
\includegraphics[width = 0.6\textwidth]{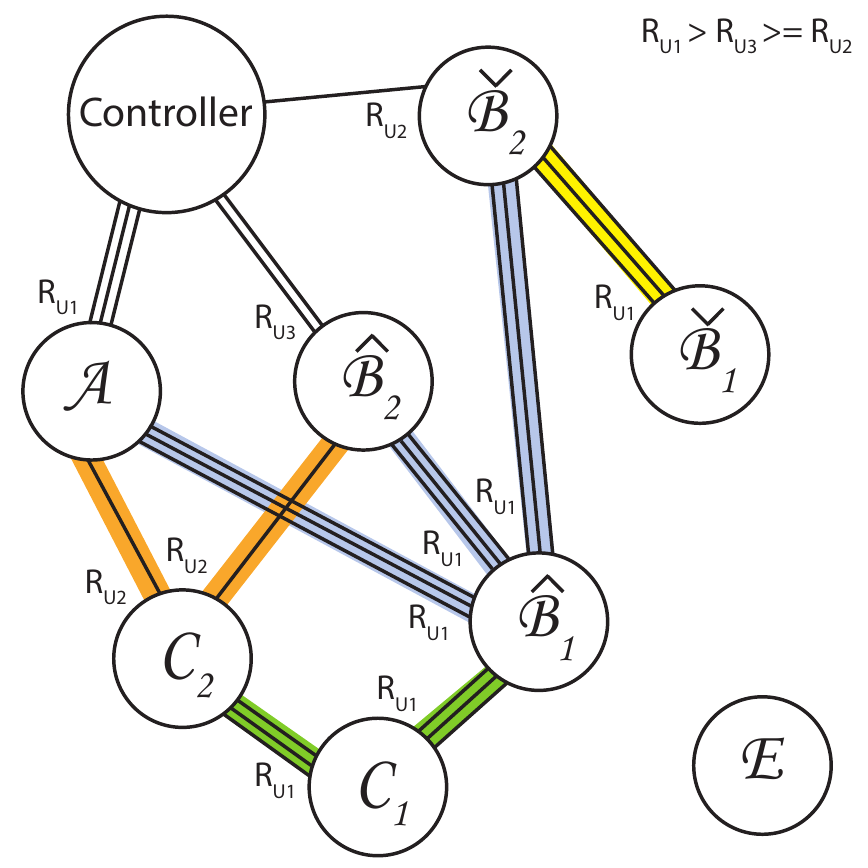}
 \caption{{Case 2: $R_{U_1} > R_{U_3} \geq R_{U_2}$. The only ways to succeed in the 2nd case of 3-hop uplink protocol are displayed. A node can succeed in Phase \upperRomannumeral{1} directly, in Phase \upperRomannumeral{2} by connecting to the controller or a node which can succeed in Phase \upperRomannumeral{2}, and in Phase \upperRomannumeral{3} by connecting directly to the nodes which have connections to the controller in Phase \upperRomannumeral{2} (thus succeeding in 2 hops) or connecting via 2 hops to the nodes which have connections to the controller (thus succeeding in 3 hops).}}
 \label{fig:3U_2}
\end{figure}
\FloatBarrier
\begin{itemize}
\item A node can succeed directly to the controller in the first hop under rate $R_{U_1}$ (is in set $\mathcal{A}$).

\item A node can succeed in the second phase of the protocol by connecting directly to the controller under the new rate, $R_{U_2}^{(a)}$ (is in set $\mathcal{B}_2$). This set is then segregated into two disjoint sets: $\widehat{\mathcal{B}}_2$ which retain links to the controller in the third phase and $\widecheck{\mathcal{B}}_2$ which lose links to the controller in the third phase.

\item A node can succeed in the second phase of the protocol by connecting in the first phase (is in set $\mathcal{B}_1$) to one of the nodes in the set $\mathcal{A} \bigcup \mathcal{B}_2$ (the set of nodes which can communicate to the controller in phase \upperRomannumeral{2}).
This ensures that the nodes which can connect to the controller in the second phase already have the message. This set is then segregated into two disjoint sets: $\widehat{\mathcal{B}}_1$ which has good links to the set which has links to the controller in the third phase (set $\mathcal{A} \bigcup \widehat{\mathcal{B}}_2$) and $\widecheck{\mathcal{B}}_1$ which does not have links to the set which has links to the controller in the third phase (set $\mathcal{A} \bigcup \widehat{\mathcal{B}}_2$). Thus, the set $\widecheck{\mathcal{B}}_1$ cannot act as relay for three-hop successes.

\item A node can succeed in the third phase in a two-hop fashion by connecting to the set $\mathcal{A} \bigcup \widehat{\mathcal{B}}_2$ under the lower phase two rate $R_{U_2}^{(a)}$ (is in set $\mathcal{C}_2$). The set $\mathcal{A} \bigcup \widehat{\mathcal{B}}_2$ is the set of nodes which can connect to the controller in the third phase. Connecting to $\mathcal{A} \bigcup \widehat{\mathcal{B}}_2$ in phase \upperRomannumeral{2} ensures that the message to be conveyed in phase \upperRomannumeral{3} has been conveyed to the relays by phase \upperRomannumeral{2}.

\item A node can succeed in the third phase in a three-hop fashion by connecting to the set $\mathcal{C}_2 \bigcup \widehat{\mathcal{B}}_1$ in the first phase under rate $R_{U_1}$ (is in set $\mathcal{C}_1$). The set $\mathcal{C}_2 \bigcup \widehat{\mathcal{B}}_1$ is the set of nodes which can connect to the set $\mathcal{A} \bigcup \widehat{\mathcal{B}}_2$ (they can connect to the controller in the third phase) in the second phase. Connecting to this set in the first phase ensures that the message to be conveyed in the third phase has been conveyed to the right relays by the second phase.
\end{itemize}

To calculate the probability of error of the three-hop uplink protocol, we will again unroll the state space in a manner similar case 1 and sum over all possible instantiations of the sets of interest that result in failure. In this case, we are interested in the event that at least one node which does not fall in sets $\mathcal{A}$, $\mathcal{B} = \mathcal{B}_1 \bigcup \mathcal{B}_2$ and $\mathcal{C}_2$ is also not in $\mathcal{C}_1$.

The probability of $A = a$ is exactly the same as we have seen before, as it relies on just point to point links to the controller, each of which fails independently with probability $p_1 = p_{U_1}$ (we use Eq.~\eqref{eq:pfail_singlelink}). This gives us $B(n, a, p_1)$.
Given $A = a$, we can calculate the probability of a node not being able to gain a connection to the controller in the second phase given there was no connection in the first phase as $q_{21} = P(C < R_{U_2}^{(a)} | C < R_{U_1}^{(a)}) = (p_2)/( p_1)$. $\mathcal{B}_2$ is the set which can connect to the controller in the second phase. Hence we calculate the probability that ${B}_2 = {b}_2$ using a binomial distribution with parameter $q_{21}$ as $B(n - a, b_2, q_{21})$.
Given $A = a$, $B_2 = b_2$, we can calculate the probability of a node in $\mathcal{B}_2$ losing connection to the controller in the third phase as $r_{32} = P(C < R_{U_3}^{(a)} | C > R_{U_2}^{(a)}) = (p_3 - p_2)/(1 - p_2)$. This set is denoted as $\widecheck{\mathcal{B}}_2$ and the set that retains the link is denoted as $\widehat{\mathcal{B}}_2$.
Hence we calculate the probability that $\widehat{B}_2 = \widehat{b}_2$ using a binomial distribution with parameter $r_{32}$ as $B(b_2, \hat{b}_2, r_{32})$.

Given $A = a$, $B_2 = b_2$, $\widehat{B}_2 = \widehat{b}_2$, we can calculate the probability of a node not succeeding in Phase \upperRomannumeral{2} in two hops as $p_{1}^{a + b_2}$, as it must fail to connect to $\mathcal{A} \bigcup \mathcal{B}_2$ in the first phase.
Hence we calculate the probability that $B_1 = b_1$ using a binomial distribution with parameter $p_{1}^{a + b_2}$ as $B(n - a - b_2, b_1, p_{1}^{a + b_2})$.
Given $A = a$, $B_2 = b_2$, $\widehat{B}_2 = \widehat{b}_2$, and $B_1 = b_1$ we can calculate the probability of a node in $\mathcal{B}_1$ being only connected to $\widecheck{\mathcal{B}}_2$ in the second phase given it connected to the set $\widecheck{\mathcal{B}}_2 \bigcup \widehat{\mathcal{B}}_2 \bigcup \mathcal{A}$ as $s_{22}[a + \hat{b}_2, a + b_2] = (1 - p_2^{a + \widehat{b}_2} )/(1 - p_2^{a + b_2})$. Hence we calculate the probability that $\widecheck{B}_1 = \widecheck{b}_1$ using a binomial distribution with parameter $s_{22}[a + \hat{b}_2, a + b_2] $ as $B(b_1, \widehat{b}_1, s_{22}[a + \hat{b}_2, a + b_2])$.

Given $A = a$, $B_1 = b_1$, $\widehat{B}_1 = \widehat{b}_1$, $B_2 = b_2$, $\widehat{B}_2 = \widehat{b}_2$, we can calculate the probability of a node not succeeding in Phase \upperRomannumeral{3} in two hops as $q_{21}^{a + \widehat{b}_2}$, as it must fail to connect to $\mathcal{A} \bigcup \widehat{\mathcal{B}}_2$ in the second phase having failed to connect in the first phase already.
Hence we calculate the probability that $C_2 = c_2$ using a binomial distribution with parameter $q_{21}^{a + \widehat{b}_2}$ as $B(n - a - b, c_2, q_{21}^{a + \widehat{b}_2})$.
Given $C_2 = c_2$, $B_1 = b_1$, $\widehat{B}_1 = \widehat{b}_1$, $B_2 = b_2$, $\widehat{B}_2 = \widehat{b}_2$ and $A = a$, the probability of a node (not in $\mathcal{A} \bigcup \mathcal{B}_1 \bigcup \mathcal{B}_2 \bigcup \mathcal{C}_2$) failing after three-hops is the probability that it cannot connect to $\mathcal{C}_2 \bigcup \widehat{\mathcal{B}}_1$ in the first phase. This is distributed Bernoulli $p_1^{\widehat{b}_1 + c_2}$, and can be written with Eq.~\eqref{eq:F} as $F(n - a - b - c_2, p_1^{\widehat{b}_1 + c_2})$.

Thus we have that given the realization $A = a$, the probability that the protocol fails under case 2: $R_{U_1} > R_{U_3} > R_{U_2}$ is given by
$$ P(\text{fail} | \text{Case 2}, A = a) = \left( \sum_{b_2 = 0}^{n - a - 1} \sum_{b_1 = 0}^{n - a - b_2 - 1} \sum_{\widehat{b}_2 = 0}^{b_2} \sum_{\widehat{b}_1 = 0}^{b_1} \sum_{c_2 = 0}^{n - a - b -1}  P(\text{fail}_2) \right)$$

where
\begin{eqnarray*}
\begin{aligned}
P(\text{fail}_2) &= F(n - a - b - c_2, p_1^{\widehat{b}_1 + c_2}) \times B(n - a - b, c_2, q_{21}^{a + \widehat{b}_2}) \times B(n - a - b_2, b_1, p_{1}^{a + b_2})\\
& \hspace{10pt} \times B(b_1, \widehat{b}_1, s_{22}[a + \hat{b}_2, a + b_2]) \times B(b_2, \hat{b}_2, r_{32}) \times B(n - a, b_2, q_{21}) \times B(n, a, p_1)
\end{aligned}
\end{eqnarray*}

\noindent\textbf{Case 3: $R_{U_3} \geq R_{U_1} > R_{U_2}$}\\
The rate of transmission in Phase \upperRomannumeral{1}, $R_{U_1}$, is determined by the time allocated for this phase, $T_{U_1}$. Let the nodes who were successful in Phase \upperRomannumeral{1} be in Set $\mathcal{A}$ (cardinality $a$). The rate in Phase \upperRomannumeral{2}, $R_{U_2}^{(a)}$ and Phase \upperRomannumeral{3}, $R_{U_3}^{(a)}$ depends on the realization of $a$, and the time allocated for the phase, $T_{U_2}$ and $T_{U_3}$. As before, $R_{U_2}^{(a)} = \frac{m\cdot (n-a)}{T_{U_2}}$, $R_{U_3}^{(a)} = \frac{m\cdot (n-a)}{T_{U_3}}$.
The probabilities of link error corresponding to each rate $R_{U_1}$, $R_{U_2}^{(a)}$ and $R_{U_3}^{(a)}$ are $p_{U_1}$, $p_{U_2}^{(a)}$ and $p_{U_3}^{(a)}$ (abbreviated to $p_1$, $p_2$ and $p_3$) respectively.
Fig.~\ref{fig:3U_3} displays an exhaustive list of ways to succeed in case 3 of the three-hop uplink protocol.

\begin{figure}[htb]
\centering
\includegraphics[width = 0.6\textwidth]{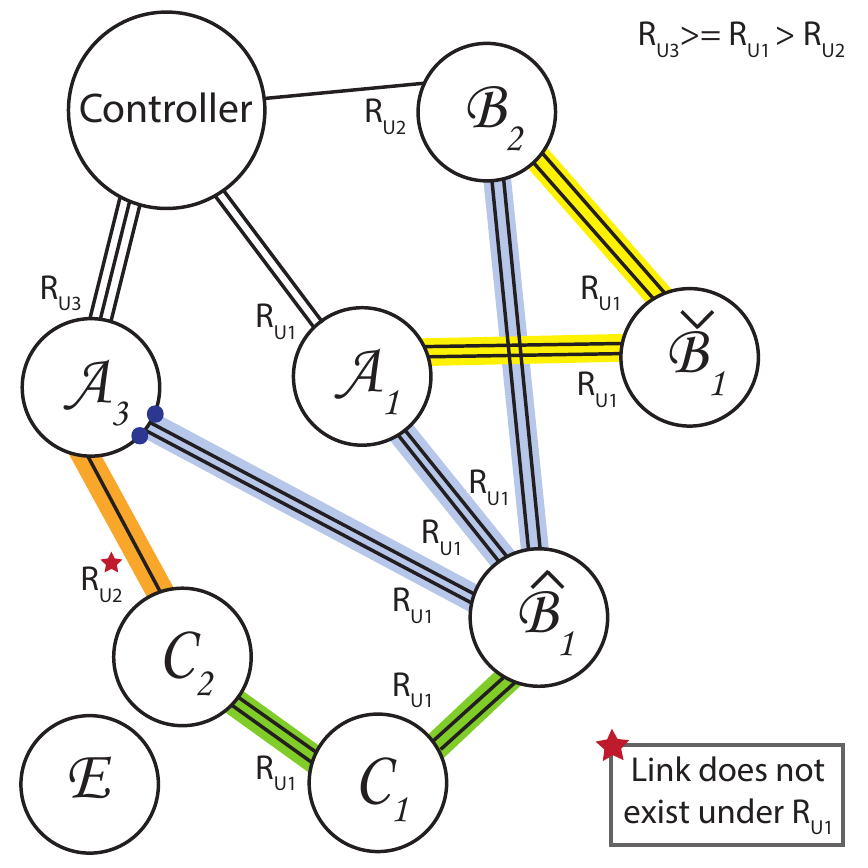}
\caption{{Case 3: $R_{U_3} \geq R_{U_1} > R_{U_2}$. The only ways to succeed in the 3rd case of 3-hop uplink protocol are displayed. A node can succeed in Phase \upperRomannumeral{1} directly, in Phase \upperRomannumeral{2} by connecting to the controller or a node which can succeed in Phase \upperRomannumeral{2}, and in Phase \upperRomannumeral{3} by connecting directly to the nodes which have connections to the controller in Phase \upperRomannumeral{2} (thus succeeding in 2 hops) or connecting via 2 hops to the nodes which have connections to the controller (thus succeeding in 3 hops).}}
\label{fig:3U_3}
\end{figure}

\begin{itemize}
\item A node can succeed directly to the controller in the first hop under rate $R_{U_1}$ (is in set $\mathcal{A}$). This set is then divided into two disjoint sets $\mathcal{A}_1$ (nodes which lose their link to the controller in phase 3) and $\mathcal{A}_3$ (nodes which retain link to the controller in phase 3) such that $\mathcal{A} = \mathcal{A}_1 \bigcup \mathcal{A}_3$.

\item A node can succeed in the second phase of the protocol by connecting directly to the controller under the new rate, $R_{U_2}^{(a)}$ (is in set $\mathcal{B}_2$).

\item A node can succeed in the second phase of the protocol by connecting in the first phase (is in set $\mathcal{B}_1$) to one of the nodes in the set $\mathcal{A} \bigcup \mathcal{B}_2$ (the set of nodes which can communicate to the controller in phase \upperRomannumeral{2}).
This ensures that the nodes which can connect to the controller in the second phase already have the message. This set is then segregated into two disjoint sets: $\widehat{\mathcal{B}}_1$ which has good links to the set which has links to the controller in the third phase (set $\mathcal{A}_3$) and $\widecheck{\mathcal{B}}_1$ which does not have links to the set which has links to the controller in the third phase (set $\mathcal{A}_3$). Thus set $\widecheck{\mathcal{B}}_1$ cannot act as relay for three-hop successes.

\item A node can succeed in the third phase in a two-hop fashion by connecting to the set $\mathcal{A}_3$ under the lower phase two rate $R_{U_2}^{(a)}$ (is in set $\mathcal{C}_2$). The set $\mathcal{A}_3$ is the set of nodes which can connect to the controller in the third phase. Connecting to $\mathcal{A}_3$ in phase \upperRomannumeral{2} ensures that the message to be conveyed in phase \upperRomannumeral{3} has been conveyed to the relays by phase \upperRomannumeral{2}.

\item A node can succeed in the third phase in a three-hop fashion by connecting to the set $\mathcal{C}_2 \bigcup \widehat{\mathcal{B}}_1$ in the first phase under rate $R_{U_1}$ (is in set $\mathcal{C}_1$). The set $\mathcal{C}_2 \bigcup \widehat{\mathcal{B}}_1$ is the set of nodes which can connect to the set $\mathcal{A}_3$ (they can connect to the controller in the third phase) in the second phase. Connecting to this set in the first phase ensures that the message to be conveyed in the third phase has been conveyed to the right relays by the second phase.
\end{itemize}

To calculate the probability of error of the three-hop uplink protocol, we will unroll the state space as before and sum over all possible instantiations of the sets of interest that result in failure. In this case, we are interested in the event that at least one node which does not fall in sets $\mathcal{A}$, $\mathcal{B} = \mathcal{B}_1 \bigcup \mathcal{B}_2$ and $\mathcal{C}_2$ is also not in $\mathcal{C}_1$.

The probability of $A = a$ is exactly the same as we have seen before, as it relies on just point to point links to the controller, each of which fails independently with probability $p_1 = p_{U_1}$ (we use Eq.~\eqref{eq:pfail_singlelink}). This gives us $B(n, a, p_1)$.
Given $A = a$, we can calculate the probability of a node not being able to gain a connection to the controller in the second phase given there was no connection in the first phase as $q_{21} = P(C < R_{U_2}^{(a)} | C < R_{U_1}^{(a)}) = (p_2)/( p_1)$. $\mathcal{B}_2$ is the set which can connect to the controller in the second phase. Hence we calculate the probability that ${B}_2 = {b}_2$ using a binomial distribution with parameter $q_{21}$ as $B(n - a, b_2, q_{21})$. None of the nodes in the set $\mathcal{B}_2$ retain the link to the controller in phase 3.
Given $A = a$ we can calculate the probability of a node in $\mathcal{A}$ losing connection to the controller in the third phase as $r_{31} = P(C < R_{U_3}^{(a)} | C > R_{U_1}^{(a)}) = (p_3 - p_1)/(1 - p_1)$. This set is denoted as $\mathcal{A}_1$ and the set that retains the link is denoted as $\mathcal{A}_3$.
Hence we calculate the probability that ${A}_3 = a_3$ using a binomial distribution with parameter $r_{31}$ as $B(a, a_3, r_{31})$.

Given $A = a$, $A_3 = a_3$, $B_2 = b_2$, we can calculate the probability of a node not succeeding in Phase \upperRomannumeral{2} in two hops as $p_{1}^{a + b_2}$, as it must fail to connect to $\mathcal{A} \bigcup \mathcal{B}_2$ in the first phase.
Hence we calculate the probability that $B_1 = b_1$ using a binomial distribution with parameter $p_{1}^{a + b_2}$ as $B(n - a - b_2, b_1, p_{1}^{a + b_2})$.
Given $A = a$, $B_2 = b_2$, $A_3 = a_3$, and $B_1 = b_1$ we can calculate the probability of a node in $\mathcal{B}_1$ being only connected to $\mathcal{A}_3$ in the second phase given it connected to the set $\mathcal{A} \bigcup \mathcal{B}_2$ as $s_{22}[a_3, a + b_2] = (1 - p_2^{a_3} )/(1 - p_2^{a + b_2})$. Hence we calculate the probability that $\widecheck{B}_1 = \widecheck{b}_1$ using a binomial distribution with parameter $s_{22}[a_3, a + b_2] $ as $B(b_1, \widehat{b}_1, s_{22}[a_3, a + b_2])$.

Given $A = a$, $A_3 = a_3$, $B_1 = b_1$, $\widehat{B}_1 = \widehat{b}_1$, $B_2 = b_2$, we can calculate the probability of a node not succeeding in Phase \upperRomannumeral{3} in two hops as $q_{21}^{a_3}$, as it must fail to connect to $\mathcal{A}_3$ in the second phase having failed to connect in the first phase already.
Hence we calculate the probability that $C_2 = c_2$ using a binomial distribution with parameter $q_{21}^{a_3}$ as $B(n - a - b, c_2, q_{21}^{a_3})$.
Given $C_2 = c_2$, $B_1 = b_1$, $\widehat{B}_1 = \widehat{b}_1$, $B_2 = b_2$, $A_3 = a_3$ and $A = a$, the probability of a node (not in $\mathcal{A} \bigcup \mathcal{B}_1 \bigcup \mathcal{B}_2 \bigcup \mathcal{C}_2$) failing after three-hops is the probability that it cannot connect to $\mathcal{C}_2 \bigcup \widehat{\mathcal{B}}_1$ in the first phase. This is distributed Bernoulli $p_1^{\widehat{b}_1 + c_2}$, and can be written with Eq.~\eqref{eq:F} as $F(n - a - b - c_2, p_1^{\widehat{b}_1 + c_2})$.

Thus we have that given the realization $A = a$, the probability that the protocol fails under case 3: $R_{U_3} \geq R_{U_1} > R_{U_2}$ is given by
$$ P(\text{fail} | \text{Case 3}, A = a) = \left( \sum_{a_3 = 0}^{a} \sum_{b_2 = 0}^{n - a - 1} \sum_{b_1 = 0}^{n - a - b_2 - 1} \sum_{\widehat{b}_1 = 0}^{b_1} \sum_{c_2 = 0}^{n - a - b - 1} P(\text{fail}_3) \right)$$

where
\begin{eqnarray*}
\begin{aligned}
P(\text{fail}_3) &= F\left(n - a - b - c_2, p_1^{\widehat{b}_1 + c_2}\right) \times B\left(n - a - b, c_2, q_{21}^{a_3}\right) \times B\left(b_1, \widehat{b}_1, s_{22}[a_3, a + b_2] \right) \times \\
& \hspace{10pt} \times B\left(n - a - b_2, b_1, p_1^{a+b_2}\right) \times B\left(a, a_3, r_{31}\right) \times B\left(n-a, b_2, q_{21}\right) \times B(n, a, p_1)
\end{aligned}
\end{eqnarray*}

\noindent\textbf{Case 4: $R_{U_3} > R_{U_2} \geq R_{U_1}$}\\
The rate of transmission in Phase \upperRomannumeral{1}, $R_{U_1}$, is determined by the time allocated for this phase, $T_{U_1}$. Let the nodes who were successful in Phase \upperRomannumeral{1} be in Set $\mathcal{A}$ (cardinality $a$). The rate in Phase \upperRomannumeral{2}, $R_{U_2}^{(a)}$ and Phase \upperRomannumeral{3}, $R_{U_3}^{(a)}$ depends on the realization of $a$, and the time allocated for the phase, $T_{U_2}$ and $T_{U_3}$. As before, $R_{U_2}^{(a)} = \frac{m\cdot (n-a)}{T_{U_2}}$, $R_{U_3}^{(a)} = \frac{m\cdot (n-a)}{T_{U_3}}$.
The probabilities of link error corresponding to each rate $R_{U_1}$, $R_{U_2}^{(a)}$ and $R_{U_3}^{(a)}$ are $p_{U_1}$, $p_{U_2}^{(a)}$ and $p_{U_3}^{(a)}$ (abbreviated to $p_1$, $p_2$ and $p_3$) respectively.
Fig.~\ref{fig:3U_4} displays an exhaustive list of ways to succeed in case 4 of the three-hop uplink protocol.

\begin{figure}[htb]
\centering
\includegraphics[width = 0.6\textwidth]{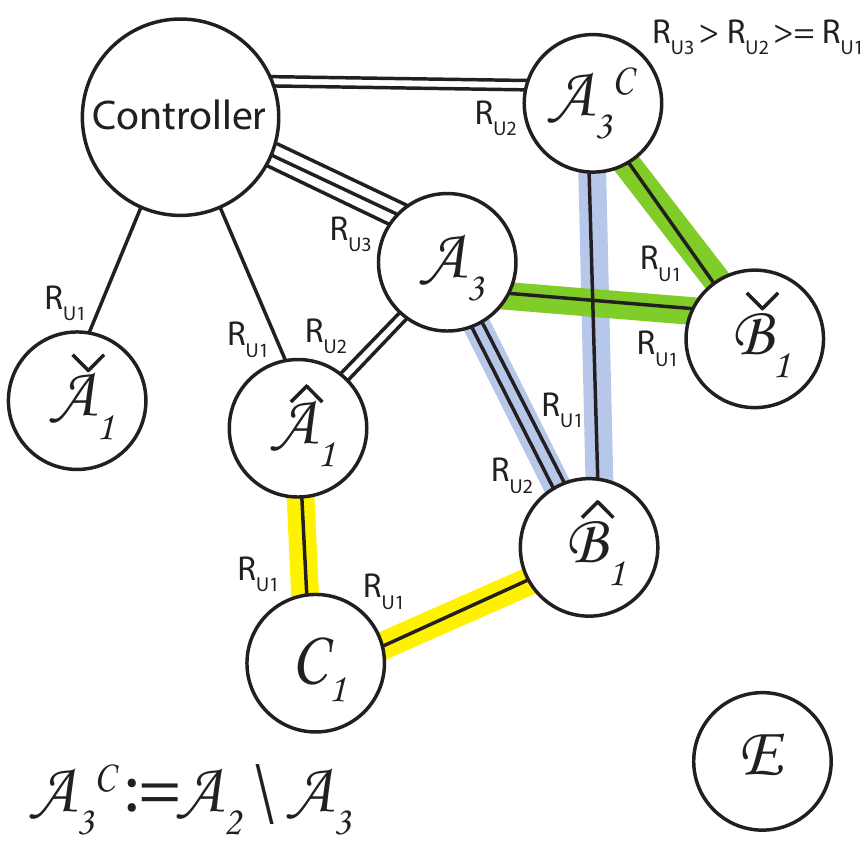}
\caption{{Case 4: $R_{U_3} > R_{U_2} \geq R_{U_1}$: The only ways to succeed in the 4th case of 3-hop uplink protocol are displayed. A node can succeed in Phase \upperRomannumeral{1} directly, in Phase \upperRomannumeral{2} by connecting to a node which can succeed in Phase \upperRomannumeral{2}, and in Phase \upperRomannumeral{3} by connecting via 2 hops to the nodes which have connections to the controller (thus succeeding in 3 hops).}}
\label{fig:3U_4}
\end{figure}
\FloatBarrier

\begin{itemize}
\item A node can succeed directly to the controller in the first hop under rate $R_{U_1}$ (is in set $\mathcal{A}$). This set is further divided into disjoint sets $\mathcal{A}_1$ (which lose connection to the controller after the first phase), $\mathcal{A}_3$ (the only set to retain connection to the controller in the third phase) and $\mathcal{A}_3^C$ (the set of nodes to retain connection to the controller in the second phase but not the third) such that $\mathcal{A} = \mathcal{A}_1 \bigcup \mathcal{A}_3 \bigcup \mathcal{A}_3^C$. Further, we divide $\mathcal{A}_1$ into disjoint sets $\widehat{\mathcal{A}}_1$ (the nodes which have a link to $\mathcal{A}_3$ in phase 2) and $\widecheck{\mathcal{A}}_1$ (the nodes which do not have a link to $\mathcal{A}_3$ in phase 2) such that $\mathcal{A}_1 = \widehat{\mathcal{A}}_1 \bigcup \widecheck{\mathcal{A}}_1$.

\item A node can succeed in the second phase of the protocol by connecting in the first phase (is in set $\mathcal{B}_1$) to one of the nodes in the set $\mathcal{A}_2 = \mathcal{A}_3 \bigcup \mathcal{A}_3^C$ (the set of nodes which can communicate to the controller in phase \upperRomannumeral{2}).
This ensures that the nodes which can connect to the controller in the second phase already have the message. This set is then segregated into two disjoint sets: $\widehat{\mathcal{B}}_1$ which has good links to the set which has links to the controller in the third phase (set $\mathcal{A}_3$) and $\widecheck{\mathcal{B}}_1$ which does not have links to the set which has links to the controller in the third phase (set $\mathcal{A}_3$). Thus set $\widecheck{\mathcal{B}}_1$ cannot act as relay for three-hop successes.

\item A node can succeed in the third phase in a three-hop fashion by connecting to the set $\widehat{\mathcal{A}}_1 \bigcup \widehat{\mathcal{B}}_1$ in the first phase under rate $R_{U_1}$ (is in set $\mathcal{C}_1$). The set $\widehat{\mathcal{A}}_1 \bigcup \widehat{\mathcal{B}}_1$ is the set of nodes which can connect to the set $\mathcal{A}_3$ (they can connect to the controller in the third phase) in the second phase. Connecting to this set in the first phase ensures that the message to be conveyed in the third phase has been conveyed to the right relays by the second phase.
\end{itemize}

To calculate the probability of error of a three-hop uplink protocol, we will unroll the state space as before and sum over all possible instantiations of the sets of interest that result in failure. In this case, we are interested in the event that at least one node which does not fall in sets $\mathcal{A}$ and $\mathcal{B}_1$ is also not in $\mathcal{C}_1$.

The probability of $A = a$ is exactly the same as we have seen before, as it relies on just point to point links to the controller, each of which fails independently with probability $p_1 = p_{U_1}$ (we use Eq.~\eqref{eq:pfail_singlelink}). This gives us $B(n, a, p_1)$.
Given $A = a$ we can calculate the probability of a node in $\mathcal{A}$ losing connection to the controller in the second phase as $r_{21} = P(C < R_{U_2}^{(a)} | C > R_{U_1}^{(a)}) = (p_2 - p_1)/(1 - p_1)$. This losing link set is denoted as $\mathcal{A}_1$ and the set that retains the link is denoted as $\mathcal{A}_2$.
Hence we calculate the probability that ${A}_2 = a_2$ using a binomial distribution with parameter $r_{21}$ as $B(a, a_2, r_{21})$.
Given $A = a$ and $A_2 = a_2$ we can calculate the probability of a node in $\mathcal{A}_2$ losing connection to the controller in the second phase as $r_{32} = P(C < R_{U_3}^{(a)} | C > R_{U_2}^{(a)}) = (p_3 - p_2)/(1 - p_2)$. This set is denoted as $\mathcal{A}_3^C$ and the set that retains the link is denoted as $\mathcal{A}_3$.
Hence we calculate the probability that ${A}_3 = a_3$ using a binomial distribution with parameter $r_{32}$ as $B(a_2, a_3, r_{32})$.

Given $A = a$, $A_2 = a_2$ and $A_3 = a_3$, we can calculate the probability of a node not succeeding in Phase \upperRomannumeral{2} in two hops as $p_{1}^{a_2}$, as it must fail to connect to $\mathcal{A}_2$ in the first phase.
Hence we calculate the probability that $B_1 = b_1$ using a binomial distribution with parameter $p_{1}^{a_2}$ as $B(n - a, b_1, p_{1}^{a_2})$.
Given $A = a$, $A_2 = a_2$, $A_3 = a_3$, and $B_1 = b_1$ we can calculate the probability of a node in $\mathcal{B}_1$ being only connected to $\mathcal{A}_3^C$ in the second phase given it connected to the set $\mathcal{A}_2$ as $s_{21}[a_3, a_2] = (1 - p_2^{a_3} )/(1 - p_1^{a_2})$. Hence we calculate the probability that $\widecheck{B}_1 = \widecheck{b}_1$ using a binomial distribution with parameter $s_{21}[a_3, a_2]$ as $B(b_1, \widehat{b}_1, s_{21}[a_3, a_2])$.
Given $A = a$, $A_2 = a_2$ and $A_3 = a_3$, we can calculate the probability of a node in $\mathcal{A}_1$ being unable to connecte to $\mathcal{A}_3$ in the second phase as $p_2^{a_3}$. The set of nodes being able to connect is denoted by $\widehat{A}_1$ and the probability that $\widehat{A}_1 = \widehat{a}_1$ is calculated using a binomial distribution with parameter $p_2^{a_3}$ as $B(a_1, - \widehat{a}_1, p_2^{a_3})$.

Given $A = a$, $A_2 = a_2$, $A_3 = a_3$, $B_1 = b_1$, $\widehat{B}_1 = \widehat{b}_1$, and $\widehat{A}_1 = \widehat{a}_1$, the probability of a node (not in $\mathcal{A} \bigcup \mathcal{B}_1$) failing after three-hops is the probability that it cannot connect to $\widehat{\mathcal{A}}_1 \bigcup \widehat{\mathcal{B}}_1$ in the first phase. This is distributed Bernoulli $p_1^{\widehat{a}_1 + \widehat{b}_1}$, and can be written with Eq.~\eqref{eq:F} as $F(n - a - b_1, p_1^{\widehat{a}_1 + \widehat{b}_1})$.

Thus we have that given the realization $A = a$, the probability that the protocol fails under case 4: $R_{U_3} > R_{U_2} > R_{U_1}$ is given by
$$ P(\text{fail} | \text{Case 4}, A = a) = \left( \sum_{a_2 = 0}^{a} \sum_{a_3 = 0}^{a_2} \sum_{\widehat{a}_1 = 0}^{a - a_2} \sum_{b_1 = 0}^{n - a - 1} \sum_{\widehat{b}_1 = 0}^{b_1} P(\text{fail}_4) \right)$$

where
\begin{eqnarray*}
\begin{aligned}
P(\text{fail}_4) &= F\left(n - a - b_1, p_1^{\widehat{a}_1 + \widehat{b}_1}\right) \times B(a_1, \widehat{a}_1, p_2^{a_3}) \times B\left(b_1, \widehat{b}_1, s_{21}[a_3, a_2]\right) \times B\left(n - a, b_1, p_1^{a_2}\right) \times  \\
& \hspace{10pt}  \times B\left(a_2, a_3, r_{32}\right) \times B\left(a, a_2, r_{21}\right) \times B(n, a, p_1)
\end{aligned}
\end{eqnarray*}

\noindent\textbf{Case 5: $R_{U_2} \geq R_{U_3} > R_{U_1}$}\\
The rate of transmission in Phase \upperRomannumeral{1}, $R_{U_1}$, is determined by the time allocated for this phase, $T_{U_1}$. Let the nodes who were successful in Phase \upperRomannumeral{1} be in Set $\mathcal{A}$ (cardinality $a$). The rate in Phase \upperRomannumeral{2}, $R_{U_2}^{(a)}$ and Phase \upperRomannumeral{3}, $R_{U_3}^{(a)}$ depends on the realization of $a$, and the time allocated for the phase, $T_{U_2}$ and $T_{U_3}$. As before, $R_{U_2}^{(a)} = \frac{m\cdot (n-a)}{T_{U_2}}$, $R_{U_3}^{(a)} = \frac{m\cdot (n-a)}{T_{U_3}}$.
The probabilities of link error corresponding to each rate $R_{U_1}$, $R_{U_2}^{(a)}$ and $R_{U_3}^{(a)}$ are $p_{U_1}$, $p_{U_2}^{(a)}$ and $p_{U_3}^{(a)}$ (abbreviated to $p_1$, $p_2$ and $p_3$) respectively.
Fig.~\ref{fig:3U_5} displays an exhaustive list of ways to succeed in case 5 of the three-hop uplink protocol.

\begin{figure}[htb]
\centering
\includegraphics[width = 0.6\textwidth]{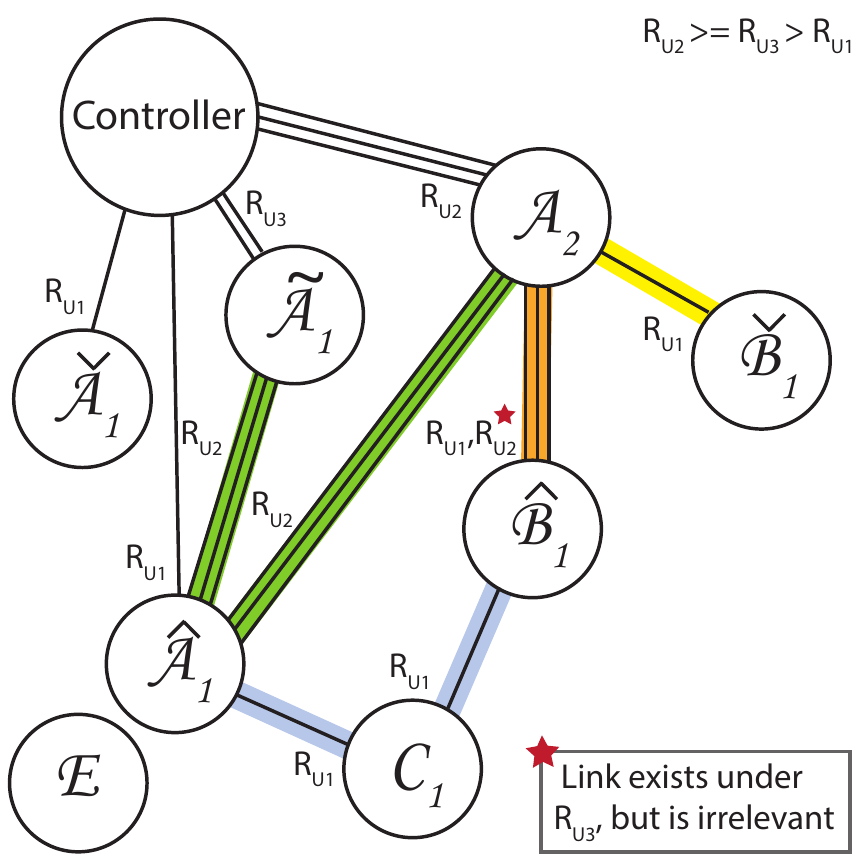}
\caption{{Case 5: $R_{U_2} \geq R_{U_3} > R_{U_1}$: The only ways to succeed in the 5th case of three-hop uplink protocol are displayed. A node can succeed in Phase \upperRomannumeral{1} directly, in Phase \upperRomannumeral{2} by connecting to a node which can succeed in Phase \upperRomannumeral{2}, and in Phase \upperRomannumeral{3} by connecting via 2 hops to the nodes which have connections to the controller (thus succeeding in 3 hops).}}
\label{fig:3U_5}
\end{figure}
\FloatBarrier

\begin{itemize}
\item A node can succeed directly to the controller in the first hop under rate $R_{U_1}$ (is in set $\mathcal{A}$). This set is further divided into disjoint sets $\mathcal{A}_1$ (which lose connection to the controller in the second phase) and $\mathcal{A}_2$ (the only set to retain connection to the controller in the second and third phase) such that $\mathcal{A} = \mathcal{A}_1 \bigcup \mathcal{A}_2$. Further, we divide $\mathcal{A}_1$ into disjoint sets $\widetilde{\mathcal{A}}_1$ (the nodes which gain back the link to the controller in the third phase), $\widehat{\mathcal{A}}_1$ (the nodes which have a link to $\mathcal{A}_2 \bigcup \widetilde{\mathcal{A}}_1$ in phase 2) and $\widecheck{\mathcal{A}}_1$ (the nodes which do not have a link to $\mathcal{A}_2 \bigcup \widetilde{\mathcal{A}}_1$ in phase 2) such that $\mathcal{A}_1 = \widecheck{\mathcal{A}}_1 \bigcup \widehat{\mathcal{A}}_1 \bigcup \widecheck{\mathcal{A}}_1$.

\item A node can succeed in the second phase of the protocol by connecting in the first phase (is in set $\mathcal{B}_1$) to one of the nodes in the set $\mathcal{A}_2$ (the set of nodes which can communicate to the controller in phase \upperRomannumeral{2}).
This ensures that the nodes which can connect to the controller in the second phase already have the message. This set is then segregated into two disjoint sets: $\widehat{\mathcal{B}}_1$ which has good links to the set which has links to the controller in the third phase (set $\mathcal{A}_3 = \mathcal{A}_2 \bigcup \widetilde{\mathcal{A}}_1$) and $\widecheck{\mathcal{B}}_1$ which does not have links to the set which has links to the controller in the third phase (set $\mathcal{A}_3$). Thus set $\widecheck{\mathcal{B}}_1$ cannot act as relay for three-hop successes.

\item A node can succeed in the third phase in a three-hop fashion by connecting to the set $\widehat{\mathcal{A}}_1 \bigcup \widehat{\mathcal{B}}_1$ in the first phase under rate $R_{U_1}$ (is in set $\mathcal{C}_1$). The set $\widehat{\mathcal{A}}_1 \bigcup \widehat{\mathcal{B}}_1$ is the set of nodes which can connect to the set $\mathcal{A}_3$ (they can connect to the controller in the third phase) in the second phase. Connecting to this set in the first phase ensures that the message to be conveyed in the third phase has been conveyed to the right relays by the second phase.
\end{itemize}

To calculate the probability of error of a three-hop uplink protocol, we will unroll the state space as before and sum over all possible instantiations of the sets of interest that result in failure. In this case, we are interested in the event that at least one node which does not fall in sets $\mathcal{A}$ and $\mathcal{B}_1$ is also not in $\mathcal{C}_1$.

The probability of $A = a$ is exactly the same as we have seen before, as it relies on just point to point links to the controller, each of which fails independently with probability $p_1 = p_{U_1}$ (we use Eq.~\eqref{eq:pfail_singlelink}). This gives us $B(n, a, p_1)$.
Given $A = a$ we can calculate the probability of a node in $\mathcal{A}$ losing connection to the controller in the second phase as $r_{21} = P(C < R_{U_2}^{(a)} | C > R_{U_1}^{(a)}) = (p_2 - p_1)/(1 - p_1)$. This set is denoted as $\mathcal{A}_1$ and the set that retains the link is denoted as $\mathcal{A}_2$.
Hence we calculate the probability that ${A}_2 = a_2$ using a binomial distribution with parameter $r_{21}$ as $B(a, a_2, r_{21})$.
Given $A = a$ and $A_2 = a_2$ we can calculate the probability of a node in $\mathcal{A}_1$ gaining back its connection to the controller in the third phase as $m_{312} = P(C < R_{U_3}^{(a)} |R_{U_1} C < R_{U_2}^{(a)}) = (p_3 - p_1)/(p_2 - p_1)$. This set is denoted by $\widetilde{\mathcal{A}}_1$ and the probability that $\widetilde{A}_1 = \widetilde{a}_1$ is calculated using a binomial distribution with parameter $m_{312}$ as $B(a - a_2, \widetilde{a}_1, m_{312})$. The set $\mathcal{A}_3 = \mathcal{A}_2 \bigcup \widetilde{\mathcal{A}}_1$.
Given $A = a$, $A_2 = a_2$ and $\widetilde{A}_1 = a_1$, we can calculate the probability of a node in $A_1 \setminus \widetilde{A}_1$ being unable to connect to $A_3$ in the second phase as $p_2^{a_2 + \widetilde{a}_1}$, as it must fail to connect to $\mathcal{A}_3$ in the second phase. The set that can connect is denoted by $\widehat{A}_1$ and the probability that $\widehat{A}_1 = \widehat{a}_1$ is calculated using a binomial distribution with parameter $p_2^{a_3}$ as $B(a - a_2 - \widetilde{a}_1, \widehat{a}_1, p_2^{a_3})$.

Given $A = a$, $A_2 = a_2$ and $A_3 = a_3$, we can calculate the probability of a node not succeeding in Phase \upperRomannumeral{2} in two hops as $p_{1}^{a_2}$, as it must fail to connect to $\mathcal{A}_2$ in the first phase.
Hence we calculate the probability that $B_1 = b_1$ using a binomial distribution with parameter $p_{1}^{a_2}$ as $B(n - a, b_1, p_{1}^{a_2})$.
Given $A = a$, $A_2 = a_2$, $A_3 = a_3$, and $B_1 = b_1$ we can calculate the probability of a node in $\mathcal{B}_1$ being connected to $\mathcal{A}_2$ in the second phase given it connected to the set $\mathcal{A}_2$ in the first phase as $s_{21}[a_2, a_2] = (1 - p_2^{a_2} )/(1 - p_1^{a_2})$. Hence we calculate the probability that $\widecheck{B}_1 = \widecheck{b}_1$ using a binomial distribution with parameter $s_{21}[a_2, a_2]$ as $B(b_1, \widehat{b}_1, s_{21}[a_2, a_2])$.

Given $A = a$, $A_2 = a_2$, $A_3 = a_3$, $B_1 = b_1$, $\widehat{B}_1 = \widehat{b}_1$, and $\widehat{A}_1 = \widehat{a}_1$, the probability of a node (not in $\mathcal{A} \bigcup \mathcal{B}_1$) failing after three-hops is the probability that it cannot connect to $\widehat{\mathcal{A}}_1 \bigcup \widehat{\mathcal{B}}_1$ in the first phase. This is distributed Bernoulli $p_1^{\widehat{a}_1 + \widehat{b}_1}$, and can be written with Eq.~\eqref{eq:F} as $F(n - a - b_1, p_1^{\widehat{a}_1 + \widehat{b}_1})$.

Thus we have that given the realization $A = a$, the probability that the protocol fails under case 5: $R_{U_2} \geq R_{U_3} > R_{U_1}$ is given by
$$ P(\text{fail} | \text{Case 5}, A = a) = \left( \sum_{a_2 = 0}^{a} \sum_{\widetilde{a} = 0}^{a - a_2} \sum_{\widehat{a}_1 = 0}^{a - a_2 - \widetilde{a}_1} \sum_{b_1 = 0}^{n - a - 1} \sum_{\widehat{b}_1 = 0}^{b_1} P(\text{fail}_5) \right)$$

where
\begin{eqnarray*}
\begin{aligned}
P(\text{fail}_5) &= F\left(n - a - b_1, p_1^{\widehat{a}_1 + \widehat{b}_1}\right) \times B\left(a - \widetilde{a}_1 - a_2, \widehat{a}_1, p_{2}^{\widetilde{a}_1 + a_2}\right) \times B\left(b_1, \widehat{b}_1, s_{21}[a_2, a_2]\right) \times\\
& \hspace{10pt} \times B\left(n - a, b_1, p_1^{a_2}\right) \times B\left(a - a_2, \widetilde{a}_1, m_{312}\right) \times B\left(a, a_2, r_{21}\right) \times B(n, a, p_1)
\end{aligned}
\end{eqnarray*}

\noindent\textbf{Case 6: $R_{U_2} > R_{U_1} \geq R_{U_3}$}\\
The rate of transmission in Phase \upperRomannumeral{1}, $R_{U_1}$, is determined by the time allocated for this phase, $T_{U_1}$. Let the nodes who were successful in Phase \upperRomannumeral{1} be in Set $\mathcal{A}$ (cardinality $a$). The rate in Phase \upperRomannumeral{2}, $R_{U_2}^{(a)}$ and Phase \upperRomannumeral{3}, $R_{U_3}^{(a)}$ depends on the realization of $a$, and the time allocated for the phase, $T_{U_2}$ and $T_{U_3}$. As before, $R_{U_2}^{(a)} = \frac{m\cdot (n-a)}{T_{U_2}}$, $R_{U_3}^{(a)} = \frac{m\cdot (n-a)}{T_{U_3}}$.
The probabilities of link error corresponding to each rate $R_{U_1}$, $R_{U_2}^{(a)}$ and $R_{U_3}^{(a)}$ are $p_{U_1}$, $p_{U_2}^{(a)}$ and $p_{U_3}^{(a)}$ (abbreviated to $p_1$, $p_2$ and $p_3$) respectively.
Fig.~\ref{fig:3U_6} displays an exhaustive list of ways to succeed in case 6 of the three-hop uplink protocol.

\begin{figure}[htb]
\centering
\includegraphics[width = 0.6\textwidth]{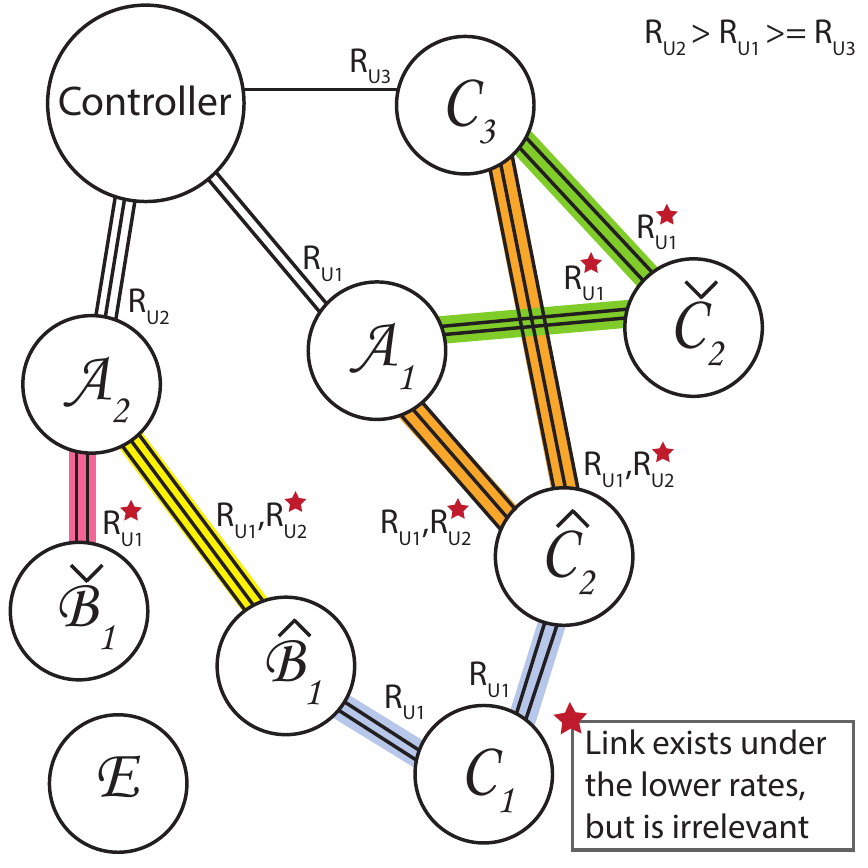}
\caption{{Case 6: $R_{U_2} > R_{U_1} \geq R_{U_3}$: The only ways to succeed in the 6th case of 3-hop uplink protocol are displayed. A node can succeed in Phase \upperRomannumeral{1} directly, in Phase \upperRomannumeral{2} by connecting to a node which can succeed in Phase \upperRomannumeral{2}, and in Phase \upperRomannumeral{3} by directly connecting to the controller or connecting to the nodes which have connections to the controller in Phase \upperRomannumeral{2} (thus succeeding in 2 hops) or connecting via 2 hops to the nodes which have connections to the controller (thus succeeding in 3 hops).}}
\label{fig:3U_6}
\end{figure}
\FloatBarrier

\begin{itemize}
\item A node can succeed directly to the controller in the first hop under rate $R_{U_1}$ (is in set $\mathcal{A}$). This set is further divided into disjoint sets $\mathcal{A}_1$ (which lose connection to the controller in the second phase) and $\mathcal{A}_2$ (retains link to controller in the second phase) such that $\mathcal{A} = \mathcal{A}_1 \bigcup \mathcal{A}_2$.

\item A node can succeed in the second phase of the protocol by connecting in the first phase (is in set $\mathcal{B}_1$) to one of the nodes in the set $\mathcal{A}_2$ (the set of nodes which can communicate to the controller in phase \upperRomannumeral{2}).
This ensures that the nodes which can connect to the controller in the second phase already have the message. This set is then segregated into two disjoint sets: $\widehat{\mathcal{B}}_1$ which has good links to the set which has links to the controller in the third phase (set $\mathcal{A}_2$) and $\widecheck{\mathcal{B}}_1$ which does not have links to the set which has links to the controller in the third phase (set $\mathcal{A}_2$). Thus set $\widecheck{\mathcal{B}}_1$ cannot act as relay for three-hop successes.

\item A node can succeed in the third phase of the protocol by connecting directly to the controller under the new rate, $R_{U_3}^{(a)}$ (is in set $\mathcal{C}_3$).

\item A node can succeed in the third phase in a two-hop fashion by connecting to the set $\mathcal{A}_1 \bigcup {\mathcal{C}}_3$ under the phase one rate of $R_{U_1}$ (is in set $\mathcal{C}_2$). As the set they connect to is the set that doesn't have a connection to the controller in the second phase but does have connection in the third phase, these nodes, succeed in phase 3. These are further divided into disjoint sets $\widehat{\mathcal{C}}_2$ (nodes that retain this link at the higher rate of $R_{U_2}^{(a)}$) and $\widecheck{\mathcal{C}}_2$ (nodes that lose this link at the higher rate of $R_{U_2}^{(a)}$) such that $\mathcal{C}_2 = \widecheck{\mathcal{C}}_2 \bigcup \widehat{\mathcal{C}}_2$.

\item A node can succeed in the third phase in a three-hop fashion by connecting to the set $\widehat{\mathcal{B}}_1 \bigcup \widehat{\mathcal{C}}_2$ in the first phase under rate $R_{U_1}$ (is in set $\mathcal{C}_1$). The set $\widehat{\mathcal{C}}_2 \bigcup \widehat{\mathcal{B}}_1$ is the set of nodes which can connect to the set $\mathcal{A} \bigcup \mathcal{C}_3$ (they can connect to the controller in the third phase) in the second phase. Connecting to this set in the first phase ensures that the message to be conveyed in the third phase has been conveyed to the right relays by the second phase.
\end{itemize}

To calculate the probability of error of the three-hop uplink protocol, we will unroll the state space as before and sum over all possible instantiations of the sets of interest that result in failure. In this case, we are interested in the event that at least one node which does not fall in sets $\mathcal{A}$, $\mathcal{B}_1$ and $\mathcal{C}_3 \bigcup \mathcal{C}_2$ is also not in $\mathcal{C}_1$.

The probability of $A = a$ is exactly the same as we have seen before, as it relies on just point to point links to the controller, each of which fails independently with probability $p_1 = p_{U_1}$ (we use Eq.~\eqref{eq:pfail_singlelink}). This gives us $B(n, a, p_1)$.
Given $A = a$ we can calculate the probability of a node in $\mathcal{A}$ losing connection to the controller in the second phase as $r_{21} = P(C < R_{U_2}^{(a)} | C > R_{U_1}^{(a)}) = (p_2 - p_1)/(1 - p_1)$. This set is denoted as $\mathcal{A}_1$ and the set that retains the link is denoted as $\mathcal{A}_2$.
Hence we calculate the probability that ${A}_2 = a_2$ using a binomial distribution with parameter $r_{21}$ as $B(a, a_2, r_{21})$.

Given $A = a$ and $A_2 = a_2$, we can calculate the probability of a node not succeeding in Phase \upperRomannumeral{2} in two hops as $p_{1}^{a_2}$, as it must fail to connect to $\mathcal{A}_2$ in the first phase.
Hence we calculate the probability that $B_1 = b_1$ using a binomial distribution with parameter $p_{1}^{a_2}$ as $B(n - a, b_1, p_{1}^{a_2})$.
Given $A = a$, $A_2 = a_2$ and $B_1 = b_1$ we can calculate the probability of a node in $\mathcal{B}_1$ being only connected to ${\mathcal{A}}_2$ in the second phase given it connected to the set ${\mathcal{A}}_2$ as $s_{21}[a_2, a_2] = (1 - p_2^{a_2} )/(1 - p_1^{a_2})$. Hence we calculate the probability that $\widecheck{B}_1 = \widecheck{b}_1$ using a binomial distribution with parameter $s_{21}[a_2, a_2] $ as $B(b_1, \widehat{b}_1, s_{21}[a_2, a_2])$.

Given $A = a$ and $B_1 = b_1$, we can calculate the probability of a node not being able to gain a connection to the controller in the third phase given there was no connection in the first two phases as $q_{31} = P(C < R_{U_3}^{(a)} | C < R_{U_1}^{(a)}) = (p_3)/( p_1)$. $\mathcal{C}_3$ is the set which can connect to the controller in the third phase. Hence we calculate the probability that ${C}_3 = {c}_3$ using a binomial distribution with parameter $q_{31}$ as $B(n - a - b_1, c_3, q_{31})$.

Given $A = a$, $A_1 = a_1$, $C_3 = c_3$, $B_1 = b_1$, $\widehat{B}_1 = \widehat{b}_1$, we can calculate the probability of a node not succeeding in Phase \upperRomannumeral{3} in two hops as $p_1^{a_1 + c_3}$, as it must fail to connect to $\mathcal{A}_1 \bigcup \mathcal{B}_3$ in the first phase.
Hence we calculate the probability that $C_2 = c_2$ using a binomial distribution with parameter $p_1^{a_1 + c_3}$ as $B(n - a - b_1 - c_3, c_2, p_1^{a_1 + c_3})$.
Given $C_2 = c_2$, $A_1 = a_1$ and $C_3 = c_3$, the probability of a node in $\mathcal{C}_2$ losing connection in the second phase is given by $s_{21}[a_1 + c_3, a_1 + c_3] = (1 - p_2^{a_1 + c_3} )/(1 - p_1^{a_1 + c_3})$. Hence we calculate the probability that $\widecheck{C}_2 = \widecheck{c}_2$ using a binomial distribution with parameter $s_{21}[a_2, a_2] $ as $B(c_2, \widehat{c}_2, s_{21}[a_1 + c_3, a_1 + c_3])$.

Given $C_2 = c_2$, $\widehat{C}_2 = \widehat{c}_2$, $B_1 = b_1$, $\widehat{B}_1 = \widehat{b}_1$ and $A = a$, the probability of a node (not in $\mathcal{A} \bigcup \mathcal{B}_1 \bigcup \mathcal{C}_3 \bigcup \mathcal{C}_3$) failing after three-hops is the probability that it cannot connect to $\widehat{\mathcal{C}}_2 \bigcup \widehat{\mathcal{B}}_1$ in the first phase. This is distributed Bernoulli $p_1^{\widehat{b}_1 + \widehat{c}_2}$, and can be written with Eq.~\eqref{eq:F} as $F(n - a - b - c_2 - c_3, p_1^{\widehat{b}_1 + \widehat{c}_2})$.

Thus we have that given the realization $A = a$, the probability that the protocol fails under case 6: $R_{U_2} > R_{U_1} \geq R_{U_3}$ is given by
$$ P(\text{fail} | \text{Case 6}, A = a) = \left( \sum_{a_2 = 0}^{a} \sum_{b_1 = 0}^{n - a - 1} \sum_{\widehat{b}_1 = 0}^{b_1} \sum_{c_3 = 0}^{n - a - b_1 - 1} \sum_{c_2 = 0}^{n - a - b_1 - c_3 - 1} \sum_{\widehat{c}_2 = 0}^{c_2} P(\text{fail}_6) \right)$$

where
\begin{eqnarray*}
\begin{aligned}
P(\text{fail}_6) &= F\left(n - a - b - c_2 - c_3, p_1^{\widehat{b}_1 + \widehat{c}_2}\right) \times B\left(c_2, \widehat{c}_2, s_{21}[a + c_3, a + c_3]\right) \times B\left(b_1, \widehat{b}_1, s_{21}[a_2, a_2]\right) \times \\
& \hspace{10pt} \times B(n - a - b_1, c_3, q_{31}) \times B(n - a - b - c_3, c_2, p_{1}^{a_1 + c_3})  \times\\
& \hspace{10pt} \times B\left(n - a, b_1, p_1^{a_2}\right) \times B\left(a, a_2, r_{21}\right) \times B(n, a, p_1).
\end{aligned}
\end{eqnarray*}

\end{IEEEproof}
\label{appendix}
\fi

\bibliographystyle{IEEEtran}
{\small{
\bibliography{IEEEabrv,Cow}}

\section*{Acknowledgments}
The authors would like to Venkat Anantharam for useful discussions. We also thank the BLISS \& BWRC students, staff, faculty and industrial sponsors and the NSF for a Graduate Research Fellowship and grants 0932410, 1321155, 1343398, and 144078.

\end{document}